\documentclass[journal]{IEEEtran}

\usepackage{ifpdf}
\usepackage{url}
\usepackage{color}
\ifpdf
\else
\fi
 
\usepackage[shortlabels]{enumitem}

\usepackage{cite}
\ifCLASSINFOpdf
  \usepackage[pdftex]{graphicx}
  \graphicspath{{../pdf/}{../jpeg/}}
  \DeclareGraphicsExtensions{.pdf,.jpeg,.png}
\else
  \usepackage[dvips]{graphicx}
  \graphicspath{{../eps/}}
  \DeclareGraphicsExtensions{.eps}
\fi

\usepackage{amsmath}
\usepackage{amsfonts}
\usepackage{algorithm}
\usepackage{algpseudocode}
\algtext*{EndWhile}
\algtext*{EndIf}
\algtext*{EndFor}

\usepackage{array}
\newcolumntype{P}[1]{>{\centering\arraybackslash}p{#1}}
\newcolumntype{U}[1]{>{\centering\arraybackslash}u{#1}}
\newcommand{\etal}{\textit{et al.}}

\usepackage{units}
\usepackage{siunitx}
\usepackage{multirow}
\usepackage{diagbox}
\usepackage{makecell}

\usepackage{xcolor}
\providecommand{\tabularnewline}{\\}

\usepackage{acro}

\DeclareAcronym{5g}{
short=5G,
long= fifth generation,
}
\DeclareAcronym{6g}{
short=6G,
long= sixth generation,
}
\DeclareAcronym{3d}{
short=3D,
long= three-dimensional,
}
\DeclareAcronym{aod}{
short=AOD,
long= angle-of-departure,
}
\DeclareAcronym{aosa}{
short=AOSA,
long= array-of-subarray,
}
\DeclareAcronym{adod}{
short=ADOD,
long= angle-difference-of-departure,
}
\DeclareAcronym{aoa}{
short=AOA,
long= angle-of-arrival,
}
\DeclareAcronym{adc}{
short=ADC,
long= analog to digital converter,
}
\DeclareAcronym{aeb}{
short=AEB,
long= angle error bound,
}
\DeclareAcronym{av}{
short=AV,
long= autonomous vehicle,
}
\DeclareAcronym{bs}{
short=BS,
long= base station,
}
\DeclareAcronym{bse}{
short=BSE,
long= beam split effect,
}
\DeclareAcronym{csi}{
short=CSI,
long= channel state information,
}
\DeclareAcronym{cfo}{
short=CFO,
long= carrier frequency offset,
}
\DeclareAcronym{ceb}{
short=CEB,
long= clock error bound,
}
\DeclareAcronym{coa}{
short=COA,
long= curvature-of-arrival,
}
\DeclareAcronym{crb}{
short=CRB,
long= Cram\'er-Rao bound,
}
\DeclareAcronym{ccrb}{
short=CCRB,
long= constrained Cram\'er-Rao bound,
}
\DeclareAcronym{cmos}{
short=CMOS,
long= complementary metal-oxide-semiconductor,
}
\DeclareAcronym{crlb}{
short=CRLB,
long= Cram\'er-Rao lower bound,
}
\DeclareAcronym{cdf}{
short=CDF,
long= cumulative distribution function,
}
\DeclareAcronym{cp}{
short=CP,
long= cyclic prefix,
}
\DeclareAcronym{dac}{
short=DAC,
long= digital to analog converter,
}
\DeclareAcronym{dfl}{
short=DFL,
long= device-free localization,
}
\DeclareAcronym{dmimo}{
short=D-MIMO,
long= distributed MIMO,
}
\DeclareAcronym{dlprs}{
short=DL-PRS,
long= downlink positioning reference signal,
}
\DeclareAcronym{d2d}{
short=D2D,
long= device-to-device,
}
\DeclareAcronym{dftsofdm}{
short=DFT-s-OFDM,
long= discrete-Fourier-transform spread OFDM,
}
\DeclareAcronym{dl}{
short=DL,
long= deep learning,
}
\DeclareAcronym{gps}{
short=GPS,
long= global positioning system,
}
\DeclareAcronym{hwi}{
short=HWI,
long= hardware impairment,
}
\DeclareAcronym{hemt}{
short=HEMT,
long= high electron mobility transistor,
}
\DeclareAcronym{hbt}{
short=HBT,
long= heterojunction bipolar transistors,
}
\DeclareAcronym{iot}{
short=IoT,
long= internet of things,
}
\DeclareAcronym{isac}{
short=ISAC,
long= integrated sensing and communication,
}
\DeclareAcronym{iqi}{
short=IQI,
long= in-phase and quadrature imbalance,
}
\DeclareAcronym{ia}{
short=IA,
long= initial access,
}
\DeclareAcronym{kpi}{
short=KPI,
long= key performance indicator,
}
\DeclareAcronym{kf}{
short=KF,
long= Kalman filter,
}
\DeclareAcronym{ekf}{
short=EKF,
long= extended Kalman filter,
}
\DeclareAcronym{ukf}{
short=UKF,
long= unscented Kalman filter,
}
\DeclareAcronym{ckf}{
short=CKF,
long= cubature Kalman filter,
}
\DeclareAcronym{pf}{
short=PF,
long= particle filter,
}
\DeclareAcronym{lb}{
short=LB,
long= lower bound,
}
\DeclareAcronym{lse}{
short=LSE,
long= least-square estimator,
}
\DeclareAcronym{lo}{
short=LO,
long= local oscillator,
}
\DeclareAcronym{mc}{
short=MC,
long= mutual coupling,
}
\DeclareAcronym{mac}{
short=MAC,
long= medium access control,
}
\DeclareAcronym{meb}{
short=MEB,
long= mapping error bound,
}
\DeclareAcronym{ml}{
short=ML,
long= machine learning,
}
\DeclareAcronym{mcrb}{
short=MCRB,
long= misspecified Cram\'er-Rao bound,
}
\DeclareAcronym{mds}{
short=MDS,
long= multidimensional scaling ,
}
\DeclareAcronym{mimo}{
short=MIMO,
long= multiple-input-multiple-output,
}
\DeclareAcronym{mm}{
short=MM,
long= mismatched model,
}
\DeclareAcronym{mpc}{
short=MPC,
long= multipath components,
}
\DeclareAcronym{mmwave}{
short=mmWave,
long= millimeter wave,
}
\DeclareAcronym{mmle}{
short=MMLE,
long= mismatched maximum likelihood estimation,
}
\DeclareAcronym{mems}{
short=MEMS,
long= micro-electro-mechanical system,
}
\DeclareAcronym{mle}{
short=MLE,
long= maximum likelihood estimation,
}
\DeclareAcronym{nlos}{
short=NLOS,
long= none-line-of-sight,
}
\DeclareAcronym{ofdm}{
short=OFDM,
long= orthogonal frequency-division multiplexing,
}
\DeclareAcronym{oeb}{
short=OEB,
long= orientation error bound,
}
\DeclareAcronym{otfs}{
short=OTFS,
long= orthogonal time-frequency space,
}
\DeclareAcronym{pdf}{
short=PDF,
long= probability density function,
}
\DeclareAcronym{papr}{
short=PAPR,
long= peak-to-average-power ratio,
}
\DeclareAcronym{pan}{
short=PAN,
long= power amplifier nonlinearity,
}
\DeclareAcronym{pa}{
short=PA,
long= power amplifier,
}
\DeclareAcronym{ps}{
short=PS,
long= phase shifter,
}
\DeclareAcronym{pn}{
short=PN,
long= phase noise,
}
\DeclareAcronym{poa}{
short=POA,
long= phase-of-arrival,
}
\DeclareAcronym{pwm}{
short=PWM,
long= planar wave model,
}
\DeclareAcronym{pdoa}{
short=PDOA,
long= phase-difference-of-arrival,
}
\DeclareAcronym{prs}{
short=PRS,
long= positioning reference signals,
}
\DeclareAcronym{peb}{
short=PEB,
long= position error bound,
}
\DeclareAcronym{rnn}{
short=RNN,
long= recurrent neural network,
}
\DeclareAcronym{rl}{
short=RL,
long= reinforcement learning,
}
\DeclareAcronym{rfc}{
short=RFC,
long= radio-frequency chain,
}
\DeclareAcronym{rf}{
short=RF,
long= radio frequency,
}
\DeclareAcronym{rfid}{
short=RFID,
long= radio frequency identification,
}
\DeclareAcronym{ris}{
short=RIS,
long= reconfigurable intelligent surface,
}
\DeclareAcronym{rss}{
short=RSS,
long= received signal strength,
}
\DeclareAcronym{rtt}{
short=RTT,
long= round-trip time,
}
\DeclareAcronym{sm}{
short=SM,
long= standard model,
}
\DeclareAcronym{sige}{
short=SiGe,
long= silicon-germanium,
}
\DeclareAcronym{spp}{
short=SPP,
long= surface plasmon polariton,
}
\DeclareAcronym{sa}{
short=SA,
long= subarray,
}
\DeclareAcronym{sota}{
short=SOTA,
long= state-of-the-art,
}
\DeclareAcronym{swm}{
short=SWM,
long= spherical wave model,
}
\DeclareAcronym{slam}{
short=SLAM,
long= simultaneous localization and mapping,
}
\DeclareAcronym{tm}{
short=TM,
long= true model,
}
\DeclareAcronym{toa}{
short=TOA,
long= time-of-arrival,
}
\DeclareAcronym{tof}{
short=TOF,
long= time-of-flight,
}
\DeclareAcronym{tdoa}{
short=TDOA,
long= time-difference-of-arrival,
}
\DeclareAcronym{thz}{
short=THz,
long= terahertz,
}
\DeclareAcronym{ue}{
short=UE,
long= user equipment,
}
\DeclareAcronym{ummimo}{
short=UM-MIMO,
long= ultra-massive multi-input-multi-output,
}
\DeclareAcronym{vlp}{
short=VLP,
long= visible light positioning,
}
\DeclareAcronym{veb}{
short=VEB,
long= velocity error bound,
}
\DeclareAcronym{vlc}{
short=VLC,
long= visible light communication,
}
\DeclareAcronym{ula}{
short=ULA,
long= uniform linear array,
}
\DeclareAcronym{upa}{
short=UPA,
long= uniform planar array,
}
\DeclareAcronym{wlan}{
short=WLAN,
long= wireless local area network,
}

\usepackage{tabu,longtable}
\usepackage{supertabular}

\newcommand{\norm}[1]{\lVert{#1}\rVert}

\usepackage{amsthm}
\newtheorem{remark}{Remark}
\usepackage{tikz}

\ifCLASSOPTIONcompsoc
 \usepackage[caption=false,font=normalsize,labelfont=sf,textfont=sf]{subfig}
\else
 \usepackage[caption=false,font=footnotesize]{subfig}
\fi

\usepackage{stfloats}
\usepackage{amssymb}

\usepackage{hyperref}

\usepackage{esint} 

\long\def\comment#1{}

\DeclareMathOperator*{\argmax}{arg\,max}
\DeclareMathOperator*{\argmin}{arg\,min}

\newfont{\bbb}{msbm10 scaled 700}

\newcommand{\ssr}[1]{{\scriptscriptstyle{\mathrm{#1}}}}
\newcommand{\ssnb}[1]{{\scriptscriptstyle{({#1})}}}

\newcommand{\vthickline}{\vrule width 0.8pt}
\newcommand{\hthickline}{\noalign{\hrule height 0.80pt}}

\newfont{\bb}{msbm10 scaled 1100}

\newcommand{\av}{{\bf a}}

\newcommand{\fv}{{\bf f}}
\newcommand{\gv}{{\bf g}}

\newcommand{\nv}{{\bf n}}
\newcommand{\ov}{{\bf o}}
\newcommand{\pv}{{\bf p}}

\newcommand{\rv}{{\bf r}}
\newcommand{\sv}{{\bf s}}
\newcommand{\tv}{{\bf t}}

\newcommand{\xv}{{\bf x}}
\newcommand{\yv}{{\bf y}}

\newcommand{\Am}{{\bf A}}
\newcommand{\Bm}{{\bf B}}

\newcommand{\Hm}{{\bf H}}

\newcommand{\Jm}{{\bf J}}

\newcommand{\Mm}{{\bf M}}

\newcommand{\Rm}{{\bf R}}

\newcommand{\Wm}{{\bf W}}

\newcommand{\Ym}{{\bf Y}}

\newcommand{\Acal}{{\cal A}}

\newcommand{\Gc}{{\cal G}}

\newcommand{\gammav}{\hbox{\boldmath$\gamma$}}

\newcommand{\etav}{\hbox{\boldmath$\eta$}}

\newcommand{\muv}{\hbox{\boldmath$\mu$}}

\newcommand{\thetav}{\hbox{\boldmath$\theta$}}
\newcommand{\tauv}{\hbox{\boldmath$\tau$}}
\newcommand{\omegav}{\hbox{\boldmath$\omega$}}
\newcommand{\xiv}{\hbox{\boldmath$\xi$}}

\newcommand{\rhov}{\hbox{\boldmath$\rho$}}

\newcommand{\vpv}{\boldsymbol{\varphi}}

\newcommand{\Sigmam}{\hbox{\boldmath$\Sigma$}}

\newcommand{\Omegam}{\hbox{\boldmath$\Omega$}}

\newcommand{\diag}{{\hbox{diag}}}

\newcommand{\trace}{{\hbox{tr}}}

\newcommand{\Abc}{\mbox{$\boldsymbol{\mathcal{A}}$}}

\newcommand{\Hbc}{\mbox{$\boldsymbol{\mathcal{H}}$}}

\makeatletter
\renewcommand\@dotsep{140}   
\makeatother

\begin{document}
\title{A Tutorial on Terahertz-Band Localization for 6G Communication Systems}

\author{Hui~Chen,~\IEEEmembership{Member,~IEEE}, Hadi~Sarieddeen,~\IEEEmembership{Member,~IEEE}, Tarig~Ballal,~\IEEEmembership{Member,~IEEE},
Henk~Wymeersch,~\IEEEmembership{Senior Member,~IEEE},
Mohamed-Slim~Alouini,~\IEEEmembership{Fellow,~IEEE}, and~Tareq~Y.~Al-Naffouri,~\IEEEmembership{Senior Member,~IEEE}
\thanks{H.~Chen and H.~Wymeersch are with the Department of Electrical Engineering, Chalmers University of Technology, 41296 Gothenburg, Sweden (e-mail: \{hui.chen; henkw\}@chalmers.se). H.~Sarieddeen, T.~Ballal, M.~S.~Alouini and~T.~Y.~Al-Naffouri are with the Division of Computer, Electrical and Mathematical Science \& Engineering, King Abdullah University of Science and Technology (KAUST), Thuwal, 23955-6900, KSA (e-mail: \{hadi.sarieddeen; tarig.ahmed; slim.alouini; tareq.alnaffouri\}@kaust.edu.sa). Most of this work is done during the Ph.D. study of H. Chen at KAUST.} 
\thanks{This work was supported by the King Abdullah University
of Science and Technology (KAUST) Office of Sponsored Research (OSR)
under Award ORA-CRG2021-4695. The work of Henk Wymeersch was supported by the European Commission through the H2020 Project Hexa-X under
Grant 101015956.}
}

 

\maketitle
\begin{abstract}
Terahertz (THz) communications are celebrated as key enablers for converged localization and sensing in future sixth-generation (6G) wireless communication systems and beyond. Instead of being a byproduct of the communication system, localization in 6G is indispensable for location-aware communications. Towards this end, we aim to identify the prospects, challenges, and requirements of THz localization techniques. We first review the history and trends of localization methods and discuss their objectives, constraints, and applications in contemporary communication systems. We then detail the latest advances in THz communications and introduce THz-specific channel and system models. Afterward, we formulate THz-band localization as a 3D position/orientation estimation problem, detailing geometry-based localization techniques and describing potential THz localization and sensing extensions. We further formulate the offline design and online optimization of THz localization systems, provide numerical simulation results, and conclude by providing {lessons learned and future research directions.} {Preliminary results illustrate that under the same transmission power and array footprint, THz-based localization outperforms {millimeter wave}-based localization. In other words, the same level of localization performance can be achieved at THz-band with less transmission power or a smaller footprint}.


\end{abstract}


\IEEEpeerreviewmaketitle

\section{Introduction}
\label{sec:introduction}
Localization is the process of estimating the {position and orientation of a target}, which is vital for a variety of applications, including location-aware communications~\cite{di2014location}, autonomous driving~\cite{bresson2017simultaneous}, industrial \ac{iot}~\cite{lohan2018benefits}, and tactile internet~\cite{antonakoglou2018toward}. Over the years, a plethora of localization techniques has been proposed. These techniques utilize different signal or measurement types that include ultrasound, visible light, \ac{rf}, inertial measurements, and hybrid signals~\cite{laoudias2018survey}. Among these modalities, \ac{rf} signals are widely used because of their ubiquity in current wireless communication systems, where abundant cellular and \ac{wlan} infrastructures provide added value to user-oriented services and network management~\cite{del2017survey}.


\subsection{Location Information: From a By-Product to an Enabler}
\label{sec:location_information_enabler}
The problem of location estimation within a communication system has been under investigation since the first generation of wireless mobile technology. However, more attention was drawn to accurate localization after the admission of the U.S. Federal Communications Commission enhanced 911 (FCC-E911) rules~\cite{del2017survey}. Furthermore, with the introduction of the \ac{gps} and the standardization of cellular communication systems, we now benefit from the accuracy of $\sim\!\!\unit[10]{cm}$ in rural areas and $\sim\! \!\unit[1]{m}$ in outdoor urban environments~\cite{del2017survey}.

In indoor environments, obtaining accurate position information from cellular networks and \ac{gps} is challenging. Consequently, WiFi- and Bluetooth-based localization methods are developed to tackle complex indoor environments, where multipath signal components and signal blockage degrade the localization accuracy. By adopting ultra-wide bandwidths, multipath signals become resolvable, which improves performance~\cite{alarifi2016ultra}. {Nevertheless, the main drawback of short-range coverage} still needs to be tackled~\cite{xiao2022overview}. Furthermore, moving to the $\unit[400-790]{THz}$ range of the electromagnetic spectrum, \ac{vlp} has been exploited for localization in \ac{vlc} systems. A corresponding assortment of systems and algorithms have been developed~\cite{keskin2018localization,zhuang2018survey}. However, issues such as blockage, limited power control, and sensitivity to the environment make \ac{vlp} system deployments challenging.

{The above-mentioned localization methods are viewed as by-products of data transmission in a communication system. However, in the \ac{5g} of wireless communication systems, localization is remodeled from location-based services to location-aware communications. Location information can thus reduce the latency and enhance the scalability and robustness of 5G communication systems~\cite{di2014location}. This trend continues towards future \ac{6g} systems, where localization and communication need to be integrated to achieve ubiquitous connectivity, high data rates, and low latency over a \ac{3d} network coverage~\cite{xiao2022overview}.}

\subsection{The Importance of Localization-Communication Interaction in the THz Band}
\label{sec:the_importance_of_loc_com_interaction}
With the increasing demands for higher data rates--up to 1 terabit-per-second (Tb/s) in 2030--the \ac{thz} frequency band ($0.1$-$\unit[10]{THz}$) is receiving noticeable attention from the research community as an ideal enabler for applications involving high-speed transmissions~\cite{elayan2019terahertz,sarieddeen2020next,akyildiz2014terahertz}. The first standard for sub-THz frequencies is already proposed in IEEE 802.15.3d, where the signal frequency/bandwidth is pushed from $\unit[73]{GHz}$/$\unit[2]{GHz}$ (5G New Radio) to $\unit[300]{GHz}$/$\unit[69]{GHz}$~\cite{petrov2020ieee}.
However, the downside of operating at high frequencies, such as the THz band, is that propagation losses increase quadratically with the carrier frequency. Molecular absorption losses, mainly due to water vapor, also affect signal propagation~\cite{akyildiz2018combating}. Potential solutions to overcome such losses are distance-aware resource allocation, beamforming via ultramassive MIMO (UM-MIMO) structures~\cite{akyildiz2018combating,lin2016terahertz}, and \acp{ris}~\cite{wu2021intelligent, basar2019wireless}.
{For distance-/angle-aware resource allocation and beamforming optimization, location information is crucial. As for RISs, the optimization of RIS coefficients also requires location information to allow passive devices\footnote{There are works that exploit the potential of active RISs~\cite{zhang2021active} and hybrid RISs~\cite{schroeder2021passive}; however, only passive RISs are discussed in this tutorial.} to control the amplitude and phase of incident signals~\cite{wu2021intelligent}, thus reshaping the channel~\cite{wu2018intelligent} in a low-complexity, energy-efficient manner.}
Therefore, localization is a prerequisite for efficient THz communications.

Conversely, the mentioned approaches (e.g., UM-MIMO, RIS) also contribute to the {localization accuracy}. The narrow beamwidth through UM-MIMO beamforming provides high angular resolution, while the wide bandwidth yields accurate delay estimation. In addition, RISs not only increase the received signal strength but also work as passive anchors providing geometrical diversity. {The abundant resources in future communication systems can be exploited for localization. From an information-sharing point of view, high-precision localization can be achieved through cooperative localization~\cite{kim2020cooperative,wymeersch2009cooperative} inside a network where \ac{d2d} communications are supported.} Furthermore, maintaining the relative positions of different neighbor devices is beneficial for efficient tracking and link re-establishment~\cite{khan2021intelligent}. As a consequence, various applications that demand high data rates and high localization accuracy that current communication systems fail to support can be satisfactorily met with the interaction between localization and communication, especially in the THz band.

\begin{table*}[t]
\scriptsize
    \caption{{Summary of Recent THz Systems and Localization Surveys/Tutorials}}
    \centering
    \renewcommand{\arraystretch}{1.2}
    \begin{tabular}{c !\vthickline c | c | m{12cm} | c }
    \hthickline
    & \textbf{Year} & \textbf{Ref} & \centering{\textbf{Main Topics}} & {\textbf{Type}}
    \\
    \hthickline
        \multirow{12}{*}{\rotatebox{90}{THz Systems\ }} 
        & 2019 & \makecell{Chen \\\etal \cite{chen2019survey}} 
        & {Reviews on the development towards THz communications and presents key technologies faced in THz wireless communication systems. Discusses potential application scenarios and technical challenges.} 
        & \makecell{THz Comm.,\\ Survey}
        \\ \cline{2-5}
        & 2020 & \makecell{Ghafoor \\\etal \cite{ghafoor2020mac}} 
        & {Surveys on terahertz \ac{mac} protocols and discusses different applications at macro- and nano- scales. Highlights the design requirements, issues, considerations, challenges, and research directions.} 
        & \makecell{\ac{mac} protocols.,\\ Survey}
        \\ \cline{2-5}
        & 2021 & \makecell{Sarieddeen \\\etal \cite{sarieddeen2021overview}} 
        & {Provides an overview of recent advances in signal processing techniques for terahertz communications, with a focus on waveform design and modulation, channel estimation, channel coding, and data detection. Motivates signal processing techniques for THz sensing and localization.} & \makecell{THz signal \\processing,\\ Tutorial}
        \\ \cline{2-5}
        & 2021 & \makecell{Han \\\etal \cite{han2021terahertz}} 
        & {Surveys on the measurement, modeling and analysis of THz wireless channels. Elaborates on open problems and future directions for 6G THz channels.} & \makecell{THz channels.,\\ Surveys}
        \\ \cline{2-5}
        & 2021 & \makecell{Lemic \\\etal \cite{lemic2021survey}} 
        & {Provides an overview of the current THz nanocommunication and nanonetworking research with the topics on supported applications, protocol for different layers, channel models, and experimentation tools.} 
        & \makecell{Nano-Comm.,\\ Survey}
        \\  \cline{2-5}
        & 2021 & \makecell{Wang \\\etal \cite{wang2021key}} & {Surveys on key technologies in 6G THz wireless communication systems covering channel modeling, multi-beam antenna design, front-end chip design, baseband signal processing algorithms, and resource management schemes.} 
        & \makecell{THz Comm.,\\ Survey}
        \\  \cline{2-5}
        & 2022 & \makecell{Chaccour \\\etal \cite{chaccour2022seven}} 
        & {Investigates seven unique defining features of THz wireless systems: 1) Quasi-opticality of the band, 2) THz- tailored wireless architectures, 3) Synergy with lower frequency bands, 4) Joint sensing and communication systems, 5) PHY- layer procedures, 6) Spectrum access techniques, and 7) Real-time network optimization.} 
        & \makecell{THz Comm. \\features, \\Survey}
        \\
        \hthickline
        \multirow{30}{*}{\rotatebox{90}{Localization\ }} 
        & 2017 & \makecell{Bresson \\\etal \cite{bresson2017simultaneous}} 
        & {Surveys on the \ac{slam} techniques and different paradigms in autonomous driving. Overviews the various large-scale experiments and discusses remaining challenges and future directions.} 
        & \makecell{SLAM, \\Survey}
        \\ \cline{2-5}
        & 2017 & \makecell{Ferreira\\ \etal \cite{ferreira2017localization}} 
        & {Describes the requirement, localization techniques and methods for indoor positioning systems specifically developed for emergency response scenarios. Reviews existing system schemes, performance and discusses future directions.} 
        & \makecell{Indoor Loc.,\\ Survey}
        \\
        \cline{2-5}
        & 2018 & \makecell{Kuutti\\ \etal \cite{kuutti2018survey}} 
        & {Evaluates the \ac{sota} vehicle localization techniques and investigates their applicability to autonomous vehicles. Discusses the benefits of vehicle-to-everything communications, in addition to vehicle sensory information.} 
        & \makecell{Vehicle Loc.,\\ Survey}
        \\
        \cline{2-5}
        & 2018 & \makecell{Laoudias\\ \etal \cite{laoudias2018survey}} 
        & {Provides current enabling technologies for localization in cellular systems and wireless local area networks. Overviews the research works for \ac{iot} and mobile scenarios and highlights future research directions.} 
        & \makecell{Network Loc.,\\Survey}
        \\
        \cline{2-5}
        & 2018 & \makecell{Peral-Rosado\\ \etal \cite{del2017survey}} 
        & {Overviews the standardized localization methods from the first to the fourth generation of cellular systems. Outlines the new research trends on \ac{5g} positioning, and the lessons learned from previous generations.} 
        & \makecell{Cellular Loc.,\\Survey}
        \\ 
        \cline{2-5}
        & 2018 & \makecell{Keskin \\\etal \cite{keskin2018localization}} 
        & {Considers the \ac{vlp} and discusses localization techniques, algorithms, system architectures and resource allocation problems.} 
        & \makecell{\ac{vlp},\\ Tutorial}
        \\
        \cline{2-5}
        & 2019 & \makecell{Shit\\ \etal \cite{shit2019ubiquitous}} 
        & {Reviews and classifies \ac{dfl} technologies. Discusses lessons learned, applications and presents current trends and future research directions.} 
        & \makecell{\ac{dfl},\\Survey}
        \\
        \cline{2-5}
        & 2019 & \makecell{Zafari \\\etal \cite{zafari2019survey}} 
        & {Surveys recent indoor localization systems with different localization techniques and radio technologies. Provides evaluation of system performance, and discusses remaining challenges for accurate indoor localization.} 
        & \makecell{Indoor Loc.,\\ Survey}
        \\
        \cline{2-5}
        & 2019 & \makecell{Saeed\\ \etal \cite{saeed2019state}} 
        & {Surveys on \ac{mds} and \ac{mds}-based localization techniques. Discusses centralized, semi-centralized and distributed methods for \ac{iot}, cognitive radio networks, and 5G networks scenarios.} 
        & \makecell{\ac{mds},\\ Survey}
        \\
        \cline{2-5}
        & 2019 & \makecell{Wen \\\etal \cite{wen2019survey}} 
        & {An overview of channel parameter estimation algorithms (subspace and compressed sensing methods), \ac{sota} localization techniques, challenges, and opportunities in the field of massive MIMO localization.} 
        & \makecell{MIMO Loc.,\\ Survey}
        \\
        \cline{2-5}
        & 2020 & \makecell{Burghal\\ \etal \cite{burghal2020comprehensive}} 
        & {Surveys ML-based localization using radio frequency signals with an emphasis on the system architectures, \ac{rf} features, ML methods, and data acquisitions.} 
        & \makecell{ML-based Loc., \\Survey}
        \\
        \cline{2-5}
        & 2020 & \makecell{Zhu\\ \etal \cite{zhu2020indoor}} 
        & {Surveys on recent indoor localization technologies and systems based on machine learning and intelligent algorithms. Summarizes and compares the \ac{sota} systems. Discusses existing challenges and potential solutions.} 
        & \makecell{Indoor, ML,\\Survey}
        \\
        \cline{2-5}
        & 2021 & \makecell{Miram\`a\\ \etal \cite{mirama2021survey}} 
        & {Surveys on the \ac{sota} ML techniques that have been adopted over the last ten years to improve the performance of pedestrian localization systems. Highlights existing issues, challenges, and possible future directions.}
        & \makecell{Pedestrain, ML,\\Survey} 
        \\
        \cline{2-5}
        & 2021 & \makecell{Motroni \\\etal \cite{motroni2021survey}} 
        & {Presents a \ac{sota} analysis on the \ac{rfid}-based technology methods under the scenario of indoor vehicle localization.} 
        & \makecell{RFID, \\Survey}
        \\
        \cline{2-5}
        & 2021 & \makecell{De Lima\\ \etal \cite{de2021convergent}} 
        & {Identifies key enabling technologies, applications, and opportunities for \ac{6g} localization. Research challenges and open questions are listed to achieve a convergent communication, sensing and localization system.} 
        & \makecell{6G Loc.,\\Survey}
        \\
        \cline{2-5}
        & 2021 & \makecell{Kanhere\\ \etal \cite{kanhere2021position}} 
        & {Describes how cm-level localization accuracy can be achieved with the use of map-based techniques and shows the potential of data fusion, machine learning and cooperative localization techniques.}  
        & \makecell{Map-based, \\ Survey}
        \\
        \cline{2-5}
        & 2022 & \makecell{Xiao \\\etal \cite{xiao2022overview}} 
        & {Surveys on wireless localization basics and \ac{sota} results, outlines promising future research directions for integrated localization and communication systems.} 
        & Tutorial
        \\ 
        \cline{2-5}
        & 2022 & \makecell{Laconte \\\etal \cite{laconte2022survey}} 
        & {Surveys on localization methods for autonomous vehicles in highway scenarios. Presents the \ac{sota} methods for main components (road inferring, position estimation, lane assessment) with the discussions of advantages and drawbacks.} 
        & \makecell{Highway Loc.,\\Survey}
        \\
        \cline{2-5}
        & 2022 & \makecell{Liu \\\etal \cite{liu2022survey}}
        & {Surveys on the current research progress (e.g., systematic classification, performance metrics and bounds, open problems and future directions) on the fundamental limits of \ac{isac}.} 
        & \makecell{\ac{isac},\\ Survey}
        \\
        \hthickline
        &  & This work 
        & {Overviews the localization basics, provides technical details on channel modeling (including \ac{ris}, \ac{aosa}, \ac{swm}), performance analysis, localization algorithms and system optimization. Highlights lessons learned and future directions.} 
        & \makecell{THz Loc.,\\ Tutorial}
        \\
        \hline
    \hthickline
    \end{tabular}
    \vspace{0.1 cm}
    \scriptsize{\raggedright \\ *Notes: {`Loc.' is short for `localization', `Comm.' is short for `communication'.} \par}
    \renewcommand{\arraystretch}{1}
    \label{tab:summary_of_surveys}
\end{table*}


\begin{figure}[t]
\centering
\includegraphics[width = 0.9\linewidth]{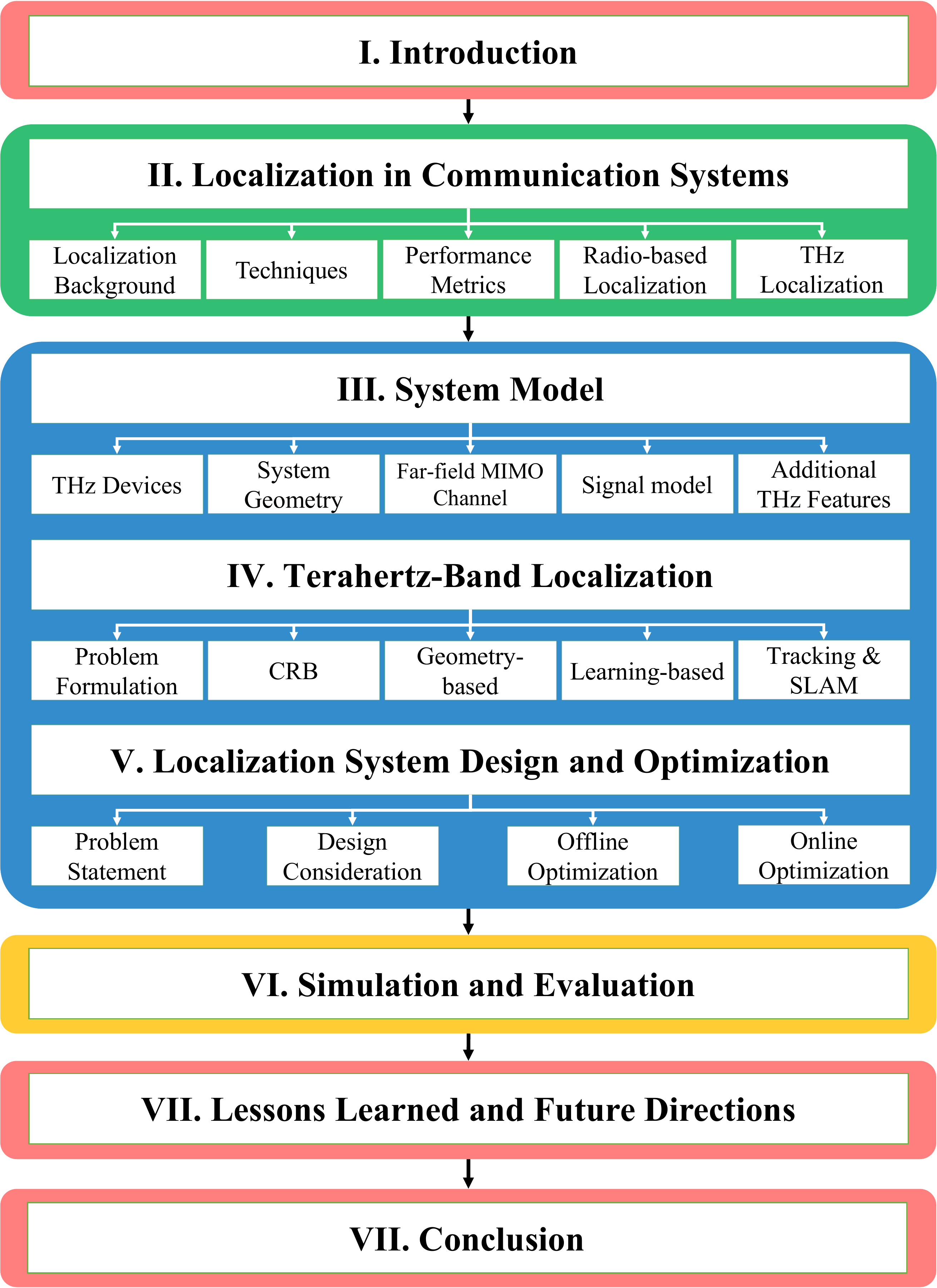}
\caption{{The overall structure of the tutorial.}}
\label{fig:structure_diagram}
\end{figure}

\subsection{Motivation and Structure of This Work}
\label{sec:motivation_and_structure}
As research on 6G wireless communications is rising~\cite{dang2020should,sarieddeen2021overview}, 
a few tutorials on 6G systems have already been published~\cite{de2021convergent, akyildiz20206g}. The 6G literature covers topics such as the role of machine-learning methods for communications~\cite{ali20206g}, broadband connectivity~\cite{rajatheva2020scoring}, and integrated localization and sensing~\cite{xiao2022overview}. {Terahertz systems, being one of the most important enablers for 6G communication, the surveys considering the key technologies~\cite{chen2019survey,wang2021key}, signal processing techniques~\cite{sarieddeen2021overview}, channel models~\cite{han2021terahertz}, nano-communications~\cite{lemic2021survey}, defining features~\cite{chaccour2022seven}, and MAC protocols~\cite{ghafoor2020mac} are available.

In terms of localization, a number of surveys exist and share the localization basics and performance metrics in common. However, their goals are totally different and their main focuses can be categorized based on the environment (indoor\cite{ferreira2017localization, zafari2019survey, zhu2020indoor, motroni2021survey}, outdoor~\cite{bresson2017simultaneous, kuutti2018survey, laconte2022survey} or both), techniques (\ac{slam}~\cite{bresson2017simultaneous}, \ac{mds}~\cite{saeed2019state}, \ac{ml}~\cite{burghal2020comprehensive, zhu2020indoor, mirama2021survey}, etc.), and signal types (radio signal~\cite{laoudias2018survey,del2017survey}, visible light~\cite{keskin2018localization}, \ac{rfid}~\cite{motroni2021survey}, etc.). The taxonomy of localization techniques will be detailed in Section~\ref{sec:localization_systems}. From the application point of view, the localization systems for autonomous driving~\cite{bresson2017simultaneous, kuutti2018survey, laconte2022survey}, emergency response~\cite{ferreira2017localization}, network tracking~\cite{laoudias2018survey}, device-free localization~\cite{shit2019ubiquitous}, and pedestrian localization~\cite{mirama2021survey} are surveyed. More recent works outlook the potential enablers in the B5G and 6G systems~\cite{de2021convergent}, show the localization potential with map-based techniques~\cite{kanhere2021position}, describe basic localization algorithms for the \ac{6g} systems~\cite{xiao2022overview}, and explore the fundamental limits of \ac{isac}. A summary of recent THz systems-related and localization-related surveys is shown in Table~\ref{tab:summary_of_surveys}. However, a comprehensive tutorial on high-frequency signal localization in OFDM-based MIMO systems, which is expected to be one of the main scenarios, is still lacking.
}

{
Unlike the aforementioned works, this tutorial is distinguished by providing a detailed system modeling, performance analysis, algorithms, and the corresponding system optimization formulations specific to the localization problem in the \ac{thz}-band. 
}
The aim of this work is to investigate the potential of THz localization and how it can be leveraged in future communication systems. Towards this end, we seek answers to the following questions:
\begin{enumerate}
    \item {What are the limitations of current communication systems (including 5G) concerning localization? What are the localization \ac{kpi} for 6G and the corresponding challenges?}
    \item What are the key properties of THz signals and systems? How can they be utilized for localization purposes?
    \item How to formulate RIS-assisted localization and sensing problems under different system assumptions?
    \item How do offline design and online optimization of a localization system achieve the desired objectives (e.g., position accuracy $\le 1$cm) under specific performance constraints (e.g., energy consumption)? 
    \item What are the key applications of THz localization? What are the main future directions?
\end{enumerate}

The rest of this paper is organized as follows. Section~\ref{sec:literature_review} reviews the general localization literature, with an emphasis on localization systems using electromagnetic signals. Section~\ref{sec:thz_system_model_and_properties} describes THz system and channel models, highlighting the realization of THz-specific features. THz localization techniques are then detailed in Section~\ref{sec:thz_signal_based_localization}, while optimized localization is investigated in Section~\ref{sec:system_design_and_optimization}. {Next, the simulation results are shown in
Section~\ref{sec:simulation_and_evaluation}, followed by the lessons learned and most prominent future research directions in Section~\ref{sec:future_directions}.} Finally, we draw concluding remarks in Section~\ref{sec:conclusion}. {The sections and main topics of this tutorial is shown in Fig.~\ref{fig:structure_diagram}, and the definitions of frequently-used abbreviations are summarized in the Abbreviations at the end of this work.}


\textit{Notations and Symbols:} Italic letters denote scalars (e.g. $a$), bold lower-case letters denote vectors (e.g. $\av$), and bold upper-case letters denote matrices (e.g. $\Am$). $(\cdot)^T$, $(\cdot)^H$, $(\cdot)^{-1}$, $\trace(\cdot)$, and $\norm{\cdot}$ represent the transpose, Hermitian transpose, inverse, trace, and $\ell$-2 norm operations, respectively; $\Am\!\odot\! \Bm$ is the Hadamard product of two matrices; $[\cdot,\ \cdot,\ \cdots, \cdot]^T\!\!=\!\! [\cdot;\ \cdot;\ \cdots; \cdot]$ denotes a column vector.



\section{Localization in Communication Systems}
\label{sec:literature_review}
{In this section, we briefly review the {localization systems} and describe localization techniques with an emphasis on geometry-based methods. We discuss {localization performance metrics} and {current localization systems} based on electromagnetic waves. Furthermore, we compare {\ac{thz} localizatition} to \ac{mmwave} localization and discuss their merits and challenges in different aspects.}

\subsection{Localization: Definition and System Taxonomy}
\label{sec:localization_systems}

The localization problem can be defined as estimating the position and orientation (with antenna arrays) of a \ac{ue} with the assist of one or multiple anchor \acp{bs} (with known position and orientation information).
{More specifically, a \ac{ue} can send (\textit{uplink}) or receive (\textit{downlink}) known pilot signals to or from a BS. The received signals are distorted by the propagation channel, which is determined by the \ac{bs}/\ac{ue} states (position and orientation) and the environment (signals can be reflected by a wall, a \ac{ris} or an object)\footnote{We call the path that a signal propagates directly from a \ac{ue} to a \ac{bs} the \textit{LOS path}; the path that a signal is reflected by a \ac{ris} is called as \textit{RIS path}. The \textit{\ac{nlos} paths} are created by either a large plane (e.g., a wall, forming a \textit{reflected path}) or an object (forming a \textit{scattered path}). The detailed channel models for different paths (LOS, RIS, and NLOS) will be discussed in Section~\ref{sec:thz_system_model_and_properties}.}.
Based on the knowledge of the pilot signals and a proper signal model, the channel can be estimated and the channel parameters (\ac{aoa}, \ac{aod} and delay) of each path that signal propagates can be extracted. Finally, \ac{ue} position and orientation can be estimated\footnote{Usually, the estimation of \ac{ue} states is called \textit{localization}, and the estimation of incidence points of the \ac{nlos} path is called \textit{mapping}.} with its relative geometry relationships with known reference anchors (e.g., \acp{bs} and \acp{ris}). More details can be found in Section~\ref{sec:localization_techniques}.} 
In this work, we focus on the uplink scenario where extensions to other scenarios are straightforward. 

A variety of localization systems have been developed in various application contexts, such as cellular networks~\cite{del2017survey, laoudias2018survey}, indoor scenarios~\cite{zafari2019survey}, 5G massive MIMO systems~\cite{wen2019survey}, and visible light systems~\cite{luo2017indoor,keskin2018localization}. These systems can be categorized based on the application scenario, wireless technology, localization technique, processed signals type, functionality, system structure, position information, information-sharing, among others. A summary of system classifications is listed in Table.~\ref{tab:localization_systems}. We next compare geometry-based and learning-based localization techniques and detail the geometry-based position/orientation estimation.

\begin{table}[t]
\footnotesize
    \caption{Taxonomy of Localization Techniques}
    \centering
    \renewcommand{\arraystretch}{1.0}
    \begin{tabular}{c !\vthickline c}
    \hthickline
    \textbf{Criteria} & \textbf{Types} \\
        \hthickline
        Application Scenario & Outdoor, Indoor\\ 
        \hline
        Wireless Technology & GPS, Cellular systems, WLAN, WiFi\\ 
        \hline
        Localization Technique & Geometry-based, Learning-based \\ 
        \hline
        Signal Type & Radio waves, LED signal, LIDAR\\ 
        \hline
        Functionality & Passive, Active \\  
        \hline
        System Structure &   Centralized, Distributed, Clustered \\ \hline
        Position Information & Absolute position, Relative position\\  \hline
        Information-sharing & Cooperative,  Non-cooperative \\   
    \hthickline
    \end{tabular}
    \renewcommand{\arraystretch}{1}
    \label{tab:localization_systems}
\end{table}


\subsection{Localization Techniques}
\label{sec:localization_techniques}
\subsubsection{Geometry-based and Learning-based Localization}
Geometry-based localization is widely used in current communication systems. Techniques such as \ac{toa}\footnote{TOA is equivalent to \ac{tof} when taking the time of transmission as a reference; we use TOA in the rest of this work.}, \ac{tdoa}, and \ac{aoa} are based on measuring the distance or angle of a \ac{ue} with respect to multiple BSs with known positions~\cite{kanhere2021position, chen2020air, ma2021maximum,chen2021doa}. 
By adopting trilateration or triangulation algorithms, the position of UEs can be calculated from these measurements. Furthermore, \ac{aod} estimation can also be obtained when an antenna array is implemented at the UE. \ac{aod} information can be used to assist \ac{adod}-based localization and estimate orientation alongside estimated positions. Other geometry information such as \ac{poa}~\cite{scherhaufl2013phase} and \ac{pdoa}~\cite{chen2020joint} can be treated as \ac{toa}/\ac{tdoa} with ambiguities, which we do not discuss here.
Estimating the position usually involves formulating an objective function that contains geometric information and solving an optimization problem with geometric constraints. Techniques in geometry-based localization are training-free, easy to analyze theoretically, and scalable to different environments.

For more complex scenarios with many non-resolvable \ac{nlos} paths, the geometry information cannot be explicitly modeled~\cite{kanhere2021position}, and {learning-based methods} are preferred. 
Machine learning is the study of computer algorithms that improve automatically through experience~\cite{mitchell1997machine}. Contrary to the practical algorithms designed for geometry-based localization, \ac{ml}-based methods require offline training. Such an offline process can sufficiently reduce online computations. However, a large volume of data from wideband \ac{mimo} systems needs to be collected for training purposes, and the trained models have to be updated to adapt to environmental variations. Typical ML algorithms and techniques such as random forest~\cite{liaw2002classification}, reinforcement learning~\cite{sutton2018reinforcement}, and deep neural networks~\cite{goodfellow2016deep} are potential solutions to maintain good \ac{kpi} network service levels~\cite{ali20206g}.

Considering the sparsity of \ac{thz} channels, we focus on geometry-based position/orientation estimation. In addition, direct extensions to tracking, and \ac{slam} are possible, as we discuss in Sec.~\ref{sec:thz_signal_based_localization}.


\begin{figure*}[t]
\centering
\includegraphics[width = 0.88\linewidth]{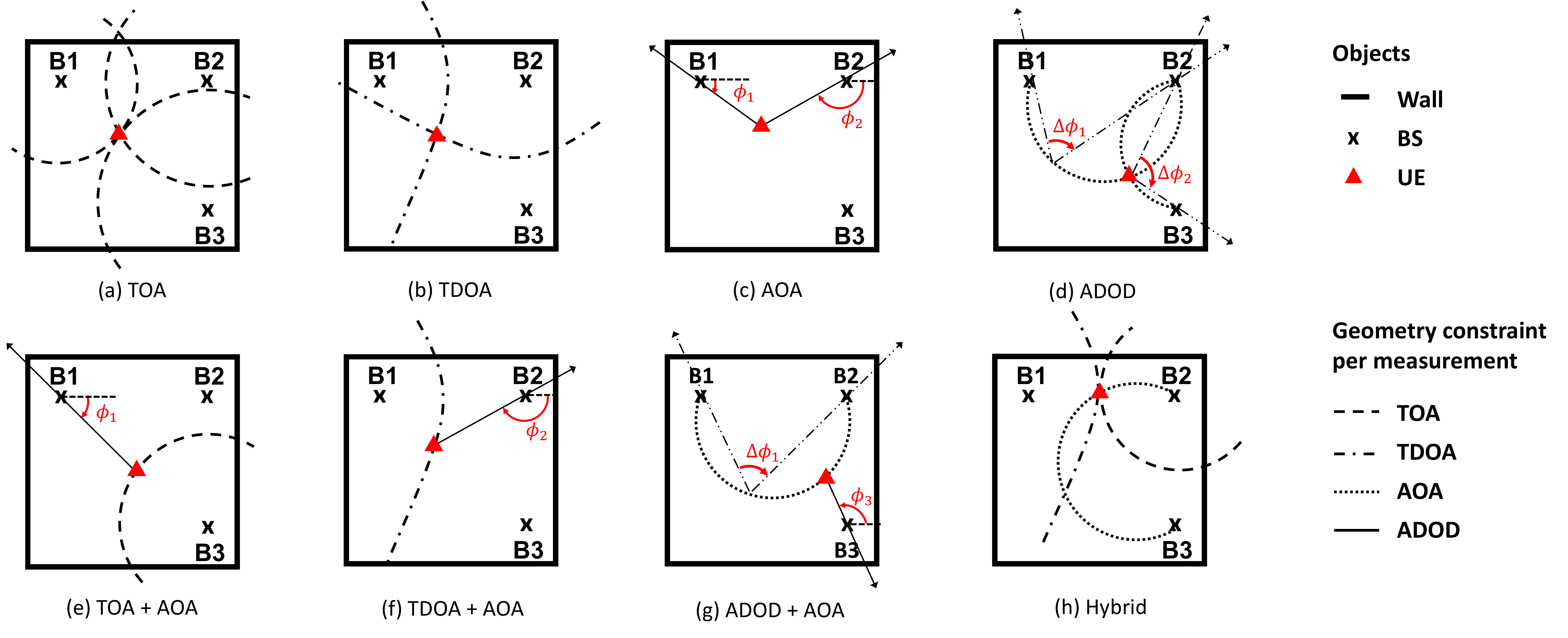}
\caption{Position estimation from different geometry information (uplink): (a)-(d) show positioning techniques utilizing four basic geometry measurements, namely, TOA, TDOA, AOA, and ADOD; (e)-(h) illustrate positioning utilizing {combinations} of geometry information.}
\label{fig:position_estimation}
\end{figure*}

\subsubsection{Position Estimation}

\begin{table}[t]
\renewcommand{\arraystretch}{1.2}
\scriptsize
    \caption{Summary of Geometry-based Localization (Uplink)}
    \centering
    \begin{tabular}{m{15 mm}|c
    !\vthickline
    m{12.3 mm}|m{10.5 mm}|m{5.7 mm}|c}
    \hthickline
        \multicolumn{2}{c !\vthickline}{} 
        &\centering \!\textbf{TOA} &\centering \!\textbf{TDOA} &\!\textbf{AOA} & \textbf{ADOD}\\ 
    \hthickline
        \!\!\!\! \#. of BS needed per Meas. & & \centering 1 & \centering 2 & \centering 1 & {2} \\
        \hline
        \centering
        \multirow{3}{*}{\!\shortstack{Geometric\\ Constraint\\ per Meas.}}
        &\centering 2D & \centering \!Circle & \centering \!Hyperbola & \centering \!\!Line & Arc\\ \cline{2-6}
        &\centering 3D & \centering \!Sphere & \!\!Hyperboloid & \centering \!\!Line & \!\!\shortstack{Surface of arc\\ revolution}\\ 
        \hline
        \multirow{2}{*}{\!\shortstack{\#. of Meas. for \\Pos. Estimation}}
        &2D & \centering 3 & \centering 3 & \centering 2 & 3\\ \cline{2-6}
        &3D & \centering 4 & \centering 4 & \centering 2 & 4\\ 
        \hline
        \multirow{3}{*}{\!\shortstack{\#. of AOD for \\Ori. Estimation}}
        & 1D & \multicolumn{4}{c}{1 (2D position needed)} \\ 
        \cline{2-6}
        & 2D & \multicolumn{4}{c}{1 (3D position needed)} \\ 
        \cline{2-6}
        & 3D &  \multicolumn{4}{c}{2 (3D position needed)} \\ 
        \hline
        \centering{System \\Requirements} &  & \centering \!{System Syn/ \\ RTT/RSS} & \!\centering {BS-BS \\ Syn} & \centering {Array \\ at BS} & {Array at UE} \\
    \hthickline
    \end{tabular}
    \renewcommand{\arraystretch}{1}
    \label{tab:geometry_based_localization}
\end{table}


Position estimation problems can be categorized as 2D and 3D. For better visualization, we illustrate the estimation of 2D position from geometry information in different uplink scenarios in Fig.~\ref{fig:position_estimation}; {four types of basic geometry information used for localization are covered}:
\begin{enumerate}[a)]
    \item \ac{toa}: The signal propagation distance can be estimated with known channel information from the \ac{rss}, which can be used for \ac{toa}-based methods. However, estimations from RSS are usually inaccurate. Alternatively, \ac{rtt} estimation by stamping the transmitting and receiving times of the signal gives the TOA information. If the system is well-synchronized, TOA can be directly inferred from the signals with a bandwidth-dependent resolution.
    
    \item \ac{tdoa}: {If only the \acp{bs} are synchronized, the estimated TOA is no longer accurate, which contains a clock offset. In this case, a reference \ac{bs} can be utilized to obtain the TDOA of the signal at other \acp{bs}.}

    \item \ac{aoa}: {Additional angle information can be obtained if an antenna array is available at the BS. The AOA (1D azimuth or 2D azimuth \& elevation depending on the dimension localization scenarios) can be used to locate the target by getting the intersection of lines.}
    
    \item \ac{aod}: When a \ac{ue} is equipped with an antenna array, the signal AOD to a specific BS can be estimated. However, due to the unknown orientation of the UE, the \ac{adod} information is more useful. The AODs at the UE are also important in orientation estimation, as will be discussed in Sec.~\ref{sec:orientation_estimation}.
\end{enumerate}
Each of the four geometry measurements provides a candidate area {of potential UE positions}, as shown in Fig.~\ref{fig:position_estimation} (a)-(d). The number of BSs needed for each measurement, {the geometric constraint induced per measurement}, and the number of measurements needed for position/orientation estimation are summarized in Table~\ref{tab:geometry_based_localization}. 

{If the candidate areas given by geometry information intersect at a unique point, the UE is localizable.
In practice, robust localization algorithms need to be designed to deal with noisy measurements. Position estimation utilizing combinations of geometry information is also possible, as shown in Fig.~\ref{fig:position_estimation} (e)-(h), and can reduce measurement uncertainties.}
Moreover, extra constraints can help reduce the number of BS needed. For example, in Fig.~\ref{fig:position_estimation} (a), the UE's position can be obtained with only `B1' and `B2' if wall constraints are considered.
The downlink scenarios can be treated similarly to the uplink ones by swapping the role of AOA and AOD.


    

\subsubsection{Orientation Estimation}
\label{sec:orientation_estimation}
The orientation of a UE is also of great interest in some application scenarios, and it can be obtained if an antenna array is available at the UE. With an estimated position, the orientation can be estimated using AOD at the UE. We classify the orientation estimation into 1D, 2D and 3D, indicating the estimation of $\alpha$, $[\alpha, \beta]$, and $[\alpha, \beta, \gamma]$ of an Euler angle vector, respectively. The relationship between the Euler angle vector and rotation matrix is detailed in Sec.~\ref{sec:system_geometry}. 

\begin{enumerate}[a)]
    \item 1D Orientation: The estimation of 1D orientation usually occurs in 2D localization scenarios with uniform linear arrays (ULAs)~\cite{shahmansoori2017position}. For example, a robot navigating a 2D area needs to know its orientation, which can be directly obtained from the estimated AOD with its position information.
    \item 2D Orientation: When the 3D position of a UE and at least one AOD angle pair (azimuth and elevation) is available, 2D orientation can be obtained~\cite{abu2018error}.
    \item 3D Orientation: If the 3D position and at least two AOD pairs are available, the 3D orientation can be estimated by solving a manifold optimization problem~\cite{alsharif2021manifold, nazari20213d}. A least-square estimator (LSE) and a maximum likelihood estimator (MLE) are proposed in~\cite{nazari20213d}; toolboxes such as Manopt~\cite{boumal2014manopt} can be used to solve this type of optimization problems.
\end{enumerate}


\subsection{Localization Performance Metrics}
\label{sec:localization_performance_metrics}
When designing a localization system, improving the position and orientation accuracy can intuitively be the primary objectives. However, other objectives such as coverage and stability are also important to ensure the system's overall performance. Several localization-related objectives are noted below:
\begin{enumerate}[a)]
    \item Accuracy: Accuracy reflects the localization performance (position and orientation estimation accuracy) a system can achieve. This is usually quantified in terms of RMSE or \ac{cdf} of measurements with an error smaller than a threshold. Given the signal and noise models, the accuracy can be lower bounded by the \ac{crb}. Although the CRB does not bound the actual performance unless the estimator is efficient, we use it as an indicator to design and optimize the localization system, as will be detailed in Sec.~\ref{sec:system_design_and_optimization}.
    \item Coverage: Because high-frequency signals attenuate drastically, the corresponding localization coverage can be defined as the range of a communication link within which the localization of a UE meets specific performance metrics. Coverage can also be defined in terms of areas (rather than ranges) for 2D and 3D scenarios.
    \item {Latency: Latency is defined as the time duration between a UE requesting location and obtaining the results. This is dictated by the duration of PRS used for localization and the processing time of the adopted localization algorithm.}
    \item {Update rate: Update rate is the time required to update a localization measurement (usually in tracking scenarios). This is determined by the latency (at most once per latency) and can be chosen depending on application scenarios.}
    \item Stability: Deafness is a crucial problem in high-frequency systems with narrow beamwidths, where beam misalignment can cause an outage (loss of tracking). The variance of localization accuracy during a certain period, especially in a mobile scenario, can be defined as system stability.
    \item Scalability: Scalability is the ability of a system to adapt to a larger number of UEs (e.g., performance as a function of UE densities). 
    \item Mobility: Mobility refers to the supported speeds of UEs in a localization system, in which the Doppler effect should be considered.

    \item System complexity: {System Complexity includes hardware and algorithm aspects. Depending on the application scenario, either of the two complexities or both should be considered.} Hardware complexity involves infrastructure deployment and hardware realization, which also determine the complexity of optimization, communication, and localization algorithms. However, in this work, we focus on the computational complexity at the algorithm level.
    
    
    
\end{enumerate}

\subsection{Current Localization Systems Using Electromagnetic Signals}
\label{sec:current_localization_systems}

We next review current electromagnetic signal-based localization systems, highlighting what can be achieved in the THz band. Based on the signal frequency band, we categorize these systems into four groups: conventional radio-frequency (CRF) systems (below \unit[30]{GHz}), mmWave systems (\unit[30-100]{GHz})\footnote{Note that the definition of mmWave signal band based on wavelength should be \unit[30-300]{GHz}. However, the upper mmWave band (\unit[100-300]{GHz}) is also called the sub-THz band, based on the deliverable D2.1~\cite{hexax_d21} of the European HEXA-X project. In this work, we consider sub-THz signals (e.g., \unit[0.1-0.3]{THz}) as THz signals.}, LED-based visible light positioning (VLP) systems (\unit[400-790]{THz}), and THz systems (\unit[0.1-10]{THz}). In this section, we shortly describe the localization using CRF, mmWave and VLP systems, and THz localization will be discussed in Sec.~\ref{sec:terahertz_localization}.

\subsubsection{Conventional Radio Frequency Systems (below \mbox{30~GHz})}
Location information is attainable in CRF communication systems with carrier frequencies below $30$ GHz. Global navigation satellite systems (GNSS) are most widely used for outdoor localization, where a meter-level accuracy can be achieved with the assistance of signals from the long-term evolution (LTE) communication systems. However, this approach does not work for indoor scenarios due to the corresponding complex environments and line-of-sight (LOS) channel blockages. Alternatively, localization systems based on ultra-wideband (UWB)~\cite{alarifi2016ultra},  WiFi~\cite{guo2019robust}, WLAN~\cite{tian2017optimization} and LoRA~\cite{lam2019rssi} are reported; the comparison between different indoor localization technologies can be found in~\cite{zafari2019survey}. Using CRF systems, we benefit from location-based services such as navigation and finding surrounding services.

\subsubsection{mmWave Systems (\mbox{30-100~GHz})}
To meet the data rate demands and overcome bandwidth scarcity, mmWave signals, combined with hybrid MIMO structures and the corresponding signal processing methods, play a fundamental role in 5G systems. The extended bandwidth realizes higher-rate communications with lower latency and better localization performance. Equipped with antenna arrays at the UE, orientation estimation becomes possible~\cite{nazari20213d,shahmansoori2017position,shahmansoori20155g}. Additionally, by exploiting the NLOS paths~\cite{abu2018error} and RISs~\cite{he2020large, wymeersch2020radio}, the localization tasks can be completed using a single BS. Such advantages make mmWave systems attractive in communication networks~\cite{di2014location} and vehicular networks~\cite{wymeersch20175g}.

\subsubsection{Visible Light Positioning Systems (\mbox{400-790~THz})}
Due to the immensely large bandwidths within the high-frequency spectrum, VLC systems offer high data transmission rates. Towards realizing VLC, laser diodes (LDs) and LEDs have emerged as two widely used types of light sources~\cite{chi2020visible}. LDs provide large bandwidths and concentrated energy for long-distance transmissions. However, precise alignments are needed to set up LD-based communication links. Although light detection and ranging
(LIDAR) systems utilize LD arrays to achieve high ranging and localization accuracies; they usually work as independent sensors and not as part of the communication system; we do not consider LIDAR in this work. LEDs used in current illumination systems provide wide coverage; nevertheless, several issues render their implementation challenging, such as blockage, limited power control, flexibility in spatial multiplexing, and sensitivity to the environment. Surveys on localization via visible light systems can be found in~\cite{keskin2018localization,luo2017indoor}.

\begin{table*}
\caption{A comparison of 5G mmWave and 6G THz (reasonable guess) from a localization perspective}
\begin{centering}
{
\scriptsize
\begin{tabular}{m{0.1cm} | m{1.2cm}
!\vthickline
m{1.4cm}|m{2.3cm}|m{2.6cm}
!\vthickline
m{1.4cm}|m{2.3cm}|m{2.7cm}}
\hthickline
\textbf{\#} & \textbf{Aspect} & \textbf{5G mmW} & \textbf{Localization Merits} & \textbf{Localization Challenges} & \textbf{6G THz} & \textbf{Localization Merits} & \textbf{Localization Challenges}\tabularnewline
\hthickline
1 & Frequency, wavelength & {\unit[30-100]{GHz}, \unit[3-10]{mm}} & Few multipath components & Range: path loss $10^{4}\times\text{distance}$ & \unit[0.1-10]{THz}, \unit[0.03-3]{mm} & Only metallic objects visible, miniaturization & Range: path loss $10^{5}-10^{7}\times\text{distance}$
\tabularnewline
\hline 
2 & Bandwidth & \unit[400]{MHz} & Distance resolution: {\unit[0.75]{m}} & {High sampling frequency} & 1-10 GHz & Distance resolution: {\mbox{\unit[3-30]{cm}}} & {ADC power consumption, large volume of data}
\tabularnewline
\hline 
3 & Array size & $10\times10$ & Angle resolution: \mbox{$\sim\!\!10^\circ$} & Moderate beam management overhead  & $100\times100$ & Angle resolution: $<\!1^\circ$, probably no multipath effect per beam & Severe beam management overhead
\tabularnewline
\hline 
4 & Array type & hybrid UPA & Reduced RX overhead, flexible TX signals, azimuth and elevation angles & Scanning time $\propto$ beamwidth & AOSA preferred & Low complexity, azimuth and elevation angles, angle-based localization without time measurements & Angle ambiguities, scanning time $\propto$ beamwidth, 
array calibration, mutual coupling
\tabularnewline
\hline 
5 & Hardware imperfection & Quantization in PSs, ADCs & Reduced complexity and power & Accuracy loss & IQI, PN, PA, ADC & Possibly location-dependent effects can be exploited & Power limitations, waveform type, unable to use standard
DSP 
\tabularnewline
\hline 
6 & Synchro- nization & $<$10 ns & Required for time-based measurements & Challenging to maintain & $<$ 1 ns & Required for time-based measurements & Extremely challenging. `Synchronization by nature solution' preferred (multipath and RIS)
\tabularnewline
\hline 
7 & \!Waveform & OFDM & Easy to account for multipath, structured signal & {Non-linear distortion caused by PAPR} & Unknown & Signal type not fundamental & DSP and signal design depends on signal type
\tabularnewline
\hline 
8 & \!Propagation effects & Few cluster model & Few clusters to resolve. LOS path can be detected  & Sensitive to LOS blockage & Few paths, BSE, near-field & Cleaner geometric channel, BSE and near-field can be exploited & The exploitation of BSE and near-field requires new DSP
\tabularnewline
\hline 
9 & \!Typical positioning method & Multi-BS, single BS \& scatterers & Simple and scalable & Synchronization and coordination between BSs, deployment complexity & Multi-BS, D-MIMO, RIS-assisted & Relaxed synchronization, angle-only positioning & New and dedicated RIS infrastructure. Careful calibration is needed. 
\tabularnewline
\hline 
\!\!10 & \!Positioning signals & DL-PRS, maybe AOD & Broadcast signals, combine angle and delay & Angle measurements based on power & PRS not defined in 6G; possibly user-specific & Better accuracy & More overhead and delay, signaling delay
\tabularnewline
\hthickline
\end{tabular}}

\vspace{1ex}
\scriptsize{\raggedright *Notes: \ac{upa}, \ac{aosa}, \ac{ps}, \ac{adc}, \ac{iqi}, \ac{pn}, \ac{pa}, \ac{bse}, \ac{dmimo}, \ac{dlprs}, \ac{prs}, {and `$\propto$' means `be proportional to'.} \par}
\par
\end{centering}
\label{table:comparison_5g_6g}
\end{table*}

\subsection{Terahertz Localization}
\label{sec:terahertz_localization}
\subsubsection{5G mmWave vs. 6G THz Localization}
THz systems are likely to be adopted as an extension to mmWave systems in heterogeneous environments \cite{kouzayha2021coverage}. Consequently, we compare the THz and mmWave systems by discussing their merits and challenges from a localization perspective. 
Moving from CRF to 5G and further to 6G systems, we expect higher frequencies, larger bandwidths, smaller footprints, and larger array sizes. High frequencies increase path loss and reduce multipath components, while large bandwidths provide high delay estimation resolutions. With a smaller wavelength, miniaturized antenna array footprints {or large array sizes with the same physical array size}\footnote{\label{fn:array_terms}We use `footprint' and `array size' to denote the physical size (e.g., $\unit[2\times 2]{cm^2}$) and the number of array antennas (e.g., $5\times5$), respectively.} become possible. 

Such changes in signal properties affect several system features. For example, hardware imperfections and synchronization issues become challenging at THz frequencies. Furthermore, the \ac{ofdm} waveform may no longer be suitable for wideband systems due to the high \ac{papr} issue and DFT-s-OFDM is a promising alternative~\cite{sahin2016flexible, tarboush2021single}. When designing localization algorithms, a geometric MIMO-based channel model should account for the resultant THz \ac{bse} and near-field conditions. 
{Moreover, due to the high path loss, THz signals need to be delicately designed to serve users with different performance requirements for an energy-efficient purpose.} A comparison between 5G mmWave and 6G THz is highlighted in Table~\ref{table:comparison_5g_6g}.
In a nutshell, we expect better localization performance in 6G THz systems. However, new challenges in hardware design, coverage, overheads, and computational complexity should be tackled.

{
\subsubsection{The Role of RIS in Localization}
Different from communication scenarios, where RIS provides high SNR to obtain a high data rate, the role of RIS is to enable or enhance localization in two aspects: working as a \textit{passive anchor} to provide geometrical diversity, and providing a \textit{near-field scenario} to exploit the \ac{coa} information.
Localization requires geometrical diversity for a satisfactory estimation performance, which can be fulfilled with multiple \acp{bs}, generally providing coverage for \acp{ue} in their convex hull. Recent emerged techniques also exploit the  \ac{mpc}\cite{witrisal2016high,ge20205g} for high-frequency signal localization. The \acp{mpc}, which are usually considered as destructive signals, can be resolved in 5G/6G systems, thereby enabling positioning and mapping~\cite{wen20205g}, and \ac{slam}~\cite{ge20205g} with even a single \ac{bs}. 
Considering the MPCs are uncontrollable and the challenges of multiple \acp{bs} deployment (e.g., high hardware cost, synchronization, and calibration error, RIS could be a potential alternative. An RIS works as a passive, customizable \ac{bs} with low energy consumption, providing additional location references and resolvable multipath measurements \cite{wymeersch2020radio}, which can either boost or enable localization \cite{keykhosravi2021siso, strinati2021reconfigurable}. In addition, no fine synchronization between RISs is needed, which simplifies the deployment on a large planar surface of the environment and can hence create a near-field scenario.
As a result, RIS will likely play a game-changing role in the future localization systems and will be discussed in detail in the later sections of this tutorial.}

\begin{table*}[t]
    \caption{{Localization KPIs (expected) for 5G/6G Systems and for Potential THz Localization Applications}}
    \centering
    \scriptsize
    \begin{tabular}{l !\vthickline c | c | c | c | c | c| c|c|c }
    \hthickline
        & 
        \textbf{Position} & \textbf{Orientation} & \textbf{Coverage} & \textbf{latency} & \textbf{Update Rate} &
        \textbf{Stability} & \textbf{Scalability} &\textbf{Mobility} &  \textbf{Data Rate (Peak)}\\
    \hthickline
        5G KPIs & \unit[10]{cm} & $\unit[10]{^\circ}$ 
        & - & $>$\unit[1]{ms} & - & -
        & $\unit[10^6]{/km^2}$ & \unit[500]{km/h} & \unit[20]{Gbps} (peak) \\
        \hline
        6G KPIs & $\unit[^*1]{cm}$ & $\unit[^*1]{^\circ}$ 
        & - & $\sim\!\unit[0.1]{ms}$ & - & -
        & $\unit[10^7]{/km^2}$ & $\unit[1000]{km/h}$ & $>$\unit[1]{Tbps}\\
    \hthickline
        {Telesurgery} & $<$\unit[1]{mm} &
        $\unit[<1]{^\circ}$ & - & - & $\ge\unit[10]{Hz}$ &
        $^*$High & $^*$Low & $^*$Low & -
        \\ \hline
        XR \& Holography & $<$\unit[1]{cm} & $\unit[1]{^\circ}$ & \unit[10]{m} & \unit[5]{ms} & $\unit[\ge100]{Hz}$ & $^*$High & $^*$Low &  $^*$Medium & \unit[\text{Orders of}]{Tbps}
        \\
        \hline
        Connected Vehicles & \unit[0.1-1]{m} & - & $\unit[>100]{m}$ & \unit[1]{ms}  & \unit[1-10]{Hz}& $^*$High & $^*$Medium & $^*$High & $\unit[>100]{Gbps}$
        \\
        \hline
        Digital Twins & \unit[1]{cm} & - & $\sim\unit[130]{m}$ & \unit[0.1-1]{ms} & $\ge\unit[10]{Hz}$ & $^*$Medium & $^*$High & $^*$Low & Depends on update rate
        \\
        \hline
        {Collaborative Robot} & $<$\unit[1]{cm} & $\unit[<1]{^\circ}$
        & - & \unit[1]{ms} & $\unit[\ge10]{Hz}$ & $^*$High & $\unit[>5]{/m^3}$ & $^*$Medium & -
        \\
        \hline 
        {Threat Localization} & \unit[1]{cm} & - &
        $\sim\unit[120]{m}$ & \unit[1]{ms}& $\ge\unit[10]{Hz}$ & $^*$Medium & $^*$Medium & $^*$Medium &  -\\ 
        \hthickline
    \end{tabular}
    
\vspace{1ex}
\scriptsize{\raggedright \ \ \ \ {Data from: Table VI in \cite{ghafoor2020mac}, Table V in \cite{chaccour2022seven}, and Table II in \cite{rajatheva2020scoring}, Figure 2 in \cite{giordani2020toward}, and Section 3.1 in \cite{hexax_d31}. The entries with a star marker ($^*$) are from on our best guess.} \par}
\par
\label{tab:localization_KPIs}
\end{table*}

\subsubsection{Applications and Key Performance Indicators (KPIs)} 
\label{sec:kpis}
{As reported in the European 6G project Hexa-X, representative use cases envisioned for 6G are categorized into five groups: sustainable development, massive twinning, telepresence, robots to cobots, and local trust zones~\cite{hexax_d12}. From the THz localization point of view, it not only improves the communication performance by aiding beamforming with location information, but also enhances physical layer security with narrow beamwidth, distance-dependent attenuation~\cite{akyildiz2018combating}, multipath exploitation~\cite{petrov2019exploiting}, and new counter-measure techniques~\cite{ma2018security}. When narrowing down to the location-aware services, potential applications that need high-accuracy localization information, such as telesurgery, XR and holography, connected vehicles, digital twins, etc., will be enabled~\cite{chaccour2022seven, xiao2022overview, ghafoor2020mac, giordani2020toward, hexax_d31}.}

Based on the objectives defined in Section~\ref{sec:localization_performance_metrics}, the \acp{kpi} which are necessary to evaluate the performance of localization schemes can be outlined. A summary of important localization KPIs (expected) and their value ranges for 5G/6G systems, and typical applications that require THz localization is presented in Table~\ref{tab:localization_KPIs}~\cite{rajatheva2020scoring, wymeersch20175g, giordani2020toward, hexax_d31}. The data rate requirements for some applications are also presented for reference. We notice that the applications that require high accuracy and data rates beyond what can be achieved in 5G communication systems are projected to thrive in 6G systems.

\subsubsection{Current Research in THz Localization}
{Researchers have started working on THz localization by addressing system structures, localization algorithms, and simulation platforms.} In~\cite{peng2016three}, an AOA estimation method based on a forward-backward algorithm is developed for dynamic indoor THz channels by measuring and developing different human movement models. Furthermore, a tracking approach is developed for time-variant channel modeling in indoor THz communications by using extended Kalman filtering~\cite{nie2017three}. 
Cooperation-aided localization approaches are also proposed to provide high estimation accuracy and alleviate the deafness problem in 2D scenarios~\cite{stratidakis2019cooperative}.
In~\cite{tan2021wideband}, a delay-phase precoding structure is proposed, and a beam zooming mechanism is adopted for THz beam tracking, demonstrating the ability to track multiple users by one \ac{rfc} and substantially reduce the beam training overhead. In~\cite{guerra2021near}, a near-field model is considered with large antenna arrays to leverage the \ac{coa} as an extra degree of freedom for inferring the source position.
Besides geometry-based methods, deep learning-based methods using \acp{rnn} for 3D THz indoor localization are proposed in~\cite{fan2020structured}. Here, a localization accuracy of \unit[0.27]{m} (mean distance error) is reported in \ac{nlos} environments, demonstrating a $\unit[60]{\%}$ enhancement over the \ac{sota} techniques.  

The mentioned localization works focus on different THz localization challenges such as misalignment, tracking, cooperation, BSE, near-field effects, and large amounts of data to be processed. However, THz localization-related research is still in its infancy, and critical issues still need to be identified and addressed despite all these efforts. The question of how THz-band signals can improve localization performance remains unanswered. 
In the next sections, we describe the THz system model, formulate localization and system optimization problems, and evaluate the potential of THz localization through simulations.

\begin{table}[t]
\scriptsize
\centering
\caption{Summary of Symbols}
\renewcommand{\arraystretch}{1.25}
\begin{tabular} {r !\vthickline l}
\hthickline
\textbf{Notation} & \textbf{Description} \\
\hthickline
${\pv_{\scriptscriptstyle{Q}}}$ & Global position, $Q\in\{\mathrm{B,R,U,N}\}$ (BS, RIS, UE, NLOS)\\
\hline
${\pv_q/\tilde \pv_q}$ & Global/local
position of the $q$th element, $q\in\{b, r, u\}$\\
\hline
${\tilde \pv_{\mathring q}}$ & Local position of the $\mathring q$th AE at SA\\
\hline
${N_\mathrm{Q}}$ & Number of elements,  $Q\in\{\mathrm{B,R,U}\}$\\
\hline
${\mathring N_Q}$ & Number of antennas per SA in AOSA structures $Q\in\{\mathrm{B,U}\}$\\
\hline
${L_\mathrm{N}}$ & Number of NLOS paths\\
\hline
${\Rm_\mathrm{Q}}$ & Rotation matrix, $Q\in\{\mathrm{B,R,U}\}$ \\
\hline
${\mathbf{o}_\mathrm{Q}}$ & Euler angles (orientation vector) $Q\in\{\mathrm{B,R,U}\}$\\
\hline
${\tv/\tilde \tv}$ & Global/local direction vector\\
\hline
${\vpv/\tilde \vpv /\mathring\vpv}$ & Global/local/beamforming angles\\
\hline
$\Hm/\Hbc$ & Channel matrix for conventional MIMO/AOSA-based MIMO\\
\hline
${\rho}e^{-j\xi}$ & Complex channel gain of each path\\
\hline
${G_\ssr{B}/G_\ssr{U}}$ & Antenna gain of the BS/UE\\
\hline
${K}$ & Number of subcarriers (for pilot signals)\\
\hline
${W}$ & Bandwidth\\
\hline
$\Gc$ & Number of transmissions \\
\hline
${B}$ & Synchronization offset\\
\hline
${\Acal}$ & Array factor of the SA\\
\hline
${K_\mathrm{a}/\mathcal{K}_N}$ & Attenuation coefficient/NLOS reflection coefficient\\
\hline
${\sv}$ & State vector\\
\hline
${\gammav}$ & Measurement vector \\
\hthickline
\end{tabular}
\renewcommand{\arraystretch}{1}
\label{table:Symbols}
\end{table}

\section{Terahertz System Model}
\label{sec:thz_system_model_and_properties}
{This section describes the proposed THz system model, emphasizing the AOSA UM-MIMO structure. We start by summarizing the latest {advances in THz devices} to justify the system model components. Then, we detail the {system geometry} and the corresponding {MIMO channel model}. We further detail the {proposed THz signal model} with AOSA structures that consist of the LOS, RIS, and NLOS channels. We end the section by discussing {additional model features} such as beam split and hardware imperfections.}

\subsection{Advances in THz Devices}
\label{sec:advances_in_thz_devices}

The fundamental advances in THz technology are still taking place at the device level, bridging the so-called ``THz gap''. Recently, multiple candidate technologies have demonstrated compact THz signal sources and detectors that achieve good power and sensitivity. In particular, recent electronic and photonic THz transceivers have achieved efficient signal generation, modulation, and radiation~\cite{kenneth2019opening, nagatsuma2016advances, sengupta2018terahertz, sengupta2019universal}.

\subsubsection{Electronic Solutions (compact, relatively high power)}
Electronic THz-band solutions \cite{hillger2020toward,rieh2020introduction} are mainly based on silicon devices \cite{hillger2018terahertz, han2019filling} which have already been utilized in mmWave systems. In particular, silicon complementary metal-oxide-semiconductor (CMOS) and silicon-germanium (SiGe) BiCMOS technologies \cite{nikpaik2017219,aghasi20170,mittleman2017perspective,heinemann2016sige} have exhibited good compatibility with fabrication processes and high compactness. However, CMOS devices have lower power handling capabilities, and their unity maximum available power gain frequency ($f_{\text{max}}$) is still limited to $\unit[320]{GHz}$. Higher frequencies are achievable with III-V-based high electron mobility transistors (HEMTs) \cite{deal2017660,leuther201420,mei2015first}, heterojunction bipolar transistors (HBTs) \cite{urteaga2017inp,bolognesi2016inp}, and Schottky diodes \cite{mehdi2017thz}. 
High array gains are required to combat the power limitations in CMOS and extend the coverage; therefore, MIMO-based systems are typical with electronic solutions. The corresponding MIMO arrays can still be compact, with AEs being proportional to the wavelength. Sufficient spatial resolution can be provided via beamforming, and conventional MIMO localization techniques can be applied.


\subsubsection{Photonic Solutions (bulky, low power, high rates)}    
Higher carrier frequencies and higher data rates are supported with photonic THz solutions~\cite{nagatsuma2016advances}. However, photonic devices are limited in power and integration capabilities, as they tend to have relatively large form factors. Optical downconversion systems \cite{nagatsuma2016advances}, photoconductive antennas \cite{huang2017globally}, quantum cascade lasers \cite{lu2016room}, and uni-traveling carrier photodiodes \cite{song2008broadband} have demonstrated operations beyond $\unit[300]{GHz}$. Furthermore, integrated hybrid electronic-photonic systems are being proposed~\cite{sengupta2018terahertz}, such as by combining photonic transmitters and III-V electronic receivers. Nevertheless, more delicate synchronization is required between transmitters and receivers in such solutions.
As a result, MIMO-based photonic solutions are challenging, especially in UEs. However, the BSs can still achieve high coverage with high-power devices that are bulky in size. Beamsweeping-based methods using mechanical rotations could be an alternative to MIMO-based solutions.


\subsubsection{Plasmonic Solutions (much smaller footprints, very high reconfigurability)}
Novel plasmonic materials, such as graphene, are also being considered as candidate THz device technologies \cite{jornet2011channel,hafez2018extremely}, supporting high reconfigurability solutions. Since the resonant wavelengths of \ac{spp} waves in plasmonics are much smaller than free space wavelengths, much more compact and flexible antenna array designs can be realized \cite{singh2020design,ferrari2015science}. Plasmonic transceivers can operate at THz frequencies without upconversion and downconversion, where the generation of energy-efficient short pulses is particularly favorable \cite{jornet2014femtosecond}.
The inherent compactness and frequency-interleaving properties~\cite{zakrajsek2017design} of plasmonic solutions make them favorable for flexible system designs and on-site reconfiguration. 
{However, the limitation on power output makes plasmonic solutions tailored for nanocommunication scenarios with a limited range (several tens of millimeters)~\cite{singh2020design, jornet2011channel}, which is not practical for localization purposes.}


\subsubsection{THz RIS Material Properties}
Since THz MIMO configurations are not yet mature, with experimental demonstrations limited to $2\times2$ MIMO~\cite{khalid2016experimental}, we argue that advances in THz-operating materials could favor THz-band RIS deployments. THz-RIS CMOS deployments are low power consuming and easy to integrate. However, they suffer from limited clock speeds and parasitic capacitance leakage \cite{venkatesh2020high,liu2014broadband}. \Acp{mems} are also considered for THz RIS \cite{zhao2018design,han2017thin}, but they are limited by switching speeds, control signaling, and relatively large footprints. Similarly, plasmonic (graphene-based) technologies are promising for RIS deployments, as they are low-power-consuming, easy to integrate, and possess simple biasing circuits \cite{nie2019intelligent}. Graphene-based metasurfaces utilize electrostatic biasing to control the chemical potential of reflecting elements, varying the complex conductivity for phase control \cite{lee2012switching}.

Compact and lightweight metasurfaces support THz signal beam steering over a wide range of angles \cite{la2019curvilinear}. Furthermore, THz-operating metasurfaces provide the option of generating orbital angular momentum and polarization conversion \cite{fu2020terahertz}. THz-operating HyperSurfaces \cite{liaskos2018new} are also gaining popularity, where a stack of virtual and physical components generates lens effects and custom reflections. Moreover, thermally- or electrically-tunable vanadium dioxide and liquid crystals can also realize efficient THz signal steering \cite{fu2020terahertz}. 
Given the THz high directionality and blockage issues, THz-operating metasurfaces that support higher reconfigurability and sensing accuracy are crucial for spatially-sensitive THz communications and localization.

Such advances in THz devices facilitate the design and realization of THz systems. However, novel calibration algorithms, beamforming optimizations, and distributed control processes (e.g., optical internetworking~\cite{liaskos2018new}) are desired to benefit from antenna arrays and RISs in a THz system. In this work, we only consider a system model with one RIS and one BS, as will be detailed in Sec.~\ref{sec:mimo_channel_model}.

\subsection{System Geometry}
\label{sec:system_geometry}
\subsubsection{Global Coordinate Systems}
Consider a MIMO system containing a BS, an RIS, and a UE as shown in Fig.~\ref{fig:system_model_geometry}.
We define the array center $\pv_\mathrm{Q}=[x_\mathrm{Q}, y_\mathrm{Q}, z_\mathrm{Q}]^T$ as the location of a device containing $N_\mathrm{Q}$ elements, where $\mathrm{Q}\in\{\mathrm{B},\mathrm{R},\mathrm{U}\}$ represents BS, RIS, and UE, respectively. Here, the \textbf{\textit{element}} is defined as the minimum communication element (e.g., an antenna of a conventional array, a \ac{sa} inside an AOSA structure, or {an RIS} element), and the position of each element is $ \pv_b$ ($\pv_r$, $\pv_m$). Assume $L_\mathrm{N}$ NLOS paths are generated in the channel, where the $l$th NLOS path corresponds to a scatterer with an unkown location $\pv_N^{\ssnb{l}}$. The scatterer could also be a reflector or a diffractor that creates signal paths.

\subsubsection{Local Coordinate Systems}
For a planar array, we define the array center to be its local coordinate origin and the array norm to be the X-axis (e.g., a planar array lies on the YZ plane). For $\mathrm{Q}\!\in\! \{\mathrm{B}, \mathrm{R}, \mathrm{U}\}$, an Euler angle (3D orientation) vector $\ov_\mathrm{Q}\!=\![\alpha_\mathrm{Q}, \beta_\mathrm{Q}, \gamma_\mathrm{Q}]$ ($ \alpha_\mathrm{Q}\!\in\!(-\pi, \pi]$, $ \beta_\mathrm{Q}\!\in\![-\pi/2, \pi/2]$, $ \gamma_\mathrm{Q} \in(-\pi, \pi]$) and a rotation sequence Z-Y-X are used to describe the array orientation in the global coordinate system.

The relationship between the 3D global position, $\pv_q$, and the 3D local\footnote{In the remainder of this work, symbols marked with a `tilde' (e.g., $\tilde \theta$) indicate parameters in the local coordinate system (in contrast to the global parameters, e.g., $\theta$), whereas symbols marked with a `ring' (e.g., $\mathring \theta$) indicate parameters related to SAs in AOSA-based systems (detailed in Table~\ref{table:Symbols}).} position of the $q$th element on the array, $\tilde \pv_q$ ($q\in \{b, r, u\}$), can be expressed as
\begin{equation}
    \pv_{q} = \Rm_\mathrm{Q} \tilde\pv_{q} +  \pv_\mathrm{Q},
\end{equation}
where $\pv_\mathrm{Q}$ is the center of the array in which $\pv_q$ is located, and $\Rm_\mathrm{Q}$ is the rotation matrix that could be obtained using an orientation vector, $\ov_\mathrm{Q}$, as
\begin{equation}
    \Rm_\mathrm{Q} = 
    \begin{bmatrix}
    c_\alpha c_\beta & c_\alpha s_\beta s_\gamma-c_\gamma s_\alpha & s_\alpha s_\gamma +c_\alpha c_\gamma s_\beta \\
    c_\beta s_\alpha  & c_\alpha c_\gamma +s_\alpha s_\beta s_\gamma  & c_\gamma s_\alpha s_\beta -c_\alpha s_\gamma \\
    -s_\beta  & c_\beta s_\gamma  & c_\beta c_\gamma
    \end{bmatrix},
    \label{eq:rotation_matrix_3_var}
\end{equation}
where $c_\alpha$ represents $\cos(\alpha_{\scriptscriptstyle{Q}})$ and $s_\alpha$ is short for $\sin(\alpha_{\scriptscriptstyle{Q}})$. Accordingly, the 3D local location can also be obtained as
\begin{equation}
    \tilde \pv_{q} = \Rm^{-1} (\pv_{q} - \pv_\mathrm{Q}).
\end{equation}

 

\subsubsection{Direction Vector and AOA/AOD}
Consider a signal transmitted from a UE located at $\pv_\mathrm{U} = [x_\mathrm{U}, y_\mathrm{U}, z_\mathrm{U}]^T$ to a BS located at 
$\pv_\mathrm{B} = [x_\mathrm{B}, y_\mathrm{B}, z_\mathrm{B}]^T$, the distance between the UE and BS array center can be calculated as
\begin{equation}
    {d_\mathrm{BU}} = \norm{\pv_\mathrm{U}- \pv_\mathrm{B}},
\end{equation}
and the {{global direction vector}} from BS to UE, $\tv_{\mathrm{BU}}$, can be expressed as
\begin{equation}
    \tv_\mathrm{BU} = -\tv_\mathrm{UB} =
    \begin{bmatrix}
    t_{\mathrm{BU}, x}\\
    t_{\mathrm{BU}, y}\\
    t_{\mathrm{BU}, z}
    \end{bmatrix}
    = \frac{ \pv_\mathrm{U}- \pv_\mathrm{B}}{d_\mathrm{BU}}.
    \label{eq:global_direction_vector}
\end{equation}
The {{local direction vector}} can then be obtained using the rotation matrix $\Rm_\mathrm{B}$ of BS and the direction vector $\tv$ of~\eqref{eq:global_direction_vector} as
\begin{equation}
    \tilde \tv_{\mathrm{BU}} = \Rm_\mathrm{B}^{-1} \tv_{\mathrm{BU}} = \Rm_\mathrm{B}^{T} \tv_{\mathrm{BU}}.
    \label{eq:dir_global_to_local}
\end{equation}

\begin{figure*}[!t]
\centering
\includegraphics[width = 0.98\linewidth]{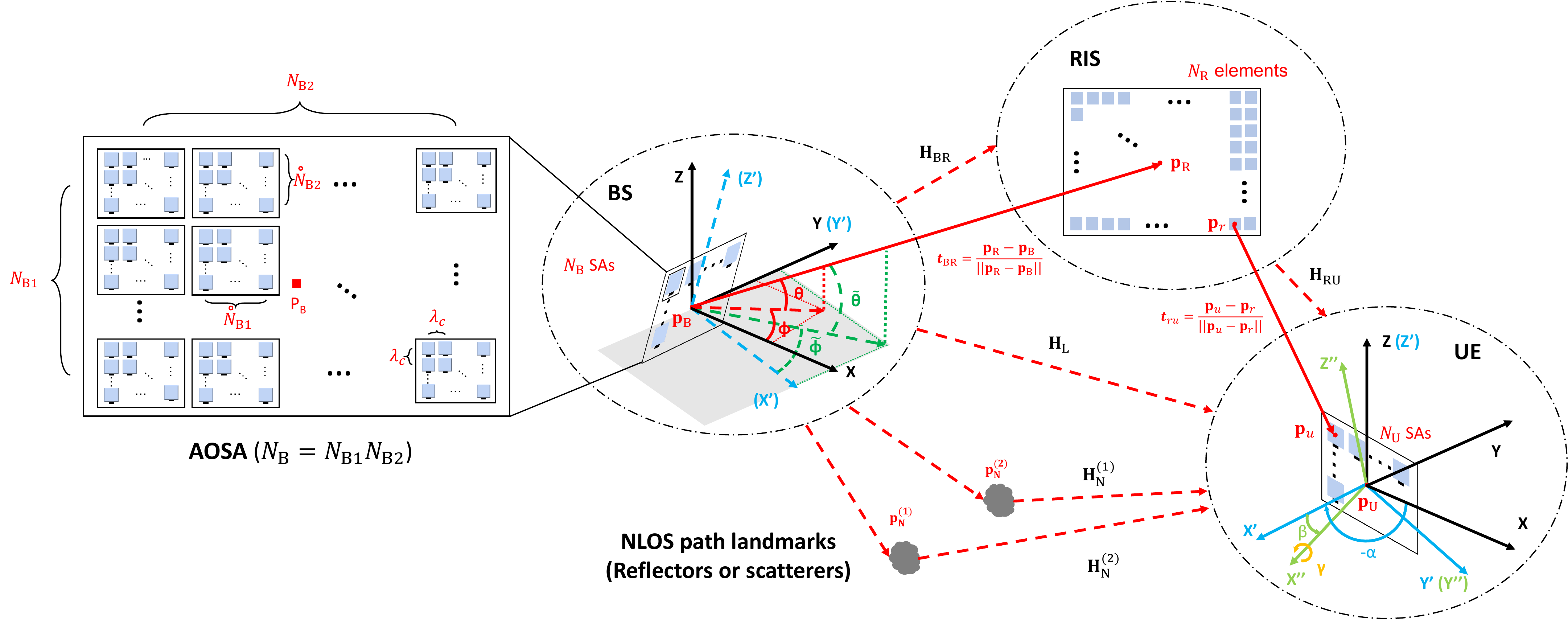}
\caption{Illustration of the proposed THz localization system model.}
\label{fig:system_model_geometry}
\end{figure*}

The AOA/AOD angle pairs of a signal are defined using an azimuth angle $\phi\in(\pi, \pi]$ (angle between the projection of the vector $\tv$ on the XY-plane and the Y-axis) and an elevation angle $\theta\in [-\pi/2, \pi/2]$ (the angle between $\tv$ and the XY-plane) as shown in Fig.~\ref{fig:system_model_geometry}. These angle pairs can be defined in both the global and local coordinate systems; however, the angles can only be measured locally at the array (e.g., using AOA/AOD estimation algorithms). We use $\tilde \vpv \!=\! [\tilde \phi, \tilde \theta]^T$ and $\vpv \!=\! [\phi, \theta]^T$ to represent local and global AOA/AOD angles, respectively. 

From the definitions of the azimuth and elevation angles, the local direction vector, $\tilde\tv_{\mathrm{BU}}$ defined in~\eqref{eq:dir_global_to_local}, can be easily expressed in terms of the AOA/AOD angles $\tilde \phi_{\mathrm{BU}}$ and $\tilde \theta_{\mathrm{BU}}$ as

\begin{equation}
    \tilde \tv_{\mathrm{BU}} = \tv(\tilde \vpv_{\mathrm{BU}}) = 
    \begin{bmatrix}
    \cos(\tilde \phi_{\mathrm{BU}})\cos(\tilde \theta_{\mathrm{BU}}) \\
    \sin(\tilde \phi_{\mathrm{BU}})\cos(\tilde \theta_{\mathrm{BU}}) \\
    \sin(\tilde \theta_{\mathrm{BU}})
    \end{bmatrix},
    \label{eq:dir_vec_from_angle}
\end{equation}
where $\tv(\vpv)$ is the function that maps the AOA/AOD angles to a direction vector. Conversely, the AOA/AOD $\vpv_{\mathrm{BU}}$ from UE to BS can be obtained from the direction vector $\tilde \tv_{\mathrm{BU}} = [\tilde t_{\mathrm{BU},x}, \tilde t_{\mathrm{BU},y}, \tilde t_{\mathrm{BU},z}]$ as
\begin{equation}
\tilde \vpv_{\mathrm{BU}} = 
    \begin{bmatrix}
        \tilde \phi_{\mathrm{BU}}\\
        \tilde \theta_{\mathrm{BU}}
    \end{bmatrix}=
    \begin{bmatrix}
        \arctan2(\tilde t_{\mathrm{BU},y},\tilde t_{\mathrm{BU},x})\\
        \arcsin(\tilde t_{\mathrm{BU},z})
    \end{bmatrix},
    \label{eq:local_doa_dod}
\end{equation}
where $\arctan2(\cdot)$ is the four-quadrant inverse tangent, and the global AOA/AOD can be similarly obtained with a global direction vector $\tv$.

We have described the signal propagation distance $d_\mathrm{BU}$, global/local direction vectors $\tv_{\mathrm{BU}}$/$\tilde \tv_{\mathrm{BU}}$, and angle pairs $\vpv_{\mathrm{BU}}$/$\tilde \vpv_{\mathrm{BU}}$ for the BS-UE LOS channel. Similar descriptions apply to the BS-RIS, RIS-UE, and the $l$th NLOS channel parameters (e.g., $d_\mathrm{BR}$, $d_\mathrm{rm}$, $d_\mathrm{BN}^{\ssnb{l}}$, and $d_\mathrm{NU}^{\ssnb{l}}$). We next describe a far-field MIMO channel model based on these geometry parameters.

\subsection{Far-field MIMO Channel Model}
\label{sec:mimo_channel_model}
An accurate channel model is essential for system configuration and performance analysis. Deterministic, statistical, and hybrid methodologies can be applied for channel modeling~\cite{han2018propagation}. For localization purposes, we start with a deterministic far-field\footnote{Far-field scenarios consider a \ac{pwm}, while near-field scenarios assume an \ac{swm}.} channel model for a multi-carrier MIMO system. 
Assume an uplink scenario, the channel matrix $\Hm \in \mathbb{C}^{N_\mathrm{B}\times N_\mathrm{U}}$ can be decomposed into three parts as
\begin{equation}
    \Hm = \Hm_\mathrm{L} + \Hm_\mathrm{R} + \Hm_{\mathrm{N}}.
\label{eq:far_field_channel_model}
\end{equation}
Here, $\Hm_\mathrm{L}$, $\Hm_\mathrm{R}$, and $\Hm_{\mathrm{N}}$ are the LOS channel matrix, RIS channel matrix, and NLOS channel matrix, respectively.


\subsubsection{BS-UE LOS Channel}
\label{sec:channel_model_los_channel}
The {passband} BS-UE LOS channel matrix $\Hm_\mathrm{L}$ can be expressed as\footnote{Note that if the direction vectors at the Tx/Rx are chosen to be identical (e.g., $\tv_T = \tv_\mathrm{R}$), $\av^T$ has to be changed to $\av^*$, as expressed in~\cite{heath2016overview}. The channel matrix can be ignored if the corresponding path does not exist.}~\cite{heath2016overview}
\begin{equation}
    \begin{split}
        \Hm_{\mathrm{L}}(t, f) = & \rho_\mathrm{L}(f) e^{-j2\pi (f\tau_{\mathrm{BU}} - \nu_{\mathrm{BU}} t)} \\
        & \times
        G_{\mathrm{B}}(\tilde\vpv_\mathrm{BU})
        G_{\mathrm{U}}(\tilde\vpv_\mathrm{UB})
        \av_{\mathrm{B}}(f, \tilde\vpv_\mathrm{BU}) \av_{\mathrm{U}}^T(f, \tilde\vpv_\mathrm{UB}),
    \label{eq:far_field_los_channel_doppler}
    \end{split}
\end{equation}
where $\rho_{\mathrm{L}}(f)$ is the path gain of the LOS path at signal frequency $f$, $\nu_\mathrm{BU}$ is the Doppler shift, $\tau_{\mathrm{BU}}$ is the signal delay (including the propagation delay $d_\mathrm{BU}/c$ and the clock offset $B$, as will be detailed later) of the LOS path, and $G_\mathrm{B}$/$G_\mathrm{U}$ and $\av_\mathrm{B}$/$\av_\mathrm{U}$ are respectively the antenna gains and steering vectors of the BS/UE, which depend on the local AOA/AOD pairs, as will be detailed shortly.

The LOS path gain, $\rho_\mathrm{L}(f)$, can be expressed as~\cite{shahmansoori2017position}
\begin{equation}
    \rho_\mathrm{L}(f) = \frac{c}{4\pi f d_\mathrm{BU}}\mathcal{K}_{\mathrm{a}}(f, d_\mathrm{BU}),
\end{equation}
where $\mathcal{K}_{\mathrm{a}}(f, d_\mathrm{BU})$ is the attenuation coefficient depending on the signal frequency and distance. For mmWave signals, $\mathcal{K}_\mathrm{a} = \mathcal{K}_{\mathrm{atm}}(f, d)$ is the atmospheric attenuation~\cite{shahmansoori2017position}. However, for THz-band signals, molecular absorption caused by water vapor and other gases increases the path loss, where $\mathcal{K}_a = e^{-\frac{1}{2}\mathcal{K}_{\mathrm{abs}}(f)d}$ is the absorption coefficient that can be obtained from the high-resolution transmission molecular absorption (HITRAN) database~\cite{gordon2017hitran2016}.

\begin{figure}[t]
\centering
\includegraphics[width = 0.98\linewidth]{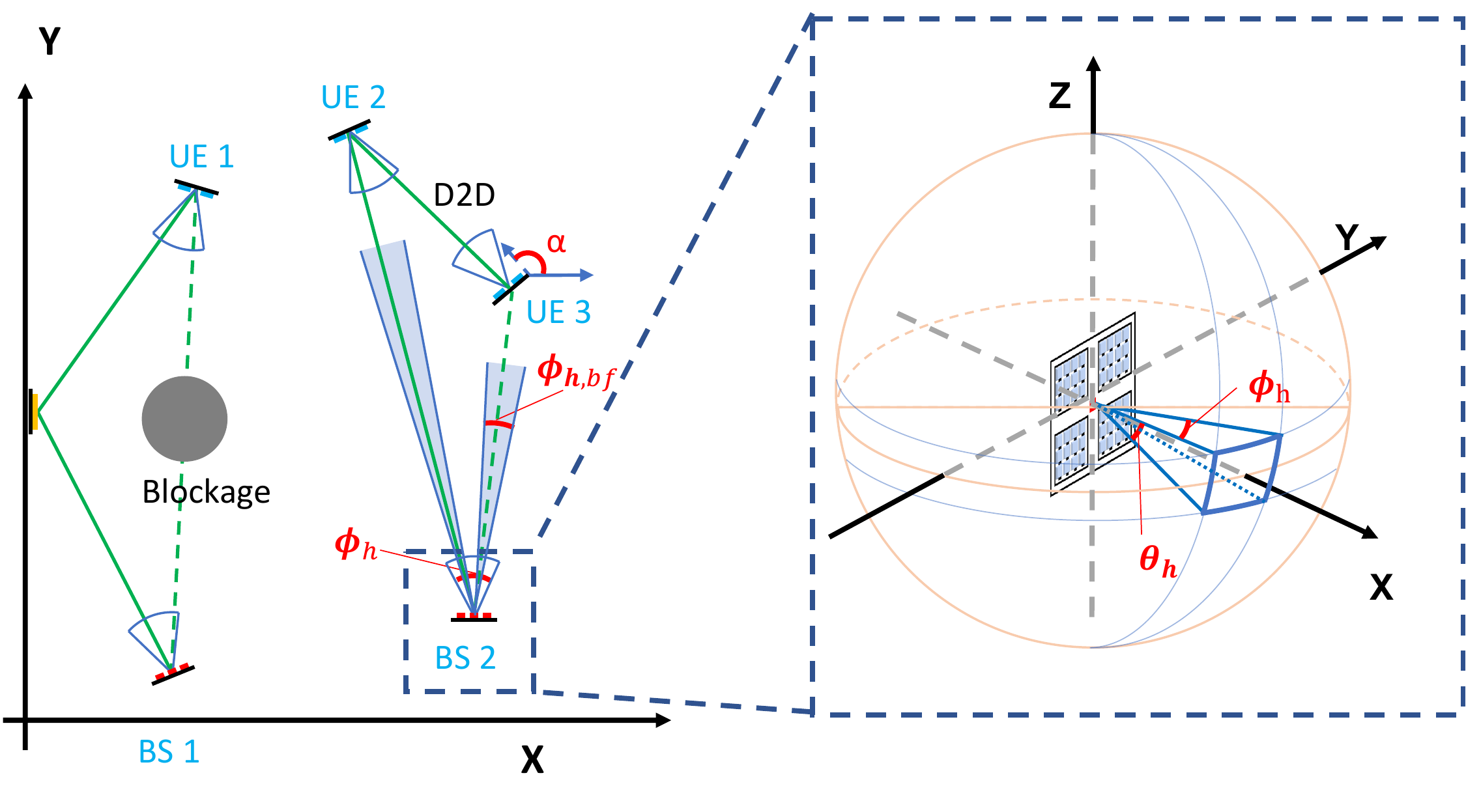}
\caption{Examplar top view of multiple BSs and UEs.}
\label{fig:system_top_view}
\end{figure}

By using an ideal sector model (ISM) \cite{lin2015indoor}, the antenna gains $G_\mathrm{B}$/$G_\mathrm{U}$ of the LOS channel at BS/UE can be obtained as
\begin{equation}
   G_\mathrm{Q}(\tilde\vpv) 
   =\left\{
    \begin{array}{c l}
    \sqrt{G_\mathrm{Q}^{0}}, & \tilde \phi\in[-\frac{\phi_\text{h}}{2},\frac{\phi_\text{h}}{2}],\tilde \theta\in[-\frac{\theta_\text{h}}{2},\frac{\theta_\text{h}}{2}], \\
    0,& \text{otherwise},
    \end{array}
\right. 
\label{eq_sector_model}
\end{equation}
where $G_\mathrm{Q}^{0}$ is the antenna gain, {$\tilde\vpv$ is the local AOA/AOD defined in~\eqref{eq:local_doa_dod},} and $\phi_\text{h}$, $\theta_\text{h}$ are the half-power beamwidth (HPBW) at E-plane and H-plane, respectively. For omnidirectional antennas, $G_\mathrm{Q}(\tilde\theta, \tilde\phi)=1$ and can hence be ignored. For highly directional antennas, however, the directivity can be approximated as $G^{0} \approx \frac{4\pi}{\theta_\text{h}\phi_\text{h}}$~\cite{xia2019link}. A top-view illustration of the antenna sector gain is shown in Fig.~\ref{fig:system_top_view}. Note that the beamforming HPBWs $\phi_{h,bf}/\theta_{h,bf}$ in Fig.~\ref{fig:system_top_view} are different from $\phi_h/\theta_h$ that are decided by the array size.
The antenna gain can also be characterized by a Gaussian beam model (GBM)~\cite{priebe2012impact}.

The frequency-dependent steering vectors $\av_{\mathrm{B}}(\tilde\vpv_\mathrm{BU})$ and $\av_{\mathrm{U}}^T(\tilde\vpv_\mathrm{UB})$ can be described as 
\begin{equation}
    \av_\mathrm{Q}(\tilde\vpv) = [a_\mathrm{Q}(1), \cdots, a_\mathrm{Q}(q),\cdots,a_\mathrm{Q}(N_\mathrm{Q})]^T,
    \label{eq:array_steering_vector}
\end{equation}
where the $q$th element can be obtained as
\begin{align}
    a_{\mathrm{Q}}(q)
    & = e^{j\frac{2\pi f}{c}\Psi_q(\tilde{\vpv})} 
    = e^{j\frac{2\pi f}{c}\tilde\pv_{q}^T \tv(\tilde\vpv)}
    \label{eq:array_steering_local_representation}\\
    & 
    = e^{j\frac{2\pi f}{c}(\Rm_\mathrm{Q}^T(\pv_q - \pv_\mathrm{Q}))^T(\Rm^T \tv(\vpv))}
    = e^{j\frac{2\pi f}{c}(\pv_q - \pv_\mathrm{Q})^T \tv(\vpv)}.
    \label{eq:array_steering_global_representation}
\end{align}
The mapping from angles to a direction vector $\tv(\vpv)$ can be found in~\eqref{eq:dir_vec_from_angle}. Equations~\eqref{eq:array_steering_local_representation} and~\eqref{eq:array_steering_global_representation} describe the steering vectors using local and global angle pairs, respectively; this steering vector applies to arrays of arbitrary layouts.

In this work, we assume identical attenuation coefficients across all subcarriers, $\mathcal{K}(f, d) = \mathcal{K}(f_c, d)$, and ignore the Doppler effect. The resultant frequency-flat fading channel at the $k$th subcarrier can be expressed from~\eqref{eq:far_field_los_channel_doppler} as
\begin{equation}
    \Hm_{L}[k] = c_k\rho_{L} e^{-j\xi_{L}} e^{-j2\pi \Delta f_k\tau_{L}} \av_{\mathrm{B}}(\tilde\vpv_\mathrm{BU}) \av_{\mathrm{U}}^T(\tilde\vpv_\mathrm{UB}),
    \label{eq:far_field_los_channel}
\end{equation}
where $c_k = \frac{f_c}{f_k}$ is the frequency ratio between the central frequency and subcarrier frequency, $f_k \!=\! f_c \!+\! \Delta f_k \!=\! f_c \!+\!\frac{(2k-1-K)W}{2K}$ is the frequency of the $k$th subcarrier, and  $\av_{\mathrm{B}}(\tilde\vpv_\mathrm{BU})$ and  $\av_{\mathrm{U}}(\tilde\vpv_\mathrm{UB})$ are the steering vectors of the $k$th subcarrier that can be obtained from~\eqref{eq:array_steering_vector} by replacing $f$ with $f_k$. The complex channel gain $\rho_\mathrm{L}e^{-j\xi_\mathrm{L}}$ and the signal delay $\tau_{\mathrm{L}}$ of the LOS path are 
\begin{align}
    \rho_{\mathrm{L}} & = \frac{\lambda_c}{4\pi  d_{\mathrm{BU}}} 
    \mathcal{K}_a(f_c, d_{\mathrm{BU}})G_{\mathrm{B}}(\tilde\vpv_\mathrm{BU})
    G_{\mathrm{U}}(\tilde\vpv_\mathrm{UB}),
    \label{eq:far_field_los_channel_rho}\\
    \xi_\mathrm{L} & = 2\pi f_c \tau_{\mathrm{L}},
    \label{eq:far_field_los_channel_xi}\\
    \tau_\mathrm{L} & = \frac{d_\mathrm{BU}}{c} + B.
    \label{eq:far_field_los_channel_tau}
\end{align}
Here, $c$ is the speed of the light and $B$ is the synchronization offset.



\begin{remark}
\label{remark_1}
By setting $B=0$, we assume the system is well-synchronized. The synchronization $B$ is identical for all the channels over a BS-UE communication link. However, for a localization system with multiple asynchronized BSs, each BS may have its own synchronization offset.
\end{remark}

\begin{remark}
\label{remark_2}
Note that even though the complex channel gains of the LOS channel ($\rho_\mathrm{L}$/$\xi_\mathrm{L}$) can be expressed using geometry information, they are usually treated as unknowns to be estimated\cite{wymeersch20185g,abu2018error,shahmansoori2017position} due to the hardware imperfection (e.g., antenna gain, phase noise) and dynamic communication environments. We call this scenario as `unknown' channel model. In addition, we define a `partially known' model as one where only $\xi$ is treated as unknown). Here, the range information can be inferred from the channel gain $\rho$, which will be discussed in Sec.~\ref{sec:pwm_swm_simulation}.
\end{remark}

\subsubsection{RIS Channel}
In the RIS channel, the transmitted signal first arrives at an RIS through an RIS-UE channel. With its amplitude and phase changed by the RIS elements, the signal is then transmitted to the receiver via a BS-RIS channel. The RIS channel matrix $\Hm_R$ can be expressed as~\cite{wu2021intelligent,wymeersc2020beyond}
\begin{equation}
    \Hm_\mathrm{R} = \Hm_{\mathrm{BR}} \Omegam \Hm_{\mathrm{RU}},
    \label{eq:ris_channel_matrix}
\end{equation}
where $\Hm_{\mathrm{BR}}\in \mathbb{C}^{N_\mathrm{B}\times N_\mathrm{R}}$ is the BS-RIS channel matrix, and $\Hm_\mathrm{RU}\in \mathbb{C}^{N_\mathrm{R}\times N_\mathrm{U}}$ is the RIS-UE channel matrix that can be similarly obtained from~\eqref{eq:far_field_los_channel}.
The coefficient matrix $\Omegam \in \mathbb{C}^{N_\mathrm{R}\times N_\mathrm{R}}$ is a diagonal matrix that can be denoted as $\Omegam \triangleq \mathrm{diag(\beta_1e^{j\omega_1}, \cdots, \beta_{N_\mathrm{R}} e^{j\omega_{N_\mathrm{R}}})}$, where $\omega_n \in[0, 2\pi)$ and $\beta_n \in[0,1]$ are the phase shift and the reflection coefficient, reflectively~\cite{wu2019intelligent}. By assuming the power radiation pattern and the RIS element gain is equal to 1, and assuming the area of each element to be $\frac{\lambda^2}{4\pi}$, the RIS channel can be modeled from~\cite{tang2020wireless} (equation (3)) as
\begin{equation}
    \begin{split}
    \Hm_\mathrm{R}[k] = & c_k^2\rho_{R} e^{-j\xi_{R}} e^{-j2\pi \Delta f_k\tau_{R}}\\
    & \times
    \av_{\mathrm{B}}(\tilde\vpv_\mathrm{BR}) \av_{\mathrm{R}}^T(\tilde\vpv_\mathrm{RB})\Omegam
    \av_{\mathrm{R}}(\tilde\vpv_\mathrm{RU}) \av_{\mathrm{U}}^T(\tilde\vpv_\mathrm{UR}),
    \label{eq:far_field_ris_channel}
    \end{split}
\end{equation}
where
\begin{align}
    \begin{split}
    \rho_{\mathrm{R}} = & \frac{\lambda_c^2}{16\pi^2 d_{\mathrm{BR}}d_{\mathrm{UR}}}\mathcal{K}_a(f_c,d_{\mathrm{BR}})
    \mathcal{K}_a(f_c,d_{\mathrm{UR}})\\
    & \times 
    G_{\mathrm{B}}(\tilde\vpv_\mathrm{BR})
    G_{\mathrm{R}}(\tilde\vpv_\mathrm{RB})
    G_{\mathrm{R}}(\tilde\vpv_\mathrm{RU})
    G_{\mathrm{U}}(\tilde\vpv_\mathrm{UR}),
    \end{split}
    \label{eq:far_field_ris_channel_amplitude}\\
    \xi_\mathrm{R} & = 2\pi f_c \tau_{\mathrm{R}}\\
    \tau_\mathrm{R} & = \frac{d_\mathrm{BR} + d_\mathrm{UR}}{c} + B.
\end{align}

\subsubsection{NLOS Channels}
\label{sec:nlos_channel}
The transmitted signal might be reflected by other objects and arrive at the receiver through NLOS paths. In addition to the channel gains, reflection coefficients should be considered. For mmWave systems, the reflection coefficients can be obtained based on the reflector geometry and the reflection loss statistics~\cite{li2014channel, fascista2019millimeter, abu2018error}. For terahertz-band systems, the NLOS channels differ in two aspects: firstly, the NLOS paths become increasingly sparse and lossy~\cite{ma2019terahertz}; secondly, the surfaces that are considered smooth at lower frequencies become rough, so NLOS paths can also be generated by scattered rays and diffracted rays~\cite{han2014multi}. The THz NLOS channels can be characterized using stochastic models~\cite{priebe2011aoa,priebe2013stochastic}, or ray-tracing models~\cite{priebe2013stochastic}. 
More details are obtained from real measurements by evaluating the partition losses for different materials under different frequencies~\cite{rappaport2019wireless,ju2019scattering,ju2021millimeter}.

For localization purposes, we consider $L_N$ resolvable reflectors between the BS and UE with NLOS coefficients $\mathcal{K}_N^\ssnb{l}$; no intra-cluster rays are considered (i.e., one ray for each reflector). We ignore second-order reflections where attenuation is more than \unit[15]{dB}~\cite{priebe2013stochastic}. The NLOS matrix can be expressed as the sum of all the $L_N$ paths, $\Hm_N = \sum_{l=1}^{L_N}\Hm_N^{\ssnb{l}}$, where the channel matrix of the $l$th NLOS path is
\begin{equation}
    \Hm_\mathrm{N}^{\ssnb{l}}[k] \!=\! c_k\rho^{\ssnb{l}}_\mathrm{N} e^{-j\xi^{\ssnb{l}}_\mathrm{N}} e^{-j2\pi \Delta f_k\tau^{\ssnb{l}}_\mathrm{N}}
    \av_{\mathrm{R}}(\tilde\vpv_\mathrm{RU}) \av_{\mathrm{U}}^T(\tilde\vpv_\mathrm{UR}),
    \label{eq:far_field_nlos_channel}
\end{equation}
and where
\begin{align}
    \rho_{\mathrm{N}} & =
    \frac{\lambda_c}{4\pi d^\ssnb{l}_{\mathrm{N}}}
    \mathcal{K}^\ssnb{l}_\mathrm{N}
    \mathcal{K}^\ssnb{l}_a(f_c, d_\mathrm{N}^\ssnb{l})
    G^\ssnb{l}_\mathrm{B}(\tilde \vpv_\mathrm{BN})
    G^\ssnb{l}_\mathrm{U}(\tilde \vpv_\mathrm{UM}),\\
    \xi_{\mathrm{N}}^{\ssnb{l}} & = 2\pi f_c \tau^{\ssnb{l}}_{\mathrm{N}},\\
    \tau_\mathrm{N}^{\ssnb{l}} & = \frac{d_\mathrm{BN}^\ssnb{l} + d_\mathrm{NU}^\ssnb{l}}{c} + B.
    \label{eq:nlos_channel_delta_tau}
\end{align}
The scattered and diffracted paths are modeled in~\cite{han2014multi}, with the major difference being the NLOS coefficient $\rho_{\mathrm{N}}$.

So far, we have described the LOS channel, RIS channel, and NLOS channels of a far-field MIMO system in~\eqref{eq:far_field_los_channel}, \eqref{eq:far_field_ris_channel}, and \eqref{eq:far_field_nlos_channel}, respectively. In the next subsection, we will describe two received signal models, namely, conventional MIMO and AOSA-based MIMO.


\subsection{Received Signal Model}
\label{sec:received_signal_model}
\subsubsection{MIMO Architectures}
In low-frequency systems, signal processing is mainly performed at the baseband. In this case, each antenna is connected to an~\ac{rfc}, and the received signal at each antenna can be obtained directly through a low-pass filter and an \ac{adc}. With the increased carrier frequency, an antenna array with a large dimension is preferred to benefit from the beamforming gain; such conventional architectures are impractical in the aspects of hardware realization and power consumption~\cite{heath2016overview}.

Hybrid MIMO structures that provide a good tradeoff between system performance and cost  are described in~\cite{heath2016overview}. The data symbols are first precoded by a baseband (digital) precoder, followed by analog beamforming using phase-shifters, switchers, or a lens antenna~\cite{heath2016overview, mendez2016hybrid}.
A fully connected array and AOSA are two possible hybrid structures. For instance, with phase-shifter-based fully-connected arrays, each available \ac{rfc} is connected to all antennas via an individual group of phase-shifters~\cite{lin2016terahertz}. In the AOSA-based structure, each RFC can only drive a portion of the antennas forming an SA. It is shown in~\cite{lin2016terahertz} that compared with the fully connected array, AOSA structures perform better in spectral efficiency and energy efficiency and hence are more suitable for terahertz UM-MIMO systems~\cite{han2018ultra,lin2017subarray,lin2015indoor,lin2016terahertz, sarieddeen2021overview, akyildiz2018combating, tarboush2021teramimo}. 
In the following, we detail the received signal models using fully-connected and AOSA-based structures.

\subsubsection{Fully-digital MIMO Model}
Assume that in a MIMO system, each antenna is connected to an independent RFC. Let $P$ be the average transmission power (in mW). Based on the proposed far-field MIMO channel from~\eqref{eq:far_field_channel_model}-\eqref{eq:nlos_channel_delta_tau}, the received signal at the $k$th subcarrier and the $g$th transmission {(OFDM symbols)}, $\yv^{\ssnb{g}}[k]$, can be expressed as
\begin{equation}
    \begin{split}
        \yv^{\ssnb{g}}[k] 
        & = \sqrt{P}\Hm[k]\xv^{\ssnb{g}}[k] + \nv^{\ssnb{g}}[k]\\
        & = \muv^\ssnb{g}[k] + \nv^{\ssnb{g}}[k],
    \end{split}
    \label{eq:conventional_signal_model}
\end{equation}
where $\muv^{\ssnb{g}}[k]\in \mathbb{C}^{N_{\mathrm{B}}\times 1}$ is the noise-free version of the received signal, $\nv^{\ssnb{g}}[k]\in \mathbb{C}^{N_{\mathrm{B}}\times 1}$ is the additive white Gaussian noise (AWGN) vector with a complex normal distribution $\mathcal{CN}(0, \sigma_n^2)$, and the normalized transmitted signal vector $\xv^{(g)}[k]\in \mathbb{C}^{N_\mathrm{U}\times1}$ ({$\Vert\xv^\ssnb{g}[k]\Vert^2\!=\!{1}$})\footnote{{If total transmission energy across all the subcarriers and transmissions is assumed, the transmitted signal is normalized as $\Vert\xv^\ssnb{g}[k]\Vert^2\!=\!\frac{1}{K\mathcal{G}}$.}} can be chosen randomly or obtained using a directional beamforming matrix with \acp{prs} as discussed in\cite{abu2018error,elzanaty2021reconfigurable,shahmansoori2017position,he2020large}. 
Although this MIMO structure is impractical when the array size is large, it helps derive the fundamental limits of MIMO localization systems and is thus widely used in localization works~\cite{wymeersch2020fisher,wen20205g,abu2018error,he2020large,elzanaty2021reconfigurable,shahmansoori2017position}. We use~\eqref{eq:conventional_signal_model} as the benchmark MIMO signal model for mmWave systems. Next, we detail the hybrid MIMO model, namely, the fully-connected hybrid model and the AOSA-based MIMO model.

\subsubsection{Fully-connected Hybrid MIMO Model}
Due to the hardware cost and complexity, it is impractical to connect each antenna with an RFC. Usually, a smaller number of $M_\mathrm{B}$/$M_\mathrm{U}$ RFCs are implemented to connect all the $N_\mathrm{B}$/$N_\mathrm{U}$ antennas at the BS/UE arrays ($M_\mathrm{B}\!\!\le\!\! N_\mathrm{B}$, $M_\mathrm{U}\!\!\le\!\! N_\mathrm{U}$). The received signal can be changed from~\eqref{eq:conventional_signal_model} as
\begin{equation}
\begin{split}
    \yv^{\ssnb{g}}[k] 
    & = \sqrt{P}\Wm_\mathrm{B}^T\Hm[k]\Wm_\mathrm{U}\xv_0^{\ssnb{g}}[k] + \Wm_\mathrm{B}^T\nv^{\ssnb{g}}[k],
\end{split}
\label{eq:fully_connected_signal_model}
\end{equation}
where $\Wm_\mathrm{B}\in \mathbb{C}^{N_\mathrm{B}\times M_\mathrm{B}}$ is the RF combiner matrix at the BS, $\Wm_\mathrm{U}\in \mathbb{C}^{N_\mathrm{U}\times M_\mathrm{U}}$ is the RF precoder matrix at the UE, and $\xv_0^{(g)}[k]\in \mathbb{C}^{M_\mathrm{U}\times 1}$ is the signal symbol vector before the precoder. Let $\xv^{(g)}[k] = \Wm_\mathrm{U}\xv_0^{\ssnb{g}}[k]$ be the transmitted signal vector, the transmission power constraint still holds as $\Vert\xv^\ssnb{g}[k]\Vert^2\!=\!{1}$. {Note that the baseband combiner/precoder matrices are not discussed here as we process the measurement from the RFCs directly for localization. All the entries of $\Wm_\mathrm{B}$ and $\Wm_\mathrm{U}$ correspond to the coefficients of phase-shifters with amplitudes $|[\Wm_\mathrm{B,rf}]_{i,j}| = \frac{1}{\sqrt{N_\mathrm{B}}}$ and $|[\Wm_\mathrm{U}]_{i,j}| = \frac{1}{\sqrt{N_\mathrm{U}}}$.}
The fully-connected MIMO structures are widely-adopted in mmWave systems. However, this type of hybrid structure is impractical and inefficient for THz systems because of the limitation of the transmit power and circuit feeding ability~\cite{lin2016terahertz}. Next, we explain the AOSA-based structure, which is more favored in THz systems.

\subsubsection{AOSA-based MIMO Model}
\label{sec:aosa_signal_model}
In AOSA-based MIMO structures, the antenna array is divided into several SAs~\cite{lin2016terahertz}. Each SA effectively represents the minimum communication element, which is driven by an independent RFC. Compared with the conventional MIMO structure, the number of RFCs needed equals the number of SAs rather than the number of antennas. {And differing from the fully-connected hybrid structure, each RFC is connected to a subset of the antennas instead of all the antennas.}
Within each SA, analog beamforming is used to focus a signal in a certain direction. Consequently, the system benefits from the beamforming gain, and optimizing phase-shifters per antenna element (AE) reduces optimizing the beamforming angles per SA. Note that all the phase-shifters in a SA can be arbitrarily optimized to perform beamforming in different directions. However, independently optimizing each phase-shifter increases the system's computational complexity and interference with other users. It is thus practical to treat each SA as a minimal communication element with a fixed beamforming angle with high carrier frequencies~\cite{lin2016terahertz,lin2015indoor}. Note that AOSA architectures can also be applied to mmWave communications, but they are more of a requirement at extremely high frequencies where a minimum beamforming gain has to be met to achieve reasonable communication distances.


Without loss of generality, we use $\mathcal{A}_\mathrm{Q}(\tilde \vpv, \mathring\vpv)$ to represent the array factor of a specific SA containing $\mathring N_\mathrm{Q}$ AEs ($Q\in\{B, U\}$). The array factor reflects the beamforming gain obtained by a specific SA with beamforming angle $\mathring \vpv$, in a specific channel with local AOA/AOD $\tilde\vpv$; we have~\cite{lin2015indoor}
\begin{equation}
\begin{split}
    \mathcal{A}_\mathrm{Q}(\tilde \vpv, \mathring\vpv) 
    & = \frac{1}{{\sqrt{\mathring N_\mathrm{Q}}}}\av_\mathrm{st}^T(\tilde \vpv)\av_\mathrm{bf}(\mathring\vpv)\\
    & = \frac{1}{{\sqrt{\mathring N_\mathrm{Q}}}}\sum^{\mathring N_\mathrm{Q}}_{\mathring q=1} e^{j\frac{2\pi f}{c}(\Psi_{\mathring q}(\tilde\vpv)-\Psi_{\mathring q}(\mathring \vpv))}.
\end{split}
\label{eq:array_factor}
\end{equation}
Here, $\Psi_{\mathring q}(\tilde{\vpv}) = \tilde\pv_{\mathring q}^T \tv(\tilde\vpv)$ is the signal delay at the $\mathring q$th element with respect to the SA center. The SA level steering vector $\av_\mathrm{st}(\tilde\vpv)$ and beamforming vector $\av_{bf}(\mathring \vpv)$ can be similarly obtained from~\eqref{eq:sa_steering_vector} as
\begin{align}
    \av_\mathrm{st}(\tilde\vpv) 
    & = [a_\mathrm{st}(1), \cdots, a_\mathrm{st}(\mathring q),\cdots,a_\mathrm{st}(\mathring N_\mathrm{Q})]^T,
    \label{eq:sa_steering_vector}\\
    a_\mathrm{st}(\mathring q) 
    & = e^{j\frac{2\pi f}{c}{\Psi}_{\mathring q}({\tilde\vpv})} = e^{j\frac{2\pi f}{c}\tilde \pv_{\mathring q}^T \tv(\tilde \vpv)},
    \label{eq:sa_steering_vector_element}
\end{align}
and 
\begin{align}
    \av_\mathrm{bf}(\mathring\vpv) & = [a_\mathrm{bf}(1), \cdots, a_\mathrm{bf}(\mathring q),\cdots,a_\mathrm{bf}(\mathring N_\mathrm{Q})]^T,
    \label{eq:sa_beamforming_vector} \\
    a_\mathrm{bf}(\mathring q) & = e^{-j\frac{2\pi f}{c}{\Psi}_{\mathring q}({\mathring\vpv})} = e^{-j\frac{2\pi f}{c}\tilde \pv_{\mathring q}^T \tv(\mathring \vpv)}.
    \label{eq:element_of_sa_beamforming_vector}
\end{align}
Note that the beamforming angle of the $q$th SA, $\mathring \vpv_q$, does not depend on the geometry information of other devices, and $\mathcal{A}(\tilde\vpv,\mathring \vpv)$ achieves the maximum beamforming gain when $\mathring \vpv \!=\! \tilde\vpv$. 
Both the steering and beamforming vectors can also be expressed using global angles and positions as in~\eqref{eq:array_steering_global_representation}.

The signal model of an AOSA-based MIMO structure can be expressed as
\begin{equation}
    \yv^{\ssnb{g}}[k] 
    \!=\! \boldsymbol{\mathcal{H}}[k]\xv^{\ssnb{g}}[k] + \nv^{\ssnb{g}}[k],
\label{eq:aosa_signal_model}
\end{equation}
where $\boldsymbol{\mathcal{H}}[k]\in \mathbb{C}^{N_\mathrm{B}\times N_\mathrm{U}}$ ($N_\mathrm{B}$ and $N_\mathrm{U}$ are the number of SAs in the AOSA structure) is the effective AOSA channel
\begin{equation}
\begin{split}
    \Hbc = & 
    \Abc_\mathrm{L}\!\odot\! \Hbc_\mathrm{L}[k] \!+\! (\Abc_\mathrm{BR}\!\odot\! \Hbc_\mathrm{BR}[k])\Omegam (\Abc_\mathrm{RU}\!\odot\! \Hbc_{RU}[k])\\ 
    & \!+\! \sum_{l=1}^{L_\mathrm{N}} \Abc^{\ssnb{l}}_\mathrm{N}\!\odot\! \Hbc_\mathrm{N}^{\ssnb{l}}[k].
\end{split}
\label{eq:AOSA_signal_model}
\end{equation}

In each path, the effective AOSA channel matrix can be expressed as the Hadamard product of an array factor matrix $\Abc_L$ (or $\Abc_\mathrm{BR}$, $\Abc_\mathrm{RU}$ $\Abc_\mathrm{L}^\ssnb{l}$) and an SA level channel matrix $\Hbc_L$ (or $\Hbc_\mathrm{BR}$, $\Hbc_\mathrm{RU}$, $\Hbc_\mathrm{N}^\ssnb{l}$) by taking the SA as the basic communication element. The SA level channel matrices ($\Hbc$) can be obtained from the AE level far-field MIMO channel matrices ($\Hm$) from~\eqref{eq:far_field_los_channel},~\eqref{eq:far_field_ris_channel}, and \eqref{eq:far_field_nlos_channel} by changing the parameters from AE to SA (e.g., number of SAs, the position of SA centers, and SA spacing). The array factor matrices describe the beamforming gains of each SA as
\begin{align}
    \Acal_{\mathrm{L},bu} & = \Acal_\mathrm{B}({\tilde\vpv_\mathrm{BU}}, \mathring \vpv_b) \Acal_\mathrm{U}({\tilde \vpv_\mathrm{UB}}, \mathring \vpv_u),
    \label{eq:array_factor_PWM_LOS}\\
    \Acal_{\mathrm{BR},br} & = \Acal_\mathrm{B}({\tilde\vpv_\mathrm{BR}}, \mathring \vpv_b),\\
    \Acal_{\mathrm{RU},ru} & = \Acal_\mathrm{U}({\tilde\vpv_\mathrm{UR}}, \mathring \vpv_u),\\
    \Acal^\ssnb{l}_{\mathrm{N},bu} & = \Acal^\ssnb{l}_\mathrm{B}({\tilde\vpv^\ssnb{l}_\mathrm{BN}}, \mathring \vpv^\ssnb{l}_b) 
    \Acal^\ssnb{l}_\mathrm{U}({\tilde \vpv^\ssnb{l}_\mathrm{UN}}, \mathring \vpv^\ssnb{l}_u).
    \label{eq:array_factor_PWM_NLOS}
\end{align}

In this subsection, we provided two receive signal models, namely, conventional fully-connected and AOSA-based MIMO models, in far-field scenarios. In what follows, we describe the extension of the signal model into near-field by assuming a spherical wave model (SWM). Other channel features are also discussed to make the signal model flexible enough for different types of signals.

\subsection{Additional Model Features}
\label{sec:additional_model_features}



\subsubsection{Near-field Channel Model}
\label{sec:near_field_channel_model}
The far-field model is considered when the range between the transceivers is much larger than the size of the array~\cite{friedlander2019localization}. {The near-field region is usually defined as the range between the Fresnel boundary $0.62\sqrt{{D^3}/{\lambda}}$ and Fraunhofer distance ${2D^2}/{\lambda}$, where $D$ is the diameter of the antenna array~\cite{guerra2021near}. With higher carrier frequencies (e.g., THz-band signals), even a small footprint can result in a larger array size (in terms of wavelength, see footnote~\ref{fn:array_terms}) and hence a larger near-field range.} In the latter case, a far-field channel model is no longer accurate. We next describe the near-field channel model.


In the near-field channel model, we also make the assumption that all the antennas receive the same signal strength (the amplitude $\rho$ applies to all array antennas). Then, what differentiates it from a far-field model is the phase change of the received signals. Take the LOS channel $\Hm_\mathrm{L}[k]$, for example, and ignore the clock offset $B$. Each element of the matrix, $h_{\mathrm{L},bu}[k]$ ($b$th row, $u$th column), under the SWM and PWM assumptions can be written as
\begin{align}
    h^{\ssr{SWM}}_{\mathrm{L},bu}[k] & = c_k\rho_\mathrm{L} e^{-j\frac{2\pi f_k}{c} d_{bu}} = c_k\rho_\mathrm{L} e^{-j\frac{2\pi f_k}{c} \norm{\pv_b-\pv_u}}, 
    \label{eq:SWM_channel_element}\\
    h^{\ssr{PWM}}_{\mathrm{L},bu}[k] & = c_k\rho_\mathrm{L} e^{-j\frac{2\pi f_k}{c} (\norm{\pv_\mathrm{B}-\pv_\mathrm{U}} - \tilde\pv_{b}^T\tv(\tilde\vpv_{\mathrm{BU}}) - \tilde \pv_{u}^T\tv(\tilde\vpv_{\mathrm{UB}}))},
    \label{eq:PWM_channel_element}
\end{align}
respectively. Note that equation~\eqref{eq:PWM_channel_element} {is identical to} the far-field channel model as~\eqref{eq:far_field_los_channel}. {By extracting the common parts for different subcarriers into complex channel gain $\rho e^{-j\xi}$, the elements of near-field channel matrices in~\eqref{eq:SWM_channel_element}} can be written from~\eqref{eq:far_field_los_channel}, \eqref{eq:far_field_ris_channel}, and \eqref{eq:far_field_nlos_channel} as
\begin{align}
    h^{\ssr{SWM}}_{\mathrm{L},bu}[k] & =  c_k\rho_{\mathrm{L}}e^{-j\xi_\mathrm{L}}e^{-j2\pi(\Delta f_k \tau_\mathrm{L} + f_k \Delta\tau_{bu})},\\
    h^{\ssr{SWM}}_{\mathrm{R},bu}[k] & =  c_k\rho_{\mathrm{R}}e^{-j\xi_\mathrm{R}}\sum_{r=1}^{N_\mathrm{R}}e^{-j2\pi(\Delta f_k \tau_\mathrm{R} + f_k \Delta\tau_{bru})},\\
    h^{{\ssnb{l}}\ssr{SWM}}_{\mathrm{N},bu}[k] & =  c_k\rho^{\ssnb{l}}_{\mathrm{N}}e^{-j\xi^{\ssnb{l}}_\mathrm{N}}e^{-j2\pi(\Delta f_k \tau^{\ssnb{l}}_\mathrm{N} + f_k \Delta\tau^{\ssnb{l}}_{bnu})},
\end{align}
where 
\begin{align}
    \Delta\tau_{bu} & = \tau_{bu} \!-\! \tau_\mathrm{L} = \frac{d_{bu}- d_\mathrm{BU}}{c},\\
    \Delta\tau_{bru} & = \tau_{bru} - \tau_\mathrm{R}= \frac{d_\mathrm{br} + d_\mathrm{rm}- d_\mathrm{BR} - d_\mathrm{RU}}{c},\\
    \Delta\tau^{\ssnb{l}}_{bnu} & = \tau^{\ssnb{l}}_{bnu} - \tau^{\ssnb{l}}_\mathrm{N}= \frac{d^{\ssnb{l}}_{bn}+d^{\ssnb{l}}_{nu}- d^{\ssnb{l}}_\mathrm{BN}-d^{\ssnb{l}}_\mathrm{NU}}{c}.
\end{align}

For the near-field AOSA-based MIMO model in~\eqref{eq:AOSA_signal_model}, we further assume that the channel model for each SA follows a PWM. The SWM is then captured by the phase differences between the SAs. This is a reasonable assumption since the size of SAs is relatively small compared to that of the whole array. Hence, the array factor matrices can be updated as
\begin{align}
    \Acal_{\mathrm{L},bu} & = \Acal_\mathrm{B}({\tilde\vpv_{bu}}, \mathring \vpv_b) \Acal_\mathrm{U}({\tilde \vpv_{ub}}, \mathring \vpv_u),
    \label{eq:array_factor_SWM_LOS}\\
    \Acal_{\mathrm{BR},br} & = \Acal_\mathrm{B}({\tilde\vpv_{br}}, \mathring \vpv_b),\\
    \Acal_{\mathrm{RU},ru} & = \Acal_\mathrm{U}({\tilde\vpv_{ur}}, \mathring \vpv_u),\\
    \Acal^\ssnb{l}_{\mathrm{N},bu} & = \Acal^\ssnb{l}_\mathrm{B}({\tilde\vpv^\ssnb{l}_{bn}}, \mathring \vpv^\ssnb{l}_b) 
    \Acal^\ssnb{l}_\mathrm{U}({\tilde \vpv^\ssnb{l}_{nu}}, \mathring \vpv^\ssnb{l}_u).
    \label{eq:array_factor_SWM_NLOS}
\end{align}
Compared with the array factors in~\eqref{eq:array_factor_PWM_LOS}-\eqref{eq:array_factor_PWM_NLOS} where the AOA/AOD pairs are calculated based on the array center, the near-field array factors need to calculate the angles for each of the SA pairs.

\subsubsection{Beam Split Effect}
\label{sec:beam_split}
The \ac{bse}, also known as the beam squint effect, is caused by frequency-independent (constant) phase shifts in analog beamforming~\cite{tan2019delay, lin2017subarray, heath2016overview}. For a narrow band system, the steering vectors calculated from~\eqref{eq:array_steering_vector} and~\eqref{eq:sa_steering_vector} are frequency-independent (i.e., $f=f_c$). However, for a wideband system, the steering vectors are frequency-dependent. When pure phase-shifters are utilized, the phase shift is constant for different subcarriers in analog beamforming. The array factor in~\eqref{eq:array_factor} is then modified by multiplying $\Psi_{\mathring q}(\tilde\vpv)$ with $1/c_k = f_k/f_c$ as
\begin{equation}
    A_\mathrm{Q}(\tilde\vpv,\mathring\vpv) = \frac{1}{\sqrt{\mathring N_\mathrm{Q}}}\sum^{\mathring N_\mathrm{Q}}_{\mathring q=1} e^{j\frac{2\pi f_c}{c}(\frac{f_k}{f_c}\Psi_{\mathring q}(\tilde\vpv)-\Psi_{\mathring q}(\mathring \vpv))}.
    \label{eq:Aeq_summation_beamsplit}
\end{equation}

The beam split is affected by three factors, namely, array size (in wavelength), bandwidth, and beamforming angle~\cite{tarboush2021teramimo}.
From equation~\eqref{eq:Aeq_summation_beamsplit}, the highest beamforming gain reaches $\sqrt{\mathring N_\mathrm{Q}}$ only at the central frequency $f_c$ where the beamforming angle vector $\mathring \vpv$ equals the steering angle vector $\tilde \vpv$, whereas the other subcarriers suffer from performance loss.
Several techniques such as true-time delays (TTD)~\cite{chu2013true} and delay-phase precoding (DPP)~\cite{tan2021wideband} can be utilized to overcome beam slit. However, we retain this feature in channel modeling to account for the systems with pure phase-shifters. Note that the \ac{bse} can also be considered in the RIS channel and the coefficient matrix $\Omegam$ in~\eqref{eq:ris_channel_matrix} will be frequency-dependent.


\subsubsection{Hardware Imperfections}
Hardware imperfections are caused by components mismatch and manufacturing defects~\cite{boulogeorgos2019analytical} and may occur in the RFC, phase-shifters, and RIS elements.
We model several types of hardware imperfections that can affect localization performance.

\begin{itemize}

    \item RFC impairments: The impairment noise caused by RFCs distorts the signal at the transmitter and the receiver~\cite{bjornson2013new}. The received signal can be modeled as
    \begin{equation}
        \yv = \Hm(\xv + \nv_t) + \nv_r + \nv,
    \end{equation}
    where $\nv_t\sim \mathcal{CN}(0, \kappa_t^2 \bar P)$, $\nv_r\sim \mathcal{CN}(0, \kappa_r^2 \bar P|h|^2)$ are the distortion noises from impairments at Tx and Rx. $\kappa_t$, $\kappa_t$, $\bar P$, ${|h|}^2$ are the Tx impairment coefficient, Rx impairment coefficient, average transmission power and instantaneous channel gain, respectively~\cite{schenk2008rf, bjornson2013new, sarieddeen2021overview}.
    
     
    \item Phase noise (PN): The presence of PN has a significant impact on localization performance, especially when the target resolution is high~\cite{zheng2020joint}. The PS can be modeled as
    \begin{equation}
        \yv = \Omegam_{\mathrm{P}}\Hm\xv + \nv,
    \end{equation}
    where $\Omegam_{\mathrm{P}} \triangleq \diag([e^{j\omega_{\mathrm{P},1}}, \cdots, e^{j\omega_{\mathrm{P},N_\mathrm{r}}}])$ contains the phase noise information and $N_r$ is the number of antennas at the receiver array. The PN vector $\omegav_\mathrm{P} = [\omega_{\mathrm{P},1}, \cdots, \omega_{\mathrm{P},N_r}]$ typically follows a zero mean jointly Gaussian distribution with $\omegav_\mathrm{P}\sim \mathcal{N}(0, \sigma_\theta^2\mathbf{I}_{N_r})$, where $\sigma_\theta^2$ (in $\unit[]{rad^2}$) is the oscillator variance.
    
    \item Quantization error: The material and hardware properties limit the accuracy and {control speed of the RIS profile (element coefficients)}~\cite{wymeersch2020radio}. For a quantized RIS element, however, the phase value $\omega$ in the coefficient matrix $\Omegam$ can only be chosen from a set of quantized values $\mathcal{Q}$ (e.g., $\mathcal{Q} = \{0, \pi/2, \pi, 3\pi/2\}$ for a 2-bit quantization). Although at the expense of accuracy loss, the power cost and system complexity can be reduced with such quantizations. Similar quantization can be added to phase-shifters (resulting in a quantized beamforming angle) and \acp{adc}.
\end{itemize}

Other sources of the impairments, such as I/Q imbalance and non-linearities, should also be modeled~\cite{schenk2008rf}. {With the introduced hardware imperfection, there will be a performance loss by using a mismatched model (ideal model without hardware imperfections) on the true data (observation of an impaired system)~\cite{fortunati2017performance}. \Ac{mcrb} can be used to derive the lower bound of using a mismatched model and related works can be found in~\cite{richmond2015parameter, roemer2020misspecified, chen2021mcrb}.}

\subsection{Summary}
In this section, we formulate an AOSA-based THz system model and highlight its unique features compared to traditional MIMO models:
\begin{itemize}
    \item We describe the system geometry to represent the relationships between the position and direction vectors, Euler angles and rotation matrices, and local and global AOAs/AODs.
    \item Building on a mmWave MIMO channel model, we detail a near-field effective AOSA channel model that reduces the complexity of UM-MIMO systems, which are potential structures in THz systems.
    \item We propose a THz signal model comprised of LOS, RIS, and NLOS channels. 
\item We discuss additional THz features such as beam split and hardware imperfections.
\end{itemize}
In the next section, we utilize the proposed signal model to formulate THz-band localization problems.

\section{Terahertz-Band Localization}
\label{sec:thz_signal_based_localization}
In this section, we first {formulate the localization problem} and present the {CRB} derivation based on our system model. Then, we describe {geometry-based localization}, such as direct localization and multi-stage localization. Afterward, we discuss potential {extensions of THz localization and sensing}, namely, learning-based localization, cooperative localization, tracking, and SLAM. Recent localization works using radio signals and their features are summarized in Table~\ref{tab:localization_work_summary}.

\subsection{Localization Problem Formulation}
\label{sec:localization_problem_formulation}
We define the localization problem as estimating the position and orientation of a UE. Different localization pilot signals will be sent depending on the prior UE state information. Usually, random pilots or pilots from a predefined codebook are used if the prior location information is unknown (e.g., a new UE {seeks to access the network}). However, if the prior information {(e.g., UE state information from prediction or other sources of observations)} is available, elaborately optimized signals can be used to achieve better performance. 
At the $g$th transmission/measurement, a signal symbol vector $\xv^\ssnb{g}$ is transmitted and a signal symbol vector $\yv^\ssnb{g}$ is observed. For an AOSA structure, beamforming angle matrices $\mathring \vpv_\mathrm{B}^\ssnb{g}$,$\mathring \vpv_\mathrm{U}^\ssnb{g}$ need to be selected for each transmission ($\mathring \vpv_\mathrm{B} = [\mathring \vpv_1;\cdots; \mathring \vpv_{N_B}]$, $\mathring \vpv_\mathrm{U} = [\mathring \vpv_1;\cdots; \mathring \vpv_{N_U}]$), which affect the equivalent array response as shown in~\eqref{eq:array_factor}. 
From the observed signal symbol vector $\hat\Ym \in \mathbb{C}^{\mathcal{G}KN_\mathrm{B}\times 1}$, which is a concatenation of {the received symbols from all the $\mathcal{G}$ transmissions, $K$ subcarriers, and $N_\mathrm{B}$ RFCs at BS}, we want to estimate the localization parameters of the UE.


\begin{table*}[t]
\scriptsize
    \caption{{Summary of Radio Signal based Localization Works}}
    \label{tab:localization_work_summary}
    \centering
    \begin{tabular}{c | c | c !\vthickline 
    c |c | c | c|
    p{1.0mm}| p{1.0mm}| p{1.0mm}| p{1.0mm}| p{1.0mm}| p{1.0mm}| p{1.0mm}| p{1.0mm}| p{1.0mm}| p{1.0mm}| p{1.0mm}| p{1.0mm}| 
    c c c}
    \hthickline
         & Year & Ref & $f_c$ & Link & System & 
        \!\rotatebox{90}{Carrier} &
        \!\rotatebox{90}{Position} & \!\rotatebox{90}{SWM} &  \!\rotatebox{90}{RIS} &  \!\rotatebox{90}{NLOS} &  \!\rotatebox{90}{Asyn} &  \!\rotatebox{90}{Orientation} & \!\rotatebox{90}{Mobile} & 
        \!\rotatebox{90}{Multi-BS} &
        \!\rotatebox{90}{Multi-UE} &
        \!\rotatebox{90}{CRB} & 
        \!\rotatebox{90}{AOSA} & 
        \!\rotatebox{90}{Beam Squint}  &
        Techniques/Features\\ 
    \hthickline

        \multirow{8}{*}{\rotatebox{90}{CRF\ }}
                
        & 2015 & \cite{han2015performance} & \unit[100]{MHz} & Downlink & SIMO & Single & \!\!3D & & & & \!$\checkmark$ & \!\!1D &  \!$\checkmark$ & \!$\checkmark$ & & \!$\checkmark$ &  & & FIM analysis \\ \cline{2-20}     
        
        & 2017 & \cite{vieira2017deep} & \unit[300]{MHz} & Uplink & SIMO & Multi & \!\!2D & &  & \!$\checkmark$ & \!$\checkmark$ &  & & & & & & & CNN \\ \cline{2-20}  
                
        & 2017 & \cite{garcia2017direct} & \unit[7]{GHz} & Uplink & SIMO & Single & \!\!2D & &  & \!$\checkmark$ & & & & & \!$\checkmark$ & &  & & Direct localization \\ \cline{2-20}
        
        & 2018 & \cite{zhao2018optimal} & - & Uplink & SIMO & Single & \!\!2D &  & & & & & & \!$\checkmark$ & \!$\checkmark$ & \!$\checkmark$ &  & & Direct localization \\ \cline{2-20}   
        
        & 2018 & \cite{studer2018channel} & \unit[2]{GHz} & Uplink & SIMO & Single & \!\!2D & &  & \!$\checkmark$ & \!$\checkmark$ &  & \!$\checkmark$ & & & &  & & Channel charting \\ \cline{2-20}
        
        & 2019 & \cite{peng2019decentralized} & - & D2D & - & - & \!\!2D &  & & & & & \!$\checkmark$ & & \!$\checkmark$ & \!$\checkmark$ & & & Cooperative, DRL \\ 
        \cline{2-20}
         
        & 2020 & \cite{wymeersc2020beyond} & \unit[28]{GHz} & Downlink & SISO & Multi & \!\!2D &  & \!$\checkmark$ & & & & & & & \!$\checkmark$ &  & & FIM analysis \\ \cline{2-20}        
        
        & 2020 & \cite{wymeersch2020fisher} & \unit[28]{GHz} & Uplink & SIMO & Multi & \!\!2D & \!$\checkmark$ & & \!$\checkmark$ & \!$\checkmark$ & & & & & \!$\checkmark$ & & & Multi-stage \\ \cline{2-20}
        
        & 2020 & \cite{elzanaty2021reconfigurable} & \unit[28]{GHz} & Uplink & MIMO & Multi & \!\!3D & \!$\checkmark$ & \!$\checkmark$ & & \!$\checkmark$ & \!\!3D & & & & \!$\checkmark$ & & & FIM analysis \\
        
\hthickline
        
        \multirow{13}{*}{\rotatebox{90}{mmWave\ }}
        
        & 2015 & \cite{shahmansoori20155g} & \unit[60]{GHz} & Both & MIMO & Single & \!\!2D & & & & & \!\!1D & &  & & \!$\checkmark$ &  & & FIM analysis \\ \cline{2-20}

        & 2017 & \cite{shahmansoori2017position} & \unit[60]{GHz} & Downlink & MIMO & Multi & \!\!2D & & & \!$\checkmark$ &  & \!\!1D & & & & \!$\checkmark$ & & & Multi-stage \\ \cline{2-20}

        & 2018 & \cite{wymeersch20185g} & - & Downlink & MIMO & Single & \!\!2D & &  & \!$\checkmark$ & \!$\checkmark$ & \!\!1D & & & & & & &Multi-stage, mapping \\ \cline{2-20}
        
        & 2018 & \cite{abu2018error} & \unit[38]{GHz} & Both & MIMO & Single & \!\!3D & & & \!$\checkmark$ & & \!\!2D & & & & \!$\checkmark$ & & & FIM analysis \\ \cline{2-20}
        
        & 2018 & \cite{guidi2018millimeter} & \unit[60]{GHz} & Two-way & SISO & Single & \!\!2D &  & & & & & & & \!$\checkmark$ & &  & & Sweeping \\ \cline{2-20}
        
        & 2018 & \cite{yassin2018mosaic} & - & Downlink & SIMO & Single & \!\!2D & &  & \!$\checkmark$ & & & & & & \!$\checkmark$ &  & & SLAM \\ \cline{2-20}        
        & 2019 & \cite{kanhere2019map} & \unit[73]{GHz} & Downlink & MISO & - & \!\!2D &  & & \!$\checkmark$ & \!$\checkmark$ & & & \!$\checkmark$ & \!$\checkmark$ & & & & Map-assist, RT-toolbox \\ \cline{2-20}
        
        & 2020 & \cite{he2020large} & \unit[60]{GHz} & Downlink & MIMO & Multi & \!\!2D & & \!$\checkmark$ & & & \!\!1D & & & & \!$\checkmark$ & & & FIM analysis \\ \cline{2-20}
        
        & 2020 & \cite{abu2020single} & \unit[38]{GHz} & Two-way & MIMO & Single & \!\!3D & &  & & \!$\checkmark$ & \!\!2D & & & & \!$\checkmark$ & & & FIM analysis \\ \cline{2-20}
        
        & 2020 & \cite{koike2020fingerprinting} & \unit[60]{GHz} & Uplink & MIMO & Multi & \!\!2D & &  & \!$\checkmark$ &  & \!\!1D & & \!$\checkmark$ & & & & & Fingerprinting \\ \cline{2-20}

        & {2021} & \!\!{\cite{keykhosravi2021siso}} & \unit[30]{GHz} & Downlink & SISO & Multi & \!\!3D & & \!$\checkmark$ & &  \!$\checkmark$ & & &  & & \!$\checkmark$ & & & {FIM analysis} \\\cline{2-20}
        
        & 2021 & \cite{guerra2021near} & \unit[30]{GHz} & Uplink & SIMO & Single &  \!\!3D & \!$\checkmark$ & & &  \!$\checkmark$ & & \!$\checkmark$ &  & & \!$\checkmark$ & & & CoA, PF \\\cline{2-20}

        & {2021} & \!\!{\cite{jiang2021beamspace}} & \unit[30]{GHz} & Both & MIMO & Multi & \!\!3D & & & \!$\checkmark$  &  & & &  & & & & & {CoA, PF} \\

\hthickline
        
        \multirow{7}{*}{\rotatebox{90}{THz\ }} 
        
        & 2017 & \cite{nie2017three} & \unit[300]{GHz} & Downlink & MISO & - & \!\!3D & & &  \!$\checkmark$ &  &  & \!$\checkmark$ & \!$\checkmark$ & & & & & Tracking, RT-toolbox \\ \cline{2-20}  
        
        & 2019 & \cite{stratidakis2019cooperative} & \unit[275]{GHz} & Uplink & SIMO & Single & \!\!2D & &  & \!$\checkmark$ &  &  & \!$\checkmark$ & \!$\checkmark$ & \!$\checkmark$ &  & & & Tracking \\ \cline{2-20}
        
        & 2020 & \cite{fan2021siabr} & \unit[100]{GHz} & Downlink & MIMO & Multi & \!\!3D & &  & \!$\checkmark$ &  &  & \!$\checkmark$ & \!$\checkmark$ & \!$\checkmark$ &  & & & RNN, RT-toolbox \\ \cline{2-20}
        
        & 2020 & \cite{batra2020indoor} & \unit[275]{GHz} & Two-way & SISO & Single & \!\!3D & &  &  &  &  & \!$\checkmark$ & \!$\checkmark$ & & & & & RFID, UAV, SLAM \\ \cline{2-20}
        
        & {2021} 
        &\!\!{\cite{kanhere2021outdoor}} & \unit[142]{GHz} & Downlink & MISO & Multi & \!\!2D & &  & \!$\checkmark$ & \!$\checkmark$ &  & \!$\checkmark$ &  & \!$\checkmark$ & & & & {Map-based, field data} \\ \cline{2-20}
        
        & {2021} & \!\!{\cite{saeidi2021thz}} & \unit[400]{GHz} & Both & MIMO & Single & \!\!2D & &  &  &  &  & &  & \!$\checkmark$ & & & & {Leaky wave antenna} \\ \cline{2-20}
        
        & 2021 & Ours & \unit[300]{GHz} & Uplink & MIMO & Multi & \!\!3D & \!$\checkmark$ & \!$\checkmark$ & \!$\checkmark$ & \!$\checkmark$ & \!\!3D &  &  &  & \!$\checkmark$ & \!$\checkmark$ & \!$\checkmark$ & AOSA, RIS, SWM \\ \cline{2-20}
\hthickline
   
    \end{tabular}
\end{table*}





We define the localization parameters as a {state parameter vector} $\sv$ and a {measurement parameter vector} $\gammav$. The state vector contains the position, orientation, and channel information (e.g., channel gain and the position of the scatters) of interest, which can be further be separated into a UE state vector $\sv_{\mathrm{U}}$ and a nuisance state vector $\sv_{\mathrm{N}}$ ($\sv_{\mathrm{U}}\cup\sv_{\mathrm{N}} = \sv$). The measurement vector contains intermediate measurements (e.g., TOA and AOA that can be obtained directly from the received signal). 
The definitions of the state and measurement vectors depend on the system structure, signal frequency, bandwidth, and localization algorithms.

To highlight the different localization parameters, we discuss the state and measurement vectors in three typical scenarios: a multi-BS CRF system, a far-field mmWave system with LOS/NLOS channels, and a near-field THz system with LOS/RIS/NLOS channels. We assume a fully digital MIMO structure for mmWave systems to exploit the fundamental limits of the localization system. However, we assume a hybrid AOSA-based structure for THz systems due to its reduced complexity and its crucial beamforming gains, without which the THz coverage would be very limited in distance. Nevertheless, the AOSA structure is not THz-specific and can also be used for mmWave systems. Therefore, the distinct models are reasonable for the corresponding frequency bands, and they are also chosen on purpose to facilitate comparing the measurement and state vectors. The assumptions on the localization models (e.g., measurement types, PWM/SWM, and RIS channels) should thus be application-scenario-specific.


\subsubsection{Multi-BS CRF System}
Localization in low-frequency systems relies mainly on geometric measurements from multiple BSs. The UE usually does not have an antenna array, so only the position can be estimated. The state vector and the measurement vector can be defined as
    \begin{align}
        \sv_{\mathrm{U}} & = [\pv_\mathrm{U}],
        \\
        \sv & = [\pv_\mathrm{U}; \rhov; \xiv; B],
        \label{eq:state_vector_low_frequency}
        \\
        \gammav & = [\rhov; \xiv; \tauv; \tilde\vpv_{\mathrm{BU}}; B],
        \label{eq:measurement_vector_low_frequency}
    \end{align}
where $\rhov$, $\xiv$, and $\tauv$ are the vectors containing the channel amplitudes, phases, and delays of all the paths, respectively (e.g., $\tauv = [\tau_1; \tau_2; \cdots]$); $\tilde\vpv_{\mathrm{BU}} = [\tilde\vpv_{1}; \tilde\vpv_{2}; \cdots]$ is the angle vector measured from the available BSs. Note that the global angle vector $\vpv_{\mathrm{BU}}$ can be obtained with a known BS rotation matrix. In addition, different types of measurements (e.g., delay and angle) may not be available all the time, and the measurement vector can only be formed using available information.
    
\subsubsection{Far-field mmWave System}
With the implementation of antenna arrays, orientation estimation becomes possible. In addition, at high carrier frequencies, the NLOS paths are fewer and more resolvable, which makes the localization of UEs and SLAM possible with even a single BS. Other channel information, such as the complex channel gain and clock offset, can also be estimated. The state and measurement vector can thus be defined as
\begin{align}
    \sv_{\mathrm{U}} & = [\pv_\mathrm{U}; \ov_\mathrm{U}],
    \\
    \sv & = [\pv_\mathrm{U}; \ov_\mathrm{U}; \rhov; \xiv; \pv_\mathrm{N}; B],
    \label{eq:state_vector_mmwave}
    \\
    \gammav & = [\rhov; \xiv; \tauv; \tilde\vpv_{\mathrm{BU}};
    \tilde\vpv_{\mathrm{UB}};
    \tilde\vpv_{\mathrm{BN}}; \tilde\vpv_{\mathrm{UN}}],
    \label{eq:measurement_vector_mmwave}
\end{align}
where $\pv_\mathrm{N} = [\pv_1; \pv_2; \cdots; \pv_{L_\mathrm{N}}]$, $\tilde\vpv_{\mathrm{BN}}=[\tilde\vpv_{\mathrm{BN}}^\ssnb{1}; \cdots; \tilde\vpv_{\mathrm{BN}}^\ssnb{L_\mathrm{N}}]$, $\tilde\vpv_{\mathrm{UN}}=[\tilde\vpv_{\mathrm{UN}}^\ssnb{1}; \cdots; \tilde\vpv_{\mathrm{UN}}^\ssnb{L_\mathrm{N}}]$, contain the position and angle information of all $L_\mathrm{N}$ NLOS paths.

Note that the state vector $\sv$ in~\eqref{eq:state_vector_mmwave} contains all unknowns (UE state parameters and channel state parameters). If we are only interested in the position/orientation of the UE, the state vector can be divided into a state vector of interest $\sv_\mathrm{U} = [\pv_\mathrm{U}; \ov_\mathrm{U}]$ and a nuisance state vector $\sv_\mathrm{N}=[\rhov; \xiv; \pv_\mathrm{N}; B]$. The corresponding CRB of the $\sv_\mathrm{U}$ can be obtained using an equivalent FIM (EFIM), as will be discussed in Sec.~\ref{sec:EFIM}.

\subsubsection{Near-field THz System with RIS}
THz systems are more likely to be in the near-field given the larger array sizes.
Moving to the THz band (or changing from PWM to SWM) does not change the state vector described in~\eqref{eq:state_vector_mmwave} too much. If SWM and RISs are considered in the system, the vectors $\sv$ and $\gammav$ can be rewritten as
    \begin{align}
        \sv_{\mathrm{U}} & = [\pv_\mathrm{U}; \ov_\mathrm{U}],
        \\
        \sv & = [\pv_\mathrm{U}; \ov_\mathrm{U}; \rhov; \xiv; \pv_\mathrm{N}; B],
        \label{eq:state_vector_Thz}
        \\
        \gammav & = [\rhov; \xiv; \tauv; \vpv_{\mathrm{BU}};
        \vpv_{\mathrm{RU}};
        \vpv_{\mathrm{BN}};
        \vpv_{\mathrm{NU}};
        \ov_\mathrm{U}].
        \label{eq:measurement_vector_Thz}
    \end{align}
Compared to~\eqref{eq:measurement_vector_mmwave}, the local angle vectors $\tilde \vpv$ are replaced by the global angle vectors $\vpv$ and the orientation of UE $\ov_\mathrm{U}$. Although $\tilde \vpv_{\mathrm{UB}}$, $\tilde \vpv_{\mathrm{UR}}$, and $\tilde \vpv_{\mathrm{UC}}$ can still be estimated, their CRLBs cannot be derived directly in near-field scenarios; this is because the calculation of antenna distances for an SWM depends on the global position instead of the local angles, as shown in~\eqref{eq:SWM_channel_element} and~\eqref{eq:PWM_channel_element}.
    
\subsubsection{Parameters in Direct Localization}
In direct localization, the parameters are estimated directly by optimizing an objective function. The measurement parameter vector is identical to the state parameter vector~\cite{elzanaty2021reconfigurable}, i.e.,
\begin{equation}
    \gammav_\mathrm{Direct} = \sv.
    \label{eq:measurement_vector_direct}
\end{equation}
It is worth noting that the vectors in~\eqref{eq:state_vector_low_frequency}-\eqref{eq:measurement_vector_direct} are not exclusive lists of all the parameters. Other parameters such as BS/RIS position and orientation errors or BS-RIS synchronization offsets can also be included. In addition, not all the parameters need to be estimated from the localization algorithm point of view. A portion of the parameters can be selected depending on the signal model, geometry model, or estimation algorithm, among others. With a smaller parameter vector size, the computational complexity can be reduced at the expense of a performance loss.


\subsection{Cram\'er-Rao Bound}
\label{sec:crb}
\subsubsection{Error Bounds}

From a localization perspective, we are more interested in a UE's position and orientation accuracy. Positioning accuracy is usually measured in terms of the mean-squared error (MSE) or the root-mean-squared error (RMSE)--the error is defined as the Euclidean distance between the estimated location and the ground truth. For orientation estimation, rotation error is defined as the angular difference between the estimation and the true angle. Both position and rotation errors are affected by the noise level. These two errors are lower-bounded by the position error bound (PEB), and the orientation error bound (OEB), which are derived from the CRB and are often used by geometry-based localization methods as effective tools to evaluate performance~\cite{shahmansoori2017position}. For more complicated techniques such as fingerprinting~\cite{koike2020fingerprinting, studer2018channel}, the bound is less tractable, and the corresponding systems often provide a tradeoff between algorithm complexity and positioning accuracy.

{The CRB analysis for pulse-based signals can be found in~\cite{shen2010fundamental}, where the analysis is based on the time-domain signal.
When working on OFDM-based systems in the frequency domain, the multipath can be easily dealt with. However, the analysis of time and frequency domain signals is fundamentally the same. From the THz systems point of view, carrier-based signals can carry localization tasks more realistic since THz-pulses are typically of very low power, which is only suitable for nanocommunication scenarios\cite{jornet2014femtosecond}.
In the rest of the work, we only discuss OFDM-based MIMO systems based on our channel model developed in Section \ref{sec:thz_system_model_and_properties}.}
Reported CRB results for mmWave MIMO systems cover both \mbox{2D} and \mbox{3D} scenarios. 
In~\cite{shahmansoori2017position}, the position and orientation bound in LOS and NLOS scenarios for a 2D 5G mmWave MIMO system are derived and verified with a multi-stage localization algorithm. The error bound for 3D scenarios is analyzed in~\cite{abu2018error}, where the differences between uplink and downlink are discussed. 
Recent works have also explored the potential of positioning with RISs, which are also called intelligent reflective surfaces (IRSs)~\cite{wu2021intelligent}, large intelligent surfaces (LISs)~\cite{he2020large}, where promising results are noted~\cite{wymeersc2020beyond,hu2018beyond,he2020large,elzanaty2021reconfigurable}. Near-field propagation conditions with RISs are further analyzed in~\cite{elzanaty2021reconfigurable}.
However, practical RIS optimization methods and RIS-assisted localization algorithms have yet to be developed. In addition, the CRB derivation and analysis of AOSA-based THz systems are still lacking, and new bounds considering THz-specific features (such as beam split, distance-dependent bandwidth, and spherical wave propagation) need to be developed.



\subsubsection{Position and Orientation Error Bounds}
\label{sec:position_and_orientation_error_bound}
Given the signal model in~\eqref{eq:conventional_signal_model}, both direct and multi-stage approaches will have the same CRB for the state vector if all available measurements (variables in $\gammav$) are considered~\cite{elzanaty2021reconfigurable}. The CRLBs for the measurement vector can also be calculated, which is useful for system design and optimization.
The CRB on the UE state vector can be written as~\cite{elzanaty2021reconfigurable}
\begin{equation}
    \mathrm{CRB} \triangleq \left[\mathbf{I}(\sv)\right]^{-1} = \left[\Jm_\mathrm{S}^T \mathbf{I}(\gammav) \Jm_\mathrm{S}\right]^{-1},
    \label{eq:CRLB_from_FIM}
\end{equation}
where $\mathbf{I}(\sv)$ is the Fisher information matrix (FIM) of the state vector that can be obtained from the FIM of measurement vector, $\mathbf{I}(\gammav)$, based on the chain rule. The Jacobian matrix, $\Jm_\mathrm{S}$, and the FIM of measurement vector, $\mathbf{I}(\gammav)$, are
\begin{align}
    \Jm_\mathrm{S} & \triangleq \frac{\partial \gammav}{\partial \sv},\\
    \mathbf{I}(\gammav) & 
    = \frac{2}{\sigma^2}\sum^{\Gc}_{g} \sum^K_{k}\mathrm{Re}\left\{
    (\frac{\partial\muv^{\ssnb{g}}[k]}{\partial\gammav})^H 
    (\frac{\partial\muv^{\ssnb{g}}[k]}{\partial\gammav})\right\},
    \label{eq:FIM_measurement}
\end{align}
where $\mathcal{G}$ is the number of measurements/transmissions. 

From the CRB, the position error bound (PEB) and orientation error bound (OEB) can be written as
\begin{align}
\mathrm{PEB} & = \sqrt{\trace([\mathrm{CRB}]_{1:3, 1:3})}
\label{eq:PEB},\\
\mathrm{OEB} & = \sqrt{\trace([\mathrm{CRB}]_{4:6, 4:6})}.
\label{eq:OEB}
\end{align}

\subsubsection{Equivalent FIM (EFIM)}
\label{sec:EFIM}
The FIM of the state vector $\mathbf{I}(\sv)$ contains the information for all the channel parameters, e.g., in~\eqref{eq:state_vector_low_frequency},~\eqref{eq:state_vector_mmwave}, and~\eqref{eq:state_vector_Thz}). Each element in the matrix $\mathbf{I}(\sv)$ is $\mathbf{I}(\sv)_{i,j} = {I}(\sv_i, \sv_j)$ ($i,j\le \mathrm{length}(\sv)$), which can be obtained from~\eqref{eq:CRLB_from_FIM}-\eqref{eq:FIM_measurement}. If we are only interested in the UE state vector $\sv_{\mathrm{U}}$, EFIM can be used. More specifically, we can rearrange $\mathbf{I}(\sv)$ into a block-diagonal structure as
\begin{equation}
    \mathbf{I}(\sv) = 
    \begin{bmatrix}
        \mathbf{I}(\sv_\mathrm{U}) & \mathbf{I}(\sv_\mathrm{U},\sv_\mathrm{N}) \\
        \mathbf{I}(\sv_\mathrm{U},\sv_\mathrm{N})^T & \mathbf{I}(\sv_\mathrm{N})
    \end{bmatrix},
\end{equation}
{where $\sv_\mathrm{U}$, $\sv_\mathrm{N}$ contain the UE state parameters and nuisance parameters from the rest of the state vector $\sv$ (more details can be found in~\cite{abu2018error}).} The EFIM of $\sv_\mathrm{U}$ is then given by
\begin{equation}
    \mathbf{I}^\ssr{E}(\sv_\mathrm{U}) = \mathbf{I}(\sv_\mathrm{U}) - \mathbf{I}(\sv_\mathrm{U},\sv_\mathrm{N})\mathbf{I}(\sv_\mathrm{N})^{-1}\mathbf{I}(\sv_\mathrm{U},\sv_\mathrm{N})^T,
\end{equation}
where $\mathbf{I}(\sv_\mathrm{U}, \sv_\mathrm{N})_{i,j} = {I}(\sv_{\mathrm{U},i}, \sv_{\mathrm{N},j})$  ($i\le \mathrm{length}(\sv_\mathrm{U}),j\le \mathrm{length}(\sv_\mathrm{N})$). The PEB and OEB can be similarly obtained from~\eqref{eq:PEB} and~\eqref{eq:OEB}.




\subsubsection{Constrained CRB for Far-field OEB}
The calculation of the OEB using~\eqref{eq:CRLB_from_FIM}-\eqref{eq:OEB} does not work for 3D orientation estimation in far-field scenarios because the orientation $\ov_\mathrm{U}$ in~\eqref{eq:state_vector_mmwave} cannot be individually mapped from the angles in~\eqref{eq:measurement_vector_mmwave}. Alternatively, the OEB can be obtained from the estimated DOD angle vector $\tilde\thetav = [\tilde\vpv_{\mathrm{UB}}; \tilde\vpv_{\mathrm{UR}}; \tilde\vpv_{\mathrm{UN}}]$ using the constrained CRB as discussed in~\cite{nazari20213d}.
When an unknown vector $\etav \in \mathbb{R}^{N_{\eta}\times 1}$ is constrained to lie on a manifold $\fv(\etav) = 0$ defined by $0 \le K_{\eta} < N_{\eta}$ non-redundant constraints. The constrained CRB $\mathbf{I}^{-1}_\mathrm{const}(\etav)$ can be expressed as~\cite{nazari20213d}
\begin{equation}
    \mathbf{I}^{-1}_\mathrm{const}(\etav) = \Mm(\Mm^T \mathbf{I}(\etav)\Mm)^{-1}\Mm^T.
\end{equation}
Here, $\mathbf{I}(\etav)$ is the {EFIM} of the unconstrained parameters and $\Mm\in\mathbb{R}^{N\times(N-K)}$ is an orthonormal basis for the null-space of the gradient matrix $\partial\fv(\etav)/\partial\etav^T$ satisfying $\Mm^T\Mm = \mathbf{I}_{N-K}$~\cite{nazari20213d}. 

For far-field orientation estimation, $\Mm$ can be chosen as~\cite{chepuri2014rigid}
\begin{equation}
    \Mm = 
    \begin{bmatrix}
        -\rv_3 & \mathbf{0}_{3\times1} & \rv_2 \\
        \mathbf{0}_{3\times1} & -\rv_3 & -\rv_1\\
        \rv_1 & \rv_2 & \mathbf{0}_{3\times1}
    \end{bmatrix},
\end{equation}
where $\rv_1$, $\rv_2$, and $\rv_3$ are the first, second, and third columns of the rotation matrix in~\eqref{eq:rotation_matrix_3_var}, respectively. By forming $\rv = \mathrm{vec}(\Rm) = [\rv_1; \rv_2; \rv_3]$. The FIM of the rotation matrix $\mathbf{I}(\rv)$ can be obtained from the FIM of the measured DOD angle vector $\mathbf{I}(\tilde\thetav)$ as
\begin{equation}
    \mathbf{I}(\rv) = \left[\left(\frac{\partial \tilde\thetav}{\partial \rv}\right)^T \mathbf{I}(\tilde\thetav) \left(\frac{\partial \tilde\thetav}{\partial \rv}\right)\right].
    \label{eq:constrained_CRLB_r_to_ori}
\end{equation}
Hence, the OEB can be obtained as
\begin{equation}
    \mathrm{OEB} = \sqrt{\mathrm{trace}([\mathbf{I}_{\mathrm{const}}(\rv)]^{-1})}.
    \label{eq:constrained_OEB}
\end{equation}

Note that the OEB in~\eqref{eq:constrained_OEB} is defined using rotation matrix ($\norm{\Rm-\hat\Rm}_F$), which is different from the OEB defined in~\eqref{eq:OEB} using Eular angles ($\norm{\ov-\hat\ov}$). However, both definitions can be used as indicators of the system orientation estimation performance.
With the derived PEB and OEB, we are able to benchmark and evaluate the designed localization estimators. Next, we describe geometry-based localization.

{
\subsubsection{CRB for an LOS Channel}
It is difficult to express the CRB considering all the paths and system parameters; however, it would be insightful to show the relationship between CRB and system parameters for a single LOS path in a closed-form. Consider a 2D uplink scenario with perfect synchronization, 
the PEB, and OEB can be expressed as~\cite{abu2016random, abu2018error}
\begin{align}
    \text{PEB}_\text{LOS} & = 
    \sqrt{\frac{N_0 W d^2}{N_\mathrm{B}N_\mathrm{U}\mathcal{G} P\lambda^{2}}
    \left(\frac{c^{2}\zeta_{\tau_\mathrm{BU}}}{W^2}
    +\frac{d^{2}\zeta_{\phi_\mathrm{BU}}}{N_{\mathrm{B}}^{2}}\right)}, 
    \label{eq:PEB_closedform_LOS}
    \\
    \text{OEB}_\text{LOS} & = 
    \sqrt{\frac{N_0 W d^2}{N_\mathrm{B}N_\mathrm{U}\mathcal{G} P\lambda^{2}}
    \left(\frac{\zeta_{\phi_\mathrm{BU}}}{N_\mathrm{B}^2}
    +\frac{\zeta_{\phi_\mathrm{UB}}}{N_{\mathrm{U}}^{2}}\right)}.
    \label{eq:OEB_closedform_LOS}
\end{align}
Where $\zeta_{\tau_\mathrm{BU}}$, $\zeta_{\phi_\mathrm{BU}}$, and $\zeta_{\phi_\mathrm{UB}}$ are the components contributing to the error bound, which are determined by delay, AOA and AOD of this LOS channel, respectively.
The common term in \eqref{eq:PEB_closedform_LOS} and \eqref{eq:OEB_closedform_LOS} shows the SNR component of the system, which is determined by the noise level ($N_0W$), path loss ($d/\lambda$), number of antennas ($N_\mathrm{B}$, $N_\mathrm{U}$), number of transmissions ($\mathcal{G}$) and the transmission power ($P$). The components inside the parenthesis in \eqref{eq:PEB_closedform_LOS} indicate that the PEB is decided by the delay and AOA, and the effect of $\zeta_{\tau_\mathrm{BU}}$, $\zeta_{\phi_\mathrm{BU}}$ on the PEB can be reduced by increasing the bandwidth and array size at the BS, respectively. Similarly, the AOA and AOD affect the OEB of the UE and the contributions of $\zeta_{\phi_\mathrm{BU}}$ and $\zeta_{\phi_\mathrm{UB}}$ can be mitigated by increasing  $N_\mathrm{B}$ and $N_\mathrm{U}$.
Note that equations describe the relationship between PEB/OEB and an ideal single LOS channel. Although the formulation is not applicable for all the scenarios (e.g., multipath, RIS channels), the relationship can be treated as a reference for deriving the scaling laws and designing localization systems.
}


\subsection{Geometry-based Algorithms}
\label{sec:geometry_based_localization}

\subsubsection{Direct Localization}
Geometry-based localization algorithms can be categorized into direct localization and multi-stage algorithms.
In direct localization, the state vector $\sv$ is estimated directly from the received signals, without estimating any intermediate parameters~\cite{garcia2017direct, wen2019survey}.
Given the transmitted signal, and assuming no prior information is available on the UE's positions, the direct localization problem can be formulated as a maximization of the likelihood function
\begin{equation}
    \begin{split}
        \hat\sv_{\mathrm{direct}}
        & = \argmax_{\sv} p(\hat\Ym| \sv)  = \argmax_{\sv} \ln (p(\hat\Ym| \sv))\\
        & = \argmin_{\sv} [(\hat\Ym - \muv(\sv))^H\Sigmam^{-1}_{\hat\Ym}(\hat\Ym - \muv(\sv))].
    \end{split}
\label{eq:direct_mle}
\end{equation}
Here, $\Sigmam_{\hat\Ym}$ is the covariance matrix of the received symbols, which can be ignored if independent identical Gaussian distribution of noise vector in~\eqref{eq:conventional_signal_model} is assumed for different antennas and measurements.



Direct localization is applicable to both quasi-synchronous (only BSs are synchronized) and asynchronous (none of the devices are synchronized) systems~\cite{keskin2018localization}. However, the computational complexity of solving the optimization problem formulated in~\eqref{eq:direct_mle} is high due to non-convexity and the large search space. Prior information is important in this case to limit the search area. 

\subsubsection{Multi-stage Localization}
{The multi-stage localization procedure divides direct localization into a geometry information estimation stage and a position/orientation estimation stage, which reduces the complexity of calculating all the unknowns from the received data. More specifically, the measurement vector $\gammav$ is first estimated and then the state vector is extracted from it~\cite{shahmansoori2017position,shahmansoori20155g}.}
Similar to~\eqref{eq:direct_mle}, a multi-stage localization problem can be formulated as
\begin{equation}
    \begin{split}
        \hat\sv_{\mathrm{multi-stage}} 
        & = \argmax_{\sv} p(\hat\gammav| \sv)\\
        & = \argmin_{\sv} [(\hat\gammav - \gammav(\sv))^H\Sigmam^{-1}_{\hat{\gamma}}(\hat{\gammav} - \gammav(\sv))],
    \end{split}
\label{eq:twostep_mle}
\end{equation}
where $\Sigmam_{\mathrm{\hat\gamma}}$ is the covariance matrix of the measurement vector and $\hat \gammav$ is the estimated measurement vector from channel estimation.
Multi-stage approaches are inherently sub-optimal~\cite{garcia2017direct} and usually inferior to direct localization. However, by considering all multipath components, multi-stage localization can reduce the performance gap with direct localization, and it is hence pursued in many works~\cite{shahmansoori2017position,shahmansoori20155g}.


In multi-stage localization, the parameter vector $\gammav$ needs to be estimated first. Each element in $\gammav$ (e.g., AOA/AOD, channel gains, and signal delay for each path) can be obtained independently or jointly. The channel gain can be estimated by solving a least-squares (LS) problem~\cite{shahmansoori2017position}. AOA/AOD can be estimated using subspace-based methods (e.g,. MUSIC)~\cite{stoica1989music}, compressed sensing (CS)~\cite{fortunati2014single}, deep learning (DL)~\cite{wan2020deep}, or Bayesian inference~\cite{peng2016three}. TOA can be estimated using correlation-based~\cite{dardari2008threshold} or energy-based methods~\cite{giorgetti2013time}. The channel parameters can also be estimated jointly using multidimensional channel parameter estimation via rotational invariance techniques (MD-ESPRIT)~\cite{jiang2021high, ge20215g}.
In general, the performance of AOA/AOD estimation depends on the array size of the device, while TOA estimation benefits from synchronization and wideband signals. {In addition, the error in different stages propagates and may affect the localization performance, which should be considered in system design for a better tradeoff of processing time and performance.}




\subsubsection{Practical Algorithms for Geometry-based Localization}
With the geometry information $\hat \gammav$ obtained from channel estimation, multi-stage localization problems can be formulated using~\eqref{eq:twostep_mle} (direct localization only requires $\hat \Ym$ {as in (73)}). An analytical closed-form solution might be obtained by setting the derivative of an objective function {equal to zero and solving for the position parameters}. However, this approach is impractical considering the non-convexity of the cost function. We discuss two practical categories of optimization algorithms: convergent iterative methods and heuristic methods~\cite{fei2016survey, ali20206g}.
\begin{itemize}
    \item Convergent iterative methods: If the gradient information from the signal model is known, gradient- or Hessian-based algorithms can be implemented. Other iterative algorithms such as alternative projection~\cite{van2004detection} and expectation-maximization~\cite{shahmansoori2017position} are also practical solutions to reduce the computational burden. Within a few iterations, such deterministic algorithms converge to an optimum of the objective function. The convergence depends on the formulation of objective functions and iteration parameters (e.g., step size), where local solutions can be reached.
    \item Heuristic/metaheuristics methods: Heuristic methods are capable of dealing with non-differential nonlinear objective functions and reaching near-optimal solutions faster. Popular algorithms include swarm intelligence, tabu search, simulated annealing, genetic algorithms, and so on~\cite{chen2020joint, kennedy2006swarm, pham2012intelligent}.
\end{itemize}

Given the sparsity of high-frequency channels and a large number of measurements (due to large bandwidth and array sizes/RFCs), we expect that multi-stage localization will be favored in THz localization. However, for applications that require high localization accuracy, a practical approach is to determine an initial position via multi-stage algorithms and then refine it using direct localization. Next, we discuss learning-based localization.

\subsection{Learning-based Algorithms}
\label{sec:learning_based_algorithms}
{

In the previous subsection, we discussed geometry-based localization. In challenging environments where geometric models cannot be formulated (e.g., many non-resolvable NLOS paths), or when geometry-based localization cannot handle the processing speed requirements of the system, learning-based methods can be used. 
In this subsection, we briefly describe the implementations of ML-based localization algorithms in two categories, namely, direct localization and multi-stage localization. Then, practical ML-based algorithms will be discussed.

\subsubsection{Direct Localization}
ML-based localization involves two phases, offline training of the model $f(\cdot)$ and online processing of the observation to obtain a position estimation $\hat \pv = f(\yv)$. During the training phase, a training data set $\mathcal{D}=\langle\mathcal{D}_\yv, \mathcal{D}_\pv\rangle$ (including $|\mathcal{D}|$ signal-position pairs $\langle \yv^{\text{train}}_{i}, \pv^{\text{train}}_{i} \rangle$, ($1\le i \le |\mathcal{D}|$, $\yv^{\text{train}}_{i} \in \mathcal{D}_\yv$, $\pv^{\text{train}}_{i} \in \mathcal{D}_\pv$) is needed to train the model $f(\cdot)$ (optimize the parameters of this function) in order to reduce the loss function $\mathcal{L}(f(\yv^{\text{train}}), \pv^{\text{train}})$. Take the mean squared error (MSE) cost function for example, we can have
\begin{equation}
    \mathcal{L}(f(\mathcal{D}_\yv), \mathcal{D}_\pv) = \sum_i^{|\mathcal{D}|} 
    \Vert f(\pv^{\text{train}}_i) - \pv^{\text{train}}_i \Vert^2.
\end{equation}
After the training, the model $f(\cdot)$ can be used to output end-to-end location information by taking the raw observation data as the input.

Fingerprinting (or pattern matching) is an approach that utilizes a database of fingerprints to find the best position match for a particular signal measurement~\cite{del2017survey}. The \ac{csi} and RSS could be used as the entries to construct the database. While RSS suffers from limited accuracy and CSI requires high computational power, spatial beam SNRs are adopted as a mid-grained intermediate channel measurement~\cite{koike2020fingerprinting}. For the retrieval process, deep learning methods such as deep neural networks (DNN) and convolutional neural networks (CNN) are valuable tools to obtain effective models for location estimations~\cite{koike2020fingerprinting, vieira2017deep}. 

In this category, all the information is maintained and will provide accurate results if the data in the implementation scenario matches the training data set. However, the drawbacks are the data collection in the training phase, and the scalability issue as one model only works for a specific scenario.

\subsubsection{Multi-stage Localization}
Similar to the geometry-based localization, the direct localization task can be decomposed into several sub-tasks (e.g., signal pre-processing, intermediate geometry parameters estimation, and localization). Each sub-task can be solved using learning-based methods with a much smaller training dataset. 
In the first stage, learning-based methods can be used to reduce the effect of the \acp{hwi} such as antenna spacing error~\cite{mateos2021end}, \ac{iqi}~\cite{wu2021low}, \ac{mc}~\cite{alzahed2019nonlinear}, and \ac{pan}~\cite{wu2021symbol}. The distorted signal due to the impairments can be recovered or compensated during the data pre-processing stage. In channel parameter estimation, learning-based methods have been implemented to estimate the angle~\cite{kase2018doa,barthelme2021machine} and delay~\cite{dvorecki2019machine, wymeersch2012machine}. In terms of the localization stage, machine learning has shown the potential to improve localization performance via NLOS identification~\cite{jiang2020uwb}, and global fusion profile~\cite{guo2017knowledge}.

Considering the high dimension of the system parameters and the complexity of the environment, the training of an end-to-end localization model may not be practical. The design of learning-based algorithms for sub-tasks reduces the training cost. These trained models are also flexible to adapt to different scenarios (e.g., a trained model in MIMO systems may not fit a MISO system, but range or angle estimation are more general). Nevertheless, the propagation of the error caused at each stage needs to be considered while adopting learning-based methods.

\subsubsection{Practical ML algorithms}
Machine learning algorithms are usually classified into supervised learning (used for solving \textit{classification} and \textit{regression} problems) and unsupervised learning (used for \textit{data clustering})~\cite{burghal2020comprehensive}. Other approaches, including semi-supervised learning, reinforcement learning, transfer learning, and federated learning, are designed to solve the issues faced by the supervised and unsupervised learning algorithms, which will be discussed as follows.

\begin{itemize}
    \item Supervised Learning: Traditional machine learning algorithms, such as random forest, support vector machine, and recent popular deep learning, belong to \textit{supervised learning}. Due to the wide application scenarios in many fields, a lot of toolboxes such as Tensorflow~\cite{abadi2016tensorflow} and PyTorch~\cite{paszke2019pytorch} make the implementation simple for the researchers. However, two challenges exist. One is the data collection of the offline phase, where sufficient real data are not easy to obtain, and synthesized data may not be accurate. Another is the selection of model parameters; for example, the number of layers and neurons, as well as model structures, make deep learning often an art rather than a science.

    \item Unsupervised learning: Without the need for well-labeled datasets, \textit{unsupervised learning} is widely used for clustering, and dimension reduction (or feature extraction). A novel framework called channel charting is proposed in~\cite{studer2018channel}, which learns CSI in a fully unsupervised manner and can map a high-dimensional point set (the channel features) into a low-dimensional point set (the channel chart). However, this category can only perform data pre-processing, and location information cannot be obtained.
    
    \item Other approaches: By combining the two above-mentioned categories, \textit{semi-supervised learning} can train the model with partially labeled data (e.g., $|\mathcal{D}_\pv|<<|\mathcal{D}_\yv|$ in the training dataset $\mathcal{D}$). For the scenario without a clear objective function (only a reward is known after taking action) \textit{reinforcement learning} is preferred, which is suitable for training without a clear cost function using online data. \textit{Transfer learning} is able to take advantage of the existing model to reduce the training time, and federated learning works in a distributed manner and hence protects user privacy. 
    More details of ML-based localization can be found in~\cite{nessa2020survey, burghal2020comprehensive, roy2021survey, zhu2020indoor, de2021convergent, mirama2021survey}.
\end{itemize}

In summary, despite the channel at THz frequencies being more deterministic than at lower frequencies, which suits geometry-based methods well, we argue that learning-based methods still have advantages in two aspects. Firstly, processing large volumes of data (due to a wide bandwidth) necessitates faster algorithms for localization, and ML algorithms are efficient at feature extraction and hence speed up the processing. Secondly, hardware impairments are severe in high-frequency systems, and the mismatch between the theoretical and actual system models affects the performance, which needs to be mitigated by learning-based methods with onsite data.

}

\subsection{Tracking and SLAM Algorithms}
\label{sec:tracking_and_slam_algorithm}
{While this paper is focused on the snapshot localization problem, this is generally part of a wider tracking \cite{schmidt1966application} or SLAM \cite{durrant2006simultaneous} routine, which the UE performs sequentially, based on its own mobility model and periodic measurements. For completeness, we briefly describe their operation in the following sections.

\subsubsection{Tracking}

In mobile applications, initial access is only needed for the first several frames or when the communication link is lost. Once initial access is completed, the UE goes into tracking mode. Mathematically, the model for a tracking problem can be expressed as
\begin{align}
    \sv_{\mathrm{U},t} & \sim  p(\sv_{\mathrm{U},t}|\sv_{\mathrm{U},t-1})\\
    \hat\gammav_t & \sim p(\gammav_t|\sv_{\mathrm{U},t}),
\end{align}
where $\sv_{\mathrm{U},t}$ is the UE state vector at time $t$, which depends on the previous state $\sv_{\mathrm{U},t-1}$ via a stochastic mobility model, and {$\hat\gammav_t$ is the measurement vector at time $t$ as defined in Sec.~\ref{sec:localization_problem_formulation}}, which depends on the UE state at time $t$. The observation contains the estimated angles and delays related to LOS and RIS paths. In addition, a prior $p(\sv_{\mathrm{U},0})$ is assumed to be given. 

Solving the tracking problem refers to determining the posterior of the state $p(\sv_{\mathrm{U},t}|\hat\gammav_{1:t})$ given all the collected measurements up till the current time. Several filters exist to solve the tracking problem, though they are all approximate (unless the mobility and measurement models are linear and Gaussian). These filters include:  
\begin{itemize}
    \item \emph{Filters based on the \ac{kf}:} the KF provides a recursive solution for linear filtering problems. For nonlinear problems, an \ac{ekf} can be used, which approximates the state distribution using a Gaussian random variable and propagates analytically through the first-order linearization~\cite{wan2000unscented}. Other extensions of the KF families include the \ac{ukf} and the \ac{ckf}, where the former addresses the approximation issues of the EKF~\cite{wan2000unscented}, and the \ac{ckf} suits high-dimensional state estimation~\cite{arasaratnam2009cubature}. The \ac{kf}-based filters generally have low complexity but are unable to cope with highly nonlinear models or multi-modal distributions. 
    \item \emph{Filters based on the \ac{pf}:} the PF is another widely-used filter that exploits the representation of an arbitrary \ac{pdf} by a set of particles~\cite{gustafsson2010particle}. PFs have the advantage of dealing with highly nonlinear and non-Gaussian models, but at the cost of high computational complexity, as the number of particles grows exponentially in the state dimensionality. 
\end{itemize}
An added advantage of tracking is that the transmitted signals and the precoders, combiners, and RIS coefficients can be optimized to account for the a priori information on the UE state. This topic will be covered in more detail in Section \ref{sec:system_design_and_optimization}.}



\subsubsection{SLAM}
{
While not considered in this work, the measurements $\hat\gammav_t$ at each time step $t$ also provide information about the location of the scatter points (\emph{landmarks} in SLAM parlance), shown in Fig.\,\ref{fig:system_model_geometry}. In turn, this knowledge can improve estimating the UE state, which is the main idea behind SLAM. SLAM has been widely applied in robotics~\cite{cadena2016past, saeedi2016multiple} and autonomous driving~\cite{bresson2017simultaneous}, where an agent locates itself and constructs the unknown map at the same time~\cite{yassin2018mosaic}. 
With the wide bandwidth and MIMO structure implemented in 5G/6G systems, this topic draws the attention of the communication community with several mmWave systems proposed. The SLAM systems can broadly be classified into two categories, infrastructure-based~\cite{yassin2018mosaic} and non-infrastructure-based systems~\cite{barneto2021radio}. In the infrastructure-based systems, the positions of UE and scatters are estimated from the signals transmitted from the BS, as mentioned in~\eqref{eq:state_vector_Thz}. In a situation where no BSs are deployed, the UE sends a sequence of signals and then processes the received signal reflected from the surrounding environments~\cite{barneto2021radio}. The SLAM problem is inherently challenging since the data association between the landmarks and measurements is unknown (i.e., which landmark generated which delay or angle measurements). }

{
In our THz context, to infer the locations of scatter points  and execute SLAM, the following modifications are needed. First of all, the channel model $\Hm_N^{\ssnb{l}}$ from \eqref{eq:far_field_nlos_channel} should be expressed as a function of the scatter location, say $\pv_\mathrm{N}^{\ssnb{l}}$~\cite{ge20205g, ge2020exploiting, kim20205g}. Secondly, the local and global data associations between the angles and delays in $\hat\gammav_t$ and the landmark locations $\pv_\mathrm{N}^{\ssnb{l}}$ should be enumerated and their likelihoods should be calculated. This calculation should account for the hidden UE state as well as the possibility of false alarms (spurious measurements) and missed detections (landmarks without measurements at the current time).  Finally, the joint posterior of the UE state and the landmark state should be computed in an iterative manner, with well-defined prediction and correction steps, accounting for all or a subset of most likely data associations.  Common methods in this field are FastSLAM~\cite{montemerlo2002fastslam}, GraphSLAM~\cite{thrun2006graph}, belief propagation  SLAM~\cite{leitinger2019belief}, and random finite set theory-based SLAM~\cite{kim20205g,mullane2011random}. These mainly differ in how the data associations are computed, how the prediction and correction steps are performed, how the UE state is represented (e.g., particles or a parametric density), and how the map is represented  (e.g., parametric, grid maps, feature maps, topological maps, semantic maps, appearance maps, and hybrid maps~\cite{saeedi2016multiple}).}

In summary, tracking in THz systems is challenging due to the narrow beamwidths resulting from beamforming with large array sizes. Adaptive beamwidth design could thus be adopted for different tracking scenarios (e.g., high speed, confident prior information). Nevertheless, the accuracy of both tracking and SLAM improves with narrow beamwidths resulting in a high angular resolution. Furthermore, with dense network deployments and wide bandwidths, an unparalleled SLAM performance can be achieved in THz systems.



\subsection{Summary}
In this section, we formulate the localization problems, describe the CRB, and detail some localization techniques. In particular:
\begin{itemize}
    \item We describe the localization parameters as a state vector and a measurement vector. Different vectors can be defined based on application scenarios and algorithm selections.
    
    \item We introduce the CRB for position and orientation estimation based on the state and measurement vectors.
    
    \item We formulate geometry-based methods, namely, direct localization and multi-stage localization, and discuss several channel estimation and localization algorithms.
    
    \item We discuss several THz localization and sensing extensions, namely, learning-based localization, tracking, and SLAM. These techniques can deal with different localization scenarios and improve localization performance. 
    
\end{itemize}

In the next section, we formulate system design and optimization problems and discuss the relationships between the variables to be optimized and the affected objectives.


\section{Localization System Design and Optimization}
\label{sec:system_design_and_optimization}
{System design and optimization are essential for determining the fundamental limits of attainable localization performance. We start by presenting the {optimization problem formulation} based on the desired system objectives. Then, we discuss the high-level {design considerations} of the system. Afterward, we detail two groups of system design problems: {offline optimization} and {online optimization}.} Finally, we conclude this section with simulations and system evaluation.

\subsection{Optimization Problem Statement}
\label{sec:optimization_problem_statement}
\subsubsection{Motivation}
Optimization is essential in communication systems to meet different objectives of signal-to-interference-plus-noise ratio (SINR), energy efficiency, maximum throughput, Etc.
For localization purposes, the PEB and OEB defined in~\ref{sec:position_and_orientation_error_bound} are used when accuracy is chosen as an objective in system design. Although this criterion is valid only when the estimator is efficient, it is still a tractable and effective tool for analyzing performance in the asymptotic region. 
Other objectives described in Sec.~\ref{sec:localization_performance_metrics} are also important in certain scenarios. However, the definition of an objective is not always straightforward and is not unique; objectives need to be defined based on the application scenario. Due to different formulations of objective functions, we have to make compromises, especially when optimizing joint communication and localization systems.

In low-frequency localization systems, \acp{prs} are broadcast by the BSs, and the corresponding system design is mainly offline (such as BS layout and antenna array design). In mmWave MIMO systems, localization performance benefits from the beamforming gain. However, beamforming requires the location knowledge of receivers. {Hence, online design of precoding and combining matrices, as well as resource allocation, are of great importance}. 
Such knowledge of transceiver locations is crucial in UM-MIMO THz systems with narrow beams. For AOSA-based THz systems, the optimization of precoder/combiner is at the SA level instead of the antenna level. Thus, in addition to the data symbols from the RFCs, the SA beamforming angles should also be well-designed.
Furthermore, the optimization of RIS coefficients and resource allocation inside a dense network (possibly in near-field scenarios) requires effective algorithms. In summary, offline and online optimizations are equally important in future communication systems. We next formulate the optimization problem and discuss the effect of different variables on system objectives.


\subsubsection{Problem Formulation}
Different localization scenarios have different performance requirements {(or objectives, such as accuracy, coverage, and so on, as defined in Sec.~\ref{sec:localization_performance_metrics})}. In most cases, these objectives are related, and tradeoffs have to be made. For example, increased coverage may increase Latency, and increased update rate may affect accuracy. A system may seek one or several objectives to be optimized while meeting other practical constraints. 

A general optimization problem formulation of THz localization systems, consisting of an objective function $\fv(\mathcal{V})$ and a constraint function $\gv(\mathcal{V})$, can be expressed as
\begin{equation}
\begin{split}
    \mathcal{V} = & \argmin_{\mathcal{V}}{\fv(\mathcal{V})},\\
    \mathrm{s.t.}\ \ &\gv(\mathcal{V}) \le 0.
\end{split}
\label{eq:optimization_general}
\end{equation}
Here, $\mathcal{V}$ is a set of variables that could be chosen from the number of devices $L_\mathrm{Q}$, positions $\pv_\mathrm{Q}$, antennas per array $N_\mathrm{Q}$ and SA $\mathring N_\mathrm{Q}$, SA spacing $\Delta$, AE spacing $\mathring\Delta$, beamforming angles $\mathring\vpv$, RIS coefficient $\Omegam$, number of transmissions $\mathcal{G}$, etc. 
Rather than choosing a single objective or constraint, multiple objective optimization (MOO) problems can also be considered, implying that $\fv(\mathcal{V})$ and $\gv(\mathcal{V})$ could comprise a set of objectives and constraints.

The objective functions depend on the system requirements for localization performance discussed in~Sec.~\ref{sec:localization_performance_metrics}, while the constraints reflect the types of variables (e.g., discrete variables or continuous variables) and the search space (e.g., positions within a specific area) of the variables to be optimized. 
In different scenarios, a parameter could either be an objective or a constraint. For example, localization accuracy can be used as an objective to be optimized, but it could also be accounted for as a constraint to be met (e.g., the minimum required accuracy) alongside other objectives to be optimized (e.g., energy efficiency).

To achieve the system objectives while sustaining the constraints, we classify the system design and optimization into offline and online. The corresponding variables and the effect on the system objectives are summarized in Table~\ref{tab:beam_level_objectives}. 
Before discussing these two categories, we detail design considerations.

\begin{table}[t]
\footnotesize
\caption{Objectives of Different Design/Optimization Considerations (and the Corresponding Variables of Optimization)}
\renewcommand{\arraystretch}{1.2}
    \centering
    \begin{tabular}{m{0.2mm}| l
    !\vthickline
    c|p{0mm}|p{0mm} |p{0mm}| p{0mm} | p{0mm} | p{0mm} | p{0mm}| p{0mm}}
    \hthickline
        & \centering{\rotatebox{0}{Considerations}} & 
        \centering{\rotatebox{0}{Variables}} & 
        {\!\!\rotatebox{90}{Accuracy}} &
        \!\!\rotatebox{90}{Coverage} & 
        \!\!\rotatebox{90}{IA Delay} & 
        \!\!\rotatebox{90}{Update Rate} & \!\!\rotatebox{90}{Stability} & \!\!\rotatebox{90}{Scalability} & \!\!\rotatebox{90}{Mobility} &
        \!\!\rotatebox{90}{Complexity}\\
    \hthickline

        \multirow{5}{*}{\!\!\rotatebox{90}{\makecell{Offline}}} 
        & \#./Pos. of BS/RIS &
        $L_\mathrm{Q}$, ${\pv_\mathrm{Q}}$
        & $\!\!\checkmark$ & $\!\!\checkmark$ & 
        &  & $\!\!\checkmark$  & $\!\!\checkmark$ 
        & $\!\!\checkmark$ &  $\!\!\checkmark$
        \\ \cline{2-11}
        & Array Size & $N_\mathrm{Q}$, $\mathring N_\mathrm{Q}$
        & $\!\!\checkmark$ & $\!\!\checkmark$ & 
        &  &  & $\!\!\checkmark$ 
        &  & $\!\!\checkmark$
        \\ \cline{2-11} 
        & Directionality & ${G_0}$ 
        & $\!\!\checkmark$ & $\!\!\checkmark$ & $\!\!\checkmark$
        &  &  & 
        & &
        \\ \cline{2-11} 
        & Quantization & ${\mathcal{Q}}$ 
        & $\!\!\checkmark$ & & 
        & & & 
        & & $\!\!\checkmark$  
        \\ \cline{2-11} 
        & Codebook & ${\mathcal{C}}$ 
        & \ & $\!\!\checkmark$ & $\!\!\checkmark$ & \ & \ & $\!\!\checkmark$ & & 
        \\ \hline
        
        \multirow{5}{*}{\!\!\rotatebox{90}{Online}} 
        & Time/ \#. of Meas. & ${T, G}$
        & $\!\!\checkmark$ &  & 
        & $\!\!\checkmark$ & $\!\!\checkmark$ & $\!\!\checkmark$
        & $\!\!\checkmark$ &
        \\ \cline{2-11}
        & Bandwidth & ${B, K}$
        & $\!\!\checkmark$ & $\!\!\checkmark$ &  
        & $\!\!\checkmark$ & & $\!\!\checkmark$ 
        & $\!\!\checkmark$ & $\!\!\checkmark$ 
        \\ \cline{2-11} 
        & Power & ${P, \sv}$ 
        & $\!\!\checkmark$ & $\!\!\checkmark$ &  
        &  & $\!\!\checkmark$ & $\!\!\checkmark$ 
        &  &
        \\ \cline{2-11} 
        & Beamforming Angles& ${\mathring {\boldsymbol{\vpv}}}$
        & $\!\!\checkmark$ & $\!\!\checkmark$ & \ 
        & \ & $\!\!\checkmark$ & $\!\!\checkmark$ 
        & $\!\!\checkmark$ &
        \\ \cline{2-11} 
        & RIS Coefficients& ${\Omega}$ 
        & $\!\!\checkmark$ & $\!\!\checkmark$ & $\!\!\checkmark$ & & $\!\!\checkmark$ & $\!\!\checkmark$ 
        & $\!\!\checkmark$ &
        \\ 
    \hthickline
        
    \end{tabular}
    \renewcommand{\arraystretch}{1}
    \label{tab:beam_level_objectives}
\end{table}


\subsection{Design Considerations}
\label{sec:design_considerations}
When designing a localization system, we consider aspects such as the selection of network structures, the cooperative strategy, and algorithms determined by the application scenarios.

\subsubsection{Network Topology}
In previous sections, we described a communication system consisting of an LOS channel, an RIS channel, and multiple NLOS channels. However, multiple BSs/RISs/UEs  (e.g., $L_\mathrm{B}/L_\mathrm{R}/L_\mathrm{U}$) should be involved as densification is one of the main features in future communication systems.
THz communication system topologies can be classified into three types: centralized, distributed, and clustered. The clustered architecture is mainly seen in nanonetwork environments where short communication distances and high energy efficiency are favored~\cite{ghafoor2020mac}. For macro scenarios where the communication distance is large, centralized and distributed structures are usually used. The centralized structures can yield better overall performance with proper scheduling, while the distributed ones protect user privacy. 

\subsubsection{Network Structure}
For macro scenarios, three structures can be considered to improve system performance:
\begin{itemize}
    \item Heterogeneous network: Future networks are likely to be heterogeneous where different wireless (and wired) protocols coexist~\cite{yastrebova2018future}. Such a multi-band network can sufficiently alleviate the deafness issue and reduce the initial access delay.
    \item RIS-assisted network: Passive RISs can reshape the channel and increase coverage. The footprints of RISs operating at THz frequencies are expected to be small {due to short wavelengths}, which can provide extra flexibility in deployment.
    \item Cell-free network: UM-MIMO systems provide beamforming gains and energy efficiency~\cite{faisal2020ultramassive}. {However, performance is limited by the THz channel due to its lower-rank and poorer, sparser structure of multipath propagation~\cite{chaccour2022seven}.} By adopting a distributed MIMO system with multiple BSs (probably with a smaller array size) without cell boundaries, UE could have a high coverage probability~\cite{ngo2017cell}, and the geometrical diversity of the BSs can also improve the localization performance.
\end{itemize}
Such infrastructure enablers can improve the localization performance, assuming proper protocols, useful network management overheads, and efficient real-time processing.

\subsubsection{Cooperative Strategies}

Although frequent communications between the UEs cause overheads and energy consumption, cooperative localization improves the localization accuracy and the localization coverage~\cite{xiao2022overview}. The corresponding performance metrics should thus be defined for a reasonable tradeoff.
In addition to cooperation between BSs, RISs, and UEs, other types of cooperation, including UAV-assisted localization~\cite{wang2019energy} and data fusion from other types of sensors such as IMUs~\cite{lu2020milliego} and cameras~\cite{zhang2019extending}, are also important.

\subsubsection{Hardware Selection}
In order to achieve a good tradeoff between hardware cost and system performance, hardware selection is involved in offline system design.
Hardware selection considers the directionality of antennas, the quantization of phase-shifters (or RIS coefficients), and the effect of hardware imperfection. In~\cite{petrov2017interference}, the effect of the antenna model, blockage, absorption, density on the interference, and SNR are analyzed {for THz systems}. This analysis provides insight into device density and antenna directionality selection for THz network design. In general, omnidirectional antennas are used at the service discovery phase, and directional antennas are used for message transmissions and localization~\cite{han2017ma}.
In addition, the amplitude and phase control of RISs are not continuous in practice, where a quantized model should be considered in system design~\cite{wu2019towards}.

\subsubsection{Signal Design}

Implementing single-carrier versus multi-carrier modulation in THz systems is still not conclusive. Wideband single-carrier modulation has low complexity and could be used in scenarios with frequency-flat channels (e.g., limited multipath components). However, due to the frequency-dependent molecular absorption loss and multipath (mainly in indoor environments), {multi-carrier} systems are still preferred at the cost of high complexity and low power efficiency. OFDM can serve as a direct off-the-shelf solution, and \ac{dftsofdm}~\cite{sahin2016flexible, tarboush2021single} can be used to reduce the PAPR effect. Other multi-carrier modulations such as \ac{otfs} modulation (suitable for highly dynamic channels)~\cite{hadani2017orthogonal}, hierarchical bandwidth modulations~\cite{hossain2019hierarchical} (mitigate the effect of molecular absorption), spatial~\cite{sarieddeen2019terahertz} and index modulations~\cite{loukil2020terahertz} (improve spectral efficiency) are also considered for certain scenarios. THz non-orthogonal multiple access is also being studied~\cite{sarieddeen2021terahertz,magbool2021terahertz}. In this work, we want to compare THz systems and mmWave systems directly; hence, OFDM modulation is assumed. 

From a communication point of view, the selection of signal parameters, such as carrier frequency, bandwidth, and packet length, affects the data rate or spectrum efficiency. These parameters are also crucial for localization to obtain specific objectives. A large bandwidth is helpful to separate paths in the delay domain, but the increased sampling rate and data size should not exceed the hardware limit.
The packets should be long enough to capture enough energy but short enough to meet delay constraints, especially in mobile scenarios. The design considerations directly affect the performance of a localization system; we evaluate some signal parameters via simulations in Sec.~\ref{sec:simulation_and_evaluation}.

\subsection{Offline Optimization}
\label{sec:offline_optimization}
In an offline design, no knowledge of the position/orientation information of UEs is available. However, the surrounding environment information could be available. The offline design includes layout optimization, array design, and codebook optimization.

\subsubsection{Layout Optimization}
If the number of BSs/RISs/UEs is determined, their positions can be optimized based on the CRB derived using a predefined codebook. Environmental information (e.g., the geometry of the detection area and position of the blockage) can also be used to optimize the layouts and achieve the best localization performance. For the BSs with antenna arrays or directional antennas, the orientation should also be optimized.

\subsubsection{Array Design}
Increasing the number of antennas in an array yields higher angular resolution and beamforming gains. However, more antennas indicate higher system complexity and power cost.
When adopting an AOSA structure, the design of SA size is also important. A large number of AE per SA increases beamforming gain and improves accuracy, but narrow beamwidths reduce coverage and cause deafness issues.

\subsubsection{Codebook Optimization}
IA is the procedure in which a new UE establishes a physical link with a BS to switch from an idle mode to a connected mode~\cite{hu2020position}. We can treat the IA procedure as localization without UE prior information. The narrow beams in THz systems make IA challenging due to deafness (transmit-receive beams do not point to each other) and blockage (channel drop caused by obstacles, device movement, or rotation)~\cite{giordani2016initial}. Hence, effective initial access procedures and dedicated codebook design are needed~\cite{li2020beam}.

The design of codebooks depends on the search strategies, which can be broadly classified into several categories:
\begin{itemize}
    \item Exhaustive search: The BS/UE transmits/receives data symbols by beamforming in different directions~\cite{alkhateeb2017initial}. 
    \item Iterative search: Hierarchical codebooks can be designed to transmit pilots over wider sectors at the beginning and then narrow down the beams to find the best angular space~\cite{noh2017multi,qi2020hierarchical,sun2019beam}. 
    \item Scene-aware search: If the position prior or environmental information is available, the beams can be learned for each partitioned area to reduce IA delay~\cite{hu2020position,chen2019two}.
\end{itemize}

For these strategies, an exhaustive search provides the best coverage and hardware feasibility, but the discovery delay grows linearly with beamforming gain~\cite{desai2014initial, barati2016initial}. Iterative search reduces the discovery delay at the expense of limited coverage. Considering the potential of THz SLAM, we expect a scene-aware search to be used.

\subsubsection{Offline Design Example}
{Consider an RIS placement problem in which we want to minimize the localization coverage area with UE's PEB greater than an error threshold $\epsilon$ (e.g., \unit[0.1]{u}), given BS locations and orientations. For each possible RIS placement (position $\pv_\mathrm{R}$ and orientation $\ov_\mathrm{R}$) and each possible UE location $\pv_\mathrm{U}$ (assuming an omnidirectional antenna), there exist a FIM $\Jm(\pv_\mathrm{U}, \etav|\pv_\mathrm{R}, \ov_\mathrm{R})$ and corresponding $\mathrm{PEB}(\pv_\mathrm{U}, \etav|\pv_\mathrm{R}, \ov_\mathrm{R})$ that can be obtained from~\eqref{eq:PEB}. Here, $\etav$ contains nuisance parameters (e.g., channel gains, clock biases), which are replaced with nominal values (e.g., $\etav(\pv_\mathrm{U}, \pv_\mathrm{R}, \ov_\mathrm{R})$ obtained from channel models). Similar assumptions need to be made for other variables such as precoders, combiners, and RIS coefficients. We can then formulate the RIS placement problem as
\begin{equation}
\begin{split}
    \mathrm{{maximize}\ \ } &|\mathcal{R}(\pv_\mathrm{R}, \ov_\mathrm{R})|\\
    \mathrm{s.t.\ \ } & \pv_\mathrm{R}\in \mathbb{R}^3 ,\ov_\mathrm{R} \in \mathrm{SO(3)},
\end{split}
\label{eq:offline_optimization_example}
\end{equation}
where $\mathcal{R}(\pv_\mathrm{R}, \ov_\mathrm{R}) = \{\pv_\mathrm{U}\!\in\! \mathbb{R}^3 | \mathrm{PEB}(\pv_\mathrm{U}, \etav|\pv_\mathrm{R}, \ov_\mathrm{R}) \le \epsilon\}$ is the localization coverage area, and $|\mathcal{R}(\pv_\mathrm{R}, \ov_\mathrm{R})|$ denotes the volume of the coverage area (e.g., a set of discrete UE positions). Such a problem is generally non-convex and grid-search techniques can be applied.}

Offline optimization in THz systems differs from low-frequency systems in several aspects. Firstly, the precoders/combiners and RIS coefficients need to be optimized first before layout optimization. Furthermore, optimization with multiple BSs/RISs is highly non-convex, and it is thus hard to obtain globally optimal solutions. Heuristic algorithms could be alternative time-saving options to get satisfactory sub-optimal results. Note that the optimization problem formulated in~\eqref{eq:offline_optimization_example} is a simplified case in which the antenna at the UE is assumed to be omnidirectional. In general, however, the orientation of the UE needs to be considered when optimizing the layout.

\subsection{Online Optimization}
\label{sec:online_optimization}
Unlike the offline design, where no prior information is available, online optimization is performed with known UE position/orientation information (or with prior information in the tracking scenario). Online optimization can be formulated as minimizing the worst-case localization performance (e.g., PEB)~\cite{bjornson2022reconfigurable}.
We consider online optimization in three aspects: resource allocation, active beamforming optimization, and RIS coefficient optimization.

\subsubsection{Resource Allocation}
Resource allocation is an essential phase in the operation of a communication network serving multiple UEs or conducting multiple tasks. This subsection focuses on three types of resources: time, bandwidth, and power (the space resource is discussed in Sec.~\ref{sec:bs_ue_beamforming} and Sec.~\ref{sec:ris_coefficients}.
\begin{itemize}
    \item {Time Resource}: For single-user communication, a tradeoff between transmission time and overhead needs to be made. Intuitively, more transmissions/measurements yield better localization accuracy at the cost of increasing the latency and overhead. The allocation strategy should also consider {UE} speed and channel coherence. The allocation of time slots (or transmissions) for multiple users aims at meeting the positioning quality-of-service (QoS) within the served area. 
    
    \item Frequency Resource: Due to the variation of vapor absorption coefficients at different frequencies, the THz spectrum is divided into a set of distance-varying spectral windows~\cite{han2018propagation}. The size of the effective bandwidth window gets smaller with increasing link distance. In THz communications, the effective bandwidth is expected to support hierarchical bandwidth modulation~\cite{hossain2019hierarchical}, optimizing device density to maximize capacity \cite{gerasimenko2019capacity}. From a localization point of view, suitable subbands and subcarriers need to be selected and assigned to the {UEs} at different distances.
   
    \item {Power Allocation}: For most applications, localization accuracy is a constraint rather than an objective to be optimized. For example, an accuracy of, say $\unit[1]{cm}$ is sufficient for a mobile user to know its location inside an office building. Hence, there is no reason to increase the transmission power to achieve an accuracy of $\unit[1]{mm}$. With proper power allocation, different performance requirements of different UEs can be met with minimal resource utilization.
\end{itemize}
Resource allocation is usually expressed as constraints for active and passive beamforming optimizations, as we describe shortly.

\subsubsection{BS/UE Beamforming Optimization (Active Beamforming)}
\label{sec:bs_ue_beamforming}
With prior location information, setting the beamforming angles to point to the receiver increases the power of the received signal, which is beneficial for communication. However, this SNR increment does not guarantee an improvement in localization performance. More practical solutions utilize the CRB as an indicator. Given the uncertainty range of the target directions, the optimal precoders for tracking the DOA and DOD are derived in~\cite{garcia2018optimal} by solving a formulated convex optimization problem. With multiple measurements, iterative location estimation and beamforming optimization can also be performed~\cite{zhou2019successive}.

By implementing AOSA structures and directional antennas, space resource allocation reduces the assignment of beams (SAs). The optimization of the analog beamforming angles directly affects the accuracy and coverage of the system. For scenarios with multiple UEs, the SAs need to be assigned wisely to different UEs, completing both the localization and communication tasks. The SA selection is especially important in localization, where accuracy depends on the array layout. For a communication network with multiple BSs, joint beamforming optimization is also needed to achieve a better overall system performance.


\subsubsection{RIS Coefficients Optimization (Passive Beamforming)}
\label{sec:ris_coefficients}
The optimization of RIS coefficients is as crucial as active beamforming at the BS/UE arrays to enhance signal gain in RIS-assisted systems. When the UE position is unknown, multiple transmissions with random symbols or beamforming angles can be used for localization purposes. With prior information of the UE position/orientation, beamforming angles at the BS/RIS/UE can be jointly optimized.

The coefficients of RIS elements can be optimized to serve communication or localization purposes~\cite{bjornson2022reconfigurable}. The elements in an RIS can be optimized to maximize the SNR at the receiver for a higher data rate. However, a high SNR does not indicate a lower CRB. By analyzing the FIM,~\cite{bjornson2022reconfigurable} optimizes the RIS elements in a 2D SISO localization system to reduce the PEB. However, the optimization algorithms for 3D MIMO systems are not yet available.

\subsubsection{Online Optimization Example}
{We again consider the case with RIS, where we aim to optimize the RIS phase profiles $\Omegam_1, \ldots, \Omegam_\mathcal{G}$ for different transmissions, given a certain precoder at the UE and a combiner at the BS. We assume that the UE location is known to be in some region with $\pv_\mathrm{U}\in \mathcal{R}^* \subset \mathbb{R}^3$.
By introducing $\omegav_g\!=\!\mathrm{diag}(\Omegam_g)\in \mathbb{C}^{N_\mathrm{R}\times N_\mathrm{R}}$ and using $\etav$ to indicate the estimated values of other nuisance parameters, we can compute the FIM $\Jm(\hat\pv_\mathrm{U}, \etav| \omegav_1, \ldots, \omegav_\mathcal{G})$ and its corresponding PEB. An online optimization can then be formulated as
\begin{equation}
    \begin{split}
        \mathrm{minimize\ \ } & \varepsilon\\
        \mathrm{s.t.\ \ } & \mathrm{PEB}(\pv_\mathrm{U}, \etav|\omegav_1, \ldots, \omegav_\mathcal{G}) \le \varepsilon\\
        & \pv_\mathrm{U}\in \mathcal{R}^*\\
        & |\omega_{r,g}|=1, \forall r,g.
    \end{split}
\end{equation}
This problem can be solved by first removing the unit norm constraint so that a convex problem can be obtained. Then the solution needs to be projected onto the appropriate manifold.}

In THz systems with wide bandwidths, more time-frequency blocks will be available. These resources need to be allocated wisely to serve a large number of users with different communication and localization performance requirements. In addition, joint beamforming optimization for BS/RIS/UE is crucial since THz systems are expected to rely heavily on RISs due to severe blockage. Furthermore, the optimization at the SA level in an AOSA structure is different from the optimization at the antenna level in traditional MIMO systems. Consequently, novel online optimization algorithms are called for in THz systems to assist in accurate tracking.

\subsection{Summary}
In this section, we formulate the optimization problems and discuss several aspects of localization system design and optimization, which can be summarized as follows:
\begin{itemize}
    \item We start by motivating system optimization and formulating the optimization problem with a set of variables related to different objective functions. 
    \item We discuss high-level system considerations such as network topology, network structure, cooperative strategy, hardware selection, and signal design, which is dependent on the application scenario.
    \item We divide system design into offline design and online optimization, and discuss the challenges for THz systems compared with low-frequency systems.
\end{itemize}
The next section provides simulations to show the effect of parameters on the system CRB. 

\section{Simulation and Evaluation}
\label{sec:simulation_and_evaluation}
In this section, we provide several simulations to evaluate the effect of system parameters on localization performance.
A $\unit[0.3]{THz}$ sub-THz system and a $\unit[60]{GHz}$ mmWave system in an uplink scenario are considered with the default parameters listed in Table~\ref{table:Simulation_parameters}. These parameters are the default setting for the rest of the simulations unless otherwise specified.
Simulations A/B/C/D (Sec.~\ref{sec:thz_mmwave_comparison} to~\ref{sec:simulation_beam_split}) consider only LOS channels, while the effects of RIS and NLOS channels are discussed later in simulations E/F/G (Sec.~\ref{sec:RIS_coefficient_vs_CRLB} to~\ref{sec:PEB_visualization}). Matlab code is available in~\cite{chenhui07c8}.

\begin{table}[t]
\scriptsize
\centering
\caption{Default Simulation Parameters}
\renewcommand{\arraystretch}{1.2}
\begin{tabular} {c !\vthickline c}
    \hthickline
    \textbf{Parameters} & \textbf{Simulation Values} \\
    \hthickline
    mmWave / THz Frequency ${f_c}$ & $\unit[60]{GHz}$ / $\unit[0.3]{THz}$\\
    \hline
    Transmission Power ${P}$ & $\unit[10]{dBm}$\\
    \hline
    Noise PSD & $\unit[-173.86]{dBm/Hz}$\\
    \hline
    Noise Figure & $\unit[13]{dBm}$\\
    \hline
    Array Footprint (BS/RIS/UE) & $\unit[2\!\!\times\!\!2]{cm^2}$ / $\unit[10\!\!\times\!\!10]{cm^2}$ / $\unit[1\!\!\times\!\!1]{cm^2}$\\
    \hline
    AE Spacing ${\mathring\Delta}$ & $\lambda_c/2$\\
    \hline
    Bandwidth ${W}$ & $\unit[100]{MHz}$\\
    \hline
    Number of Transmissions $\mathcal{G}$ & 10\\
    \hline
    Synchronization Offset $B$ & {\unit[10]{us} (for Simulation D/E/F/G)} \\
    \hline
    {Number of Subcarriers $K$} & 10\\
    \hline
    Signal Wave Model & SWM (near-field)\\
    \hline
    Localization Scenario & 2D Position, 1D Orientation\\
\hthickline
    mmWave Array Dim ${N_\mathrm{Q}}$ (BS/RIS/UE)&  
    $4\!\!\times\!\!4$/ $20\!\!\times\!\!20$ / $2\!\!\times\!\!2$ \\
    \hline    
    THz Array Dim ${N_\mathrm{Q}}\mathring{N}_\mathrm{Q}$ (BS/RIS/UE)&
    $20\!\!\times\!\!20$/ $100\!\!\times\!\!100$ / $10\!\!\times\!\!10$ \\
    \hline
    THz SA Dim ${\mathring N_\mathrm{B}}$ /$\mathring N_\mathrm{R}$/ ${\mathring N_\mathrm{U}}$ &  
    $5\!\!\times\!\!5$ / $1\!\!\times\!\!1$ / $5\!\!\times\!\!5$\\
    \hline
    Position {${\pv_\mathrm{B}}$ / ${\pv_\mathrm{R}}$ / ${\pv_\mathrm{U}}$} & 
    $[0, 0, 0]^T$ / $[5, 5, 0]^T$ / $[10, 0, 0]^T$\\
    \hline
    Orientation ${\ov_\mathrm{B}}$ / ${\ov_\mathrm{R}}$ / ${\ov_\mathrm{U}}$ &  
    $[0, 0, 0]^T$ / $[-\frac{\pi}{2}, 0, 0]^T$ / $[\frac{5\pi}{6}, 0, 0]^T$\\
    \hthickline
\end{tabular}
\renewcommand{\arraystretch}{1}
\label{table:Simulation_parameters}
\end{table}

\subsection{A Comparison between mmWave and THz Systems}
\label{sec:thz_mmwave_comparison}
We first compare the PEB and OEB between two systems with different array configurations.
To make a fair comparison, we fix variables such as transmission power, time, and maximum footprint. A fully connected antenna array is adopted in the mmWave system, while an AOSA structure is used for the THz system. All the RFCs send different random data symbols with normalized energy.

\begin{figure}[t]
\begin{minipage}[b]{0.92\linewidth}
  \centering
  \centerline{\includegraphics[width = 0.95\linewidth]{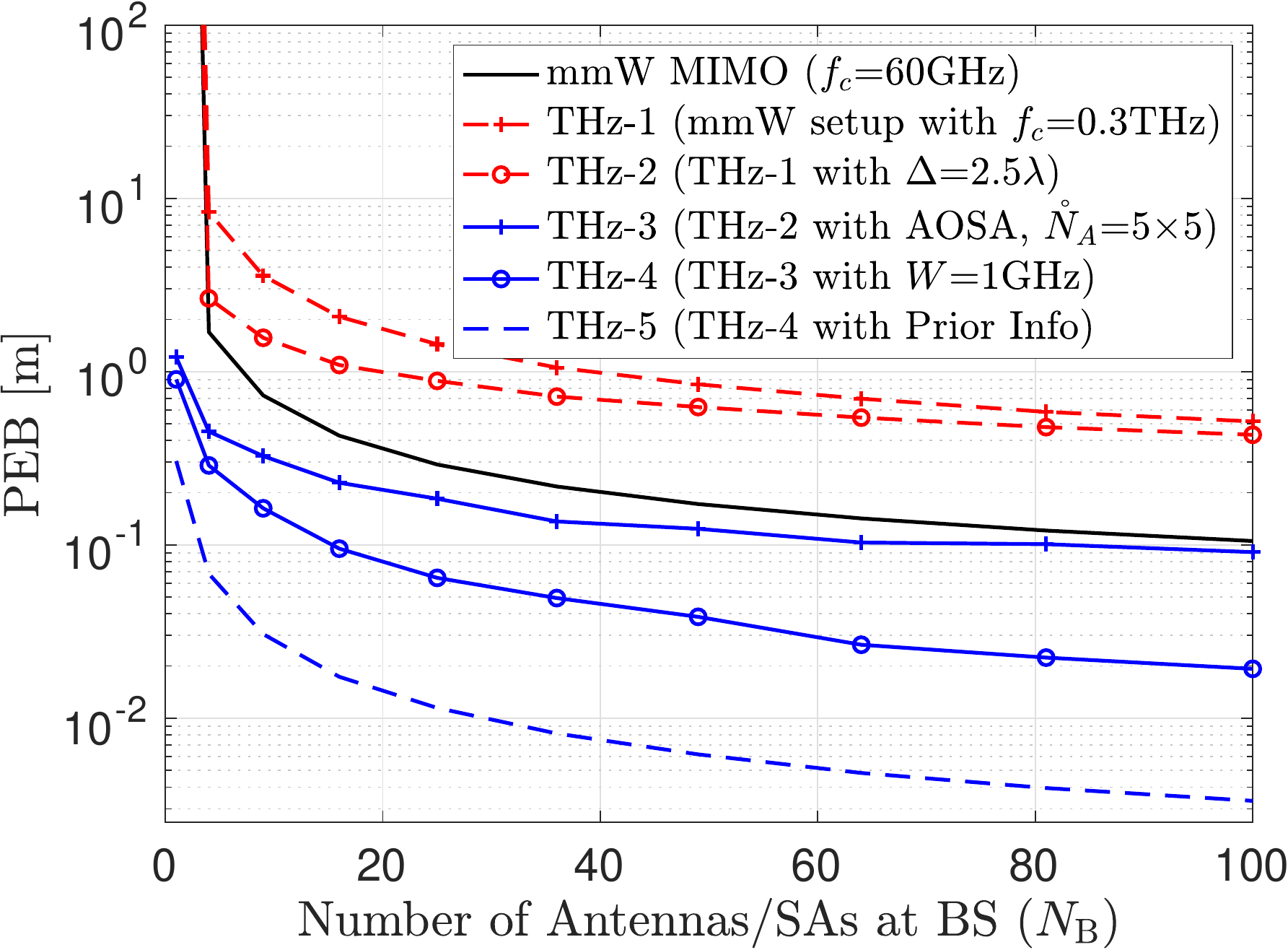}}
  \centerline{(a) PEB} \medskip
\end{minipage}
\hfill
\vspace{2mm}
\begin{minipage}[b]{0.92\linewidth}
  \centering
  \centerline{\includegraphics[width = 0.95\linewidth]{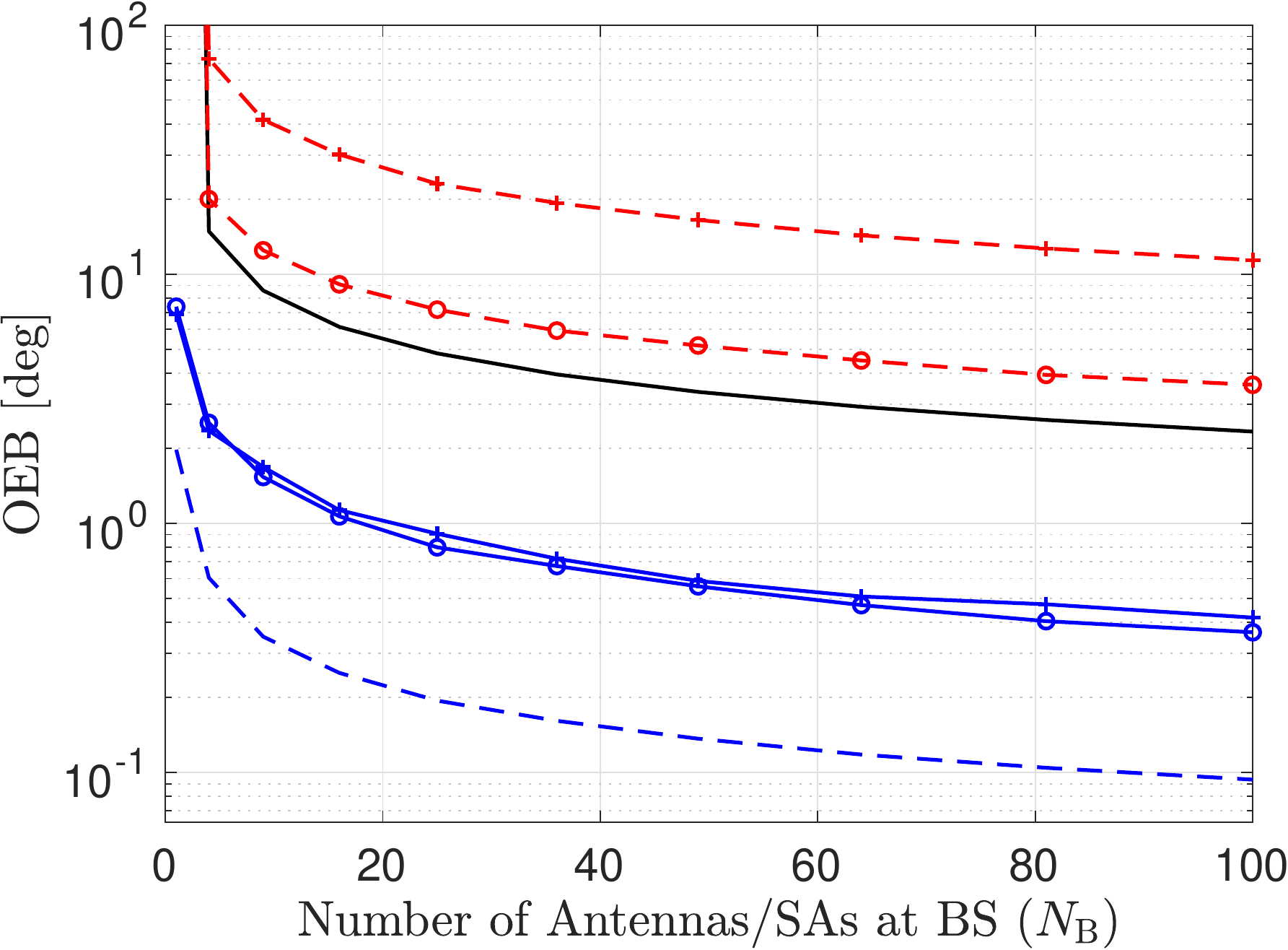}}
\centerline{(b) OEB}\medskip
\end{minipage}
\begin{minipage}[b]{0.99\linewidth}
  \centering
  \centerline{\includegraphics[width = 0.9\linewidth]{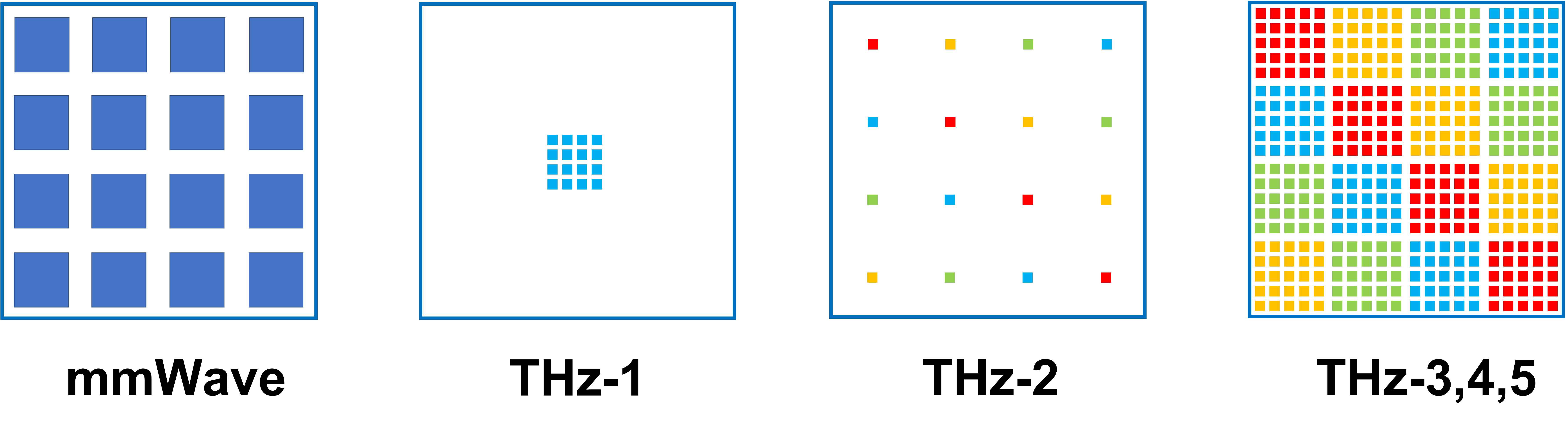}}
\centerline{(c) Illustration of different arrays ($N_\mathrm{B}=4\times4$)}\medskip
\end{minipage}
\caption{PEB/OEB vs. the number of BS antennas/SAs. (a) PEB; (b) OEB; (c) Illustration of different array configurations. By moving the carrier frequency from \unit[60]{Ghz} to \unit[300]{GHz}, the error bound increases due to high path losses. Furthermore, increasing the antenna spacing from $0.5\lambda$ to $2.5\lambda$ (maintaining the same footprint as the mmWave system) improves the performance. Adopting an AOSA structure with an SA size of $5\times5$ introduces a beamforming gain, and the error bounds outperform the benchmark mmWave system. The bounds can be even lower with a larger bandwidth (\unit[1]{GHz} rather than \unit[100]{MHz}) and prior information (e.g., setting the beamforming angle of the BS to the direction of a UE).}
\label{fig:sim_THz_mmwave_comparison}
\end{figure}

{
In this comparison, the system is assumed to be synchronized~\footnote{If only an LOS channel is considered in a far-field model, synchronization between the BS and the UE is needed for delay estimation. However, if more paths are available, e.g., extra LOS paths provided by other BSs, RIS, or NLOS paths, synchronization is not a requirement.~{One typical example is TDOA-based localization.}}, while for simulations D/E/F/G a synchronization offset is assumed. We also assume that no prior information of the UE is available, and hence the beamforming angles at the SA (AOSA systems) are set randomly as $\tilde \phi, \tilde \theta \in (-90^\circ, 90^\circ)$ for different transmissions.}
We use CRB (PEB/OEB) to evaluate the fundamental limit of the localization systems as shown in Fig.~\ref{fig:sim_THz_mmwave_comparison}.
{This figure illustrates the potential of THz localization (lower PEB and OEB), where $\sim \!\!5$ ($\sim \!\!20$) times better positioning performance without (with) prior information is expected to be achieved with the same power and footprint compared with mmWave systems.} In other words, to achieve the same performance as that of mmWave localization systems, less resources (e.g., power, footprint) are needed. However, to fully exploit the potential of THz localization systems, multiple transmissions and prior knowledge are required to solve the deafness issue. We evaluate the effect of transmission numbers on the CRB in the next subsection.

\subsection{The Effect of Transmission Numbers on CRB}
\begin{figure}[t]
\centering
\includegraphics[width = 0.92\linewidth]{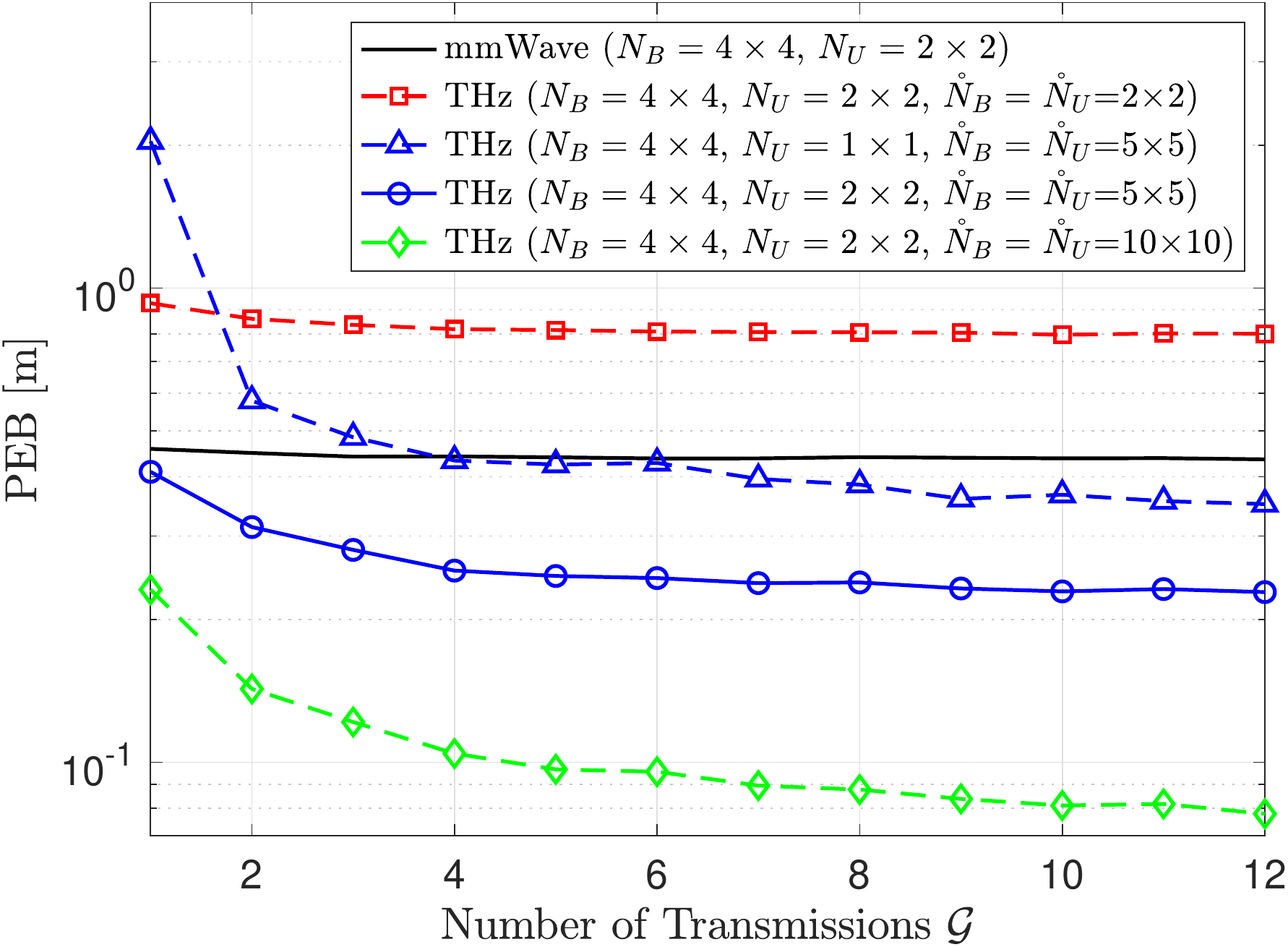}
\caption{PEB vs. number of transmissions with a fixed total transmission energy. 
The number of transmissions has a minor effect on the benchmark mmWave systems. However, multiple transmissions are needed for the AOSA structures adopting analog beamformings, especially when the SA dimension is large. ($N$ is the number of SA, and $\mathring N$ is the number of AE per SA.)}
\label{fig:sim_transmissions}
\end{figure}

We simulate the effect of transmission numbers ($\mathcal{G}$) on the PEB with different SA dimensions using fixed total transmission energy and the number of RFCs. Although the AOSA structure can also be adopted in mmWave systems, we use a fully digital array with $4\times4$ antennas for benchmarking purposes. For THz systems, we simulate different SA dimensions with $\mathring N_\mathrm{B} \!=\!\mathring N_\mathrm{U}\!=\!2\times2$, $\mathring N_\mathrm{B}\!=\!\mathring N_\mathrm{U}\!=\!5\times5$, and $\mathring N_\mathrm{B}\!=\!\mathring N_\mathrm{U}\!=\!10\times10$, respectively. The results are shown in Fig.~\ref{fig:sim_transmissions}. 

We notice that the number of transmissions has a minor effect on the benchmark mmWave systems~{because of the large beamwidth generated from a small array size}. By incorporating the AOSA structures, the computational and hardware cost can be reduced. However, multiple transmissions are needed to obtain a lower bound, where the $\mathcal{G}$ needed for the bound to converge increases with the SA dimension. With more transmissions, the deafness issues can be solved, and a UE can thus be located more accurately. In the next subsection, we discuss the effects of PWM/SWM and Syn/Asyn on the CRB of the system.



\subsection{The Evaluation of PWM/SWM for Different Channel Models}
\label{sec:pwm_swm_simulation}
\begin{figure}[t]
\centering
\centerline{\includegraphics[width = 0.95\linewidth]{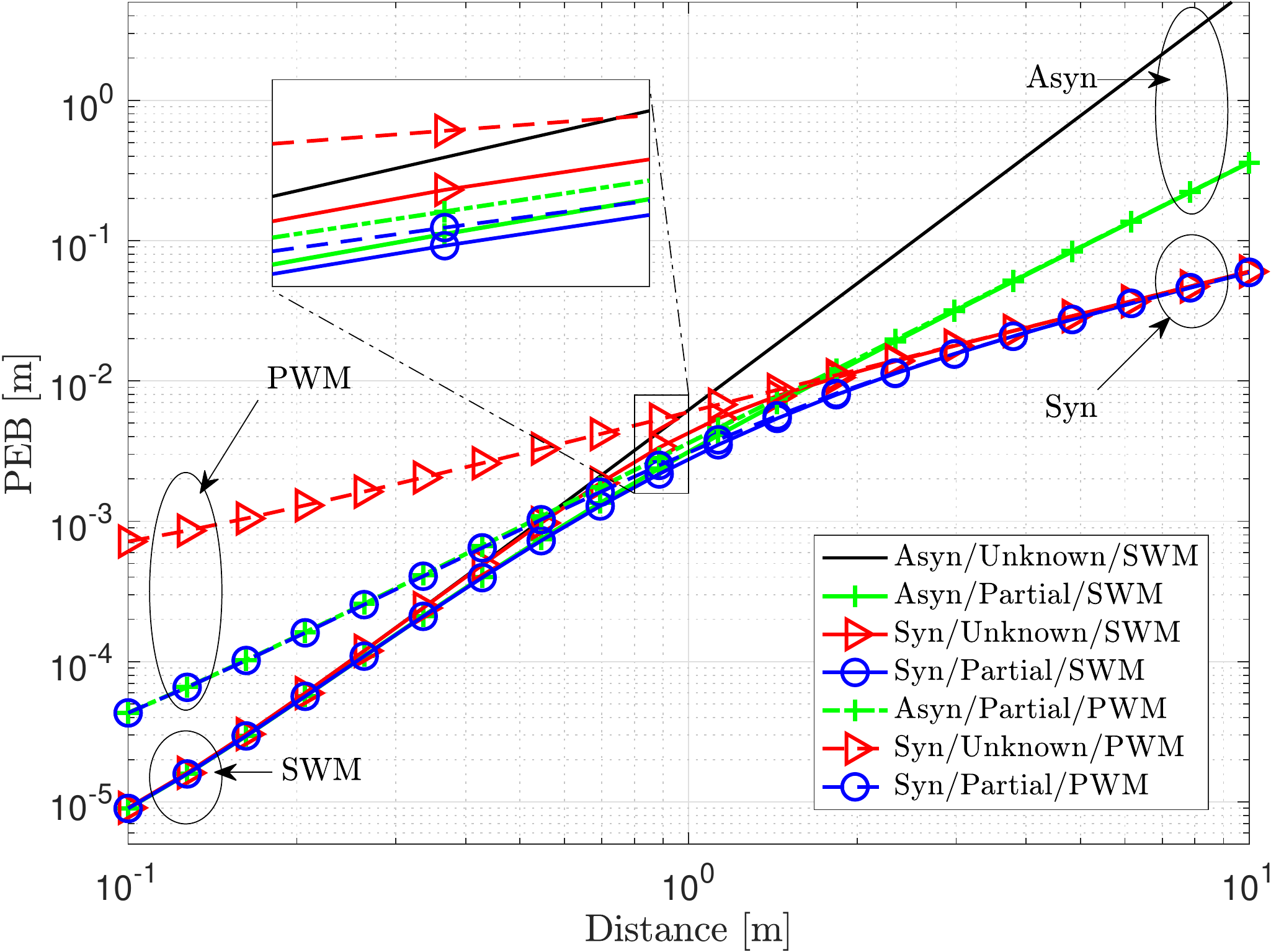}}
\caption{PEB vs. distance for different localization models assumptions. The PEBs of the systems using different signal models (e.g., SWM and PWM) have a mismatch when the distance from the BS is smaller than \unit[1]{m}. In addition, a synchronized system performs better, especially when the distance is large.}
\label{fig:sim_SWM_PWM}
\end{figure}

We have discussed several system assumptions, namely synchronized or asynchronized systems ({Remark~\ref{remark_1}}), unknown or partially-known channel gains (Remark~\ref{remark_2}), and SWM/PWM (Sec.~\ref{sec:near_field_channel_model}). These assumptions affect the channel realization, system model, and CRB. The PEB for Asyn/Unknown/PWM are not included because the position cannot be estimated in this scenario. For simulation purposes, we assume a single-antenna UE and evaluate the PEBs for different channel models by changing the position of the UE on the x-axis $x_\mathrm{U}$ with fixed $y_\mathrm{U}= z_\mathrm{U} =0$.
The simulation results are shown in Fig.~\ref{fig:sim_SWM_PWM}. 

We can see that the boundary between the far-field and near-field is at around \unit[1]{m}. With the increasing distance, the PEBs of SWM and PWM models are converging. However, the SWM is a more accurate signal model (at the expense of high computational complexity), which is advantageous when the UE is close to the BS. In addition, the SWM can help in synchronization {(by exploiting the curvature of arrival)} and the PEBs of asynchronized and synchronized systems converge in the near-field.

\subsection{Evaluation of the Beam Split Effect}
\label{sec:simulation_beam_split}
\begin{figure}[t]
\centering
\begin{minipage}[b]{0.92\linewidth}
  \centerline{\includegraphics[width = 0.99\linewidth]{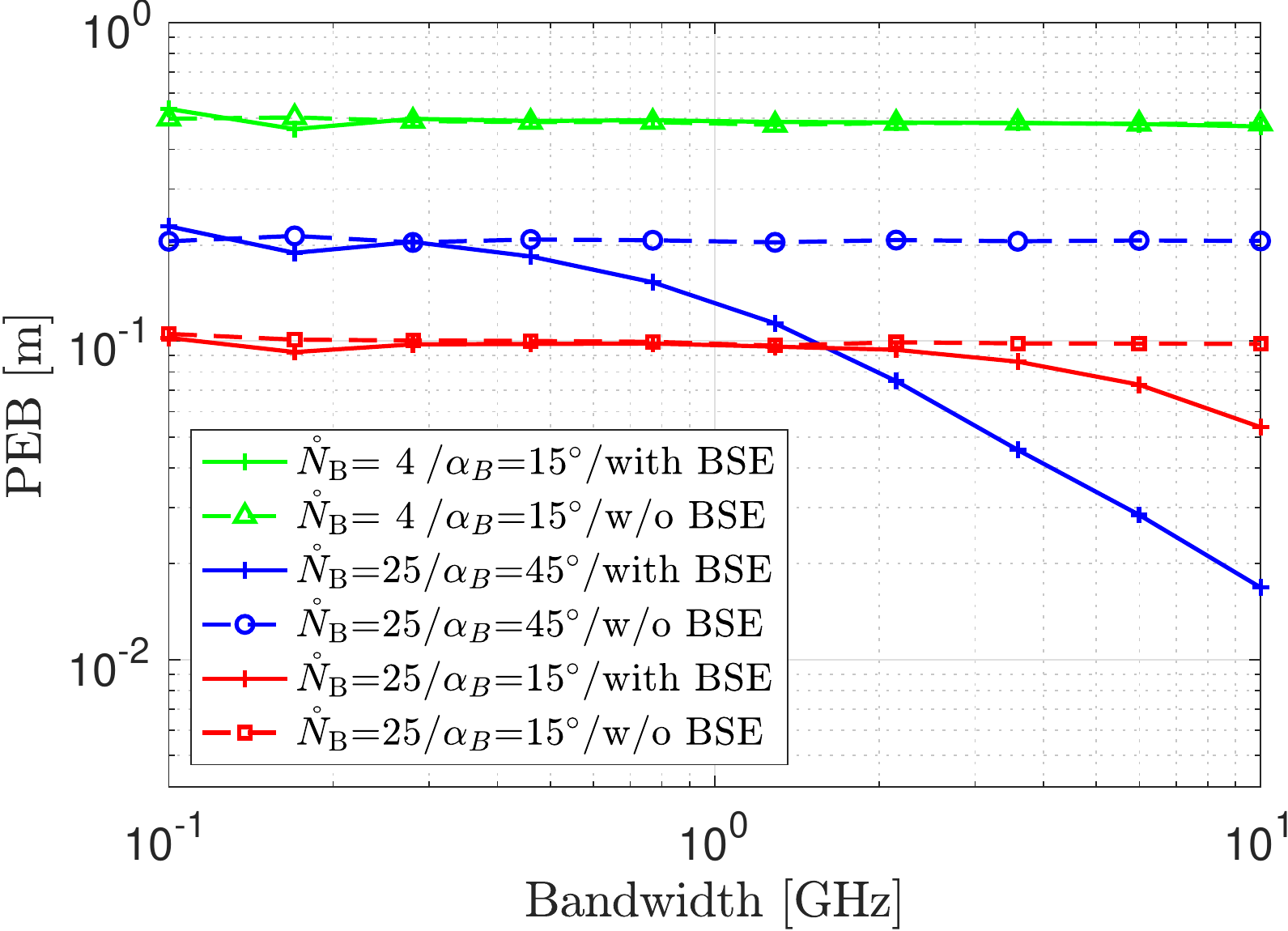}}
\end{minipage}
\hfill
\caption{CRB vs. bandwidth for the systems with/without the beam split effect. The mismatch between different models (solid vs. dashed curves) is larger with increased beamwidth, array size (blue dot vs. green triangle), and angles (blue dot vs. red square).}
\label{fig:sim_beam_split_effect}
\end{figure}

To evaluate the effect of beam split on the CRB, we assume that prior position information is known, and hence the beamforming angle at the BS is set as pointing to the UE. {The orientation of the BS is set as $15^\circ$ and $45^\circ$, and an asynchronized UE is located at $[2, 0, 0]^T$}. The PEBs for channel models with and without (w/o) the beam split effect are shown in Fig.~\ref{fig:sim_beam_split_effect}. Beam split affects the signal gain and is not preferred in communications. However, this `split' phenomenon may provide extra geometry information {(e.g., beams at different subcarriers are pointing to different directions)} that can lower the error bound, especially in wide bandwidth systems. 

So far, our discussions are limited to LOS channel. In the following, we discuss the effect of RIS and NLOS channels on localization performance.

\subsection{The Effect of RIS on CRB}
\label{sec:RIS_coefficient_vs_CRLB}
\begin{figure}[t]
\centering
\includegraphics[width = 0.92\linewidth]{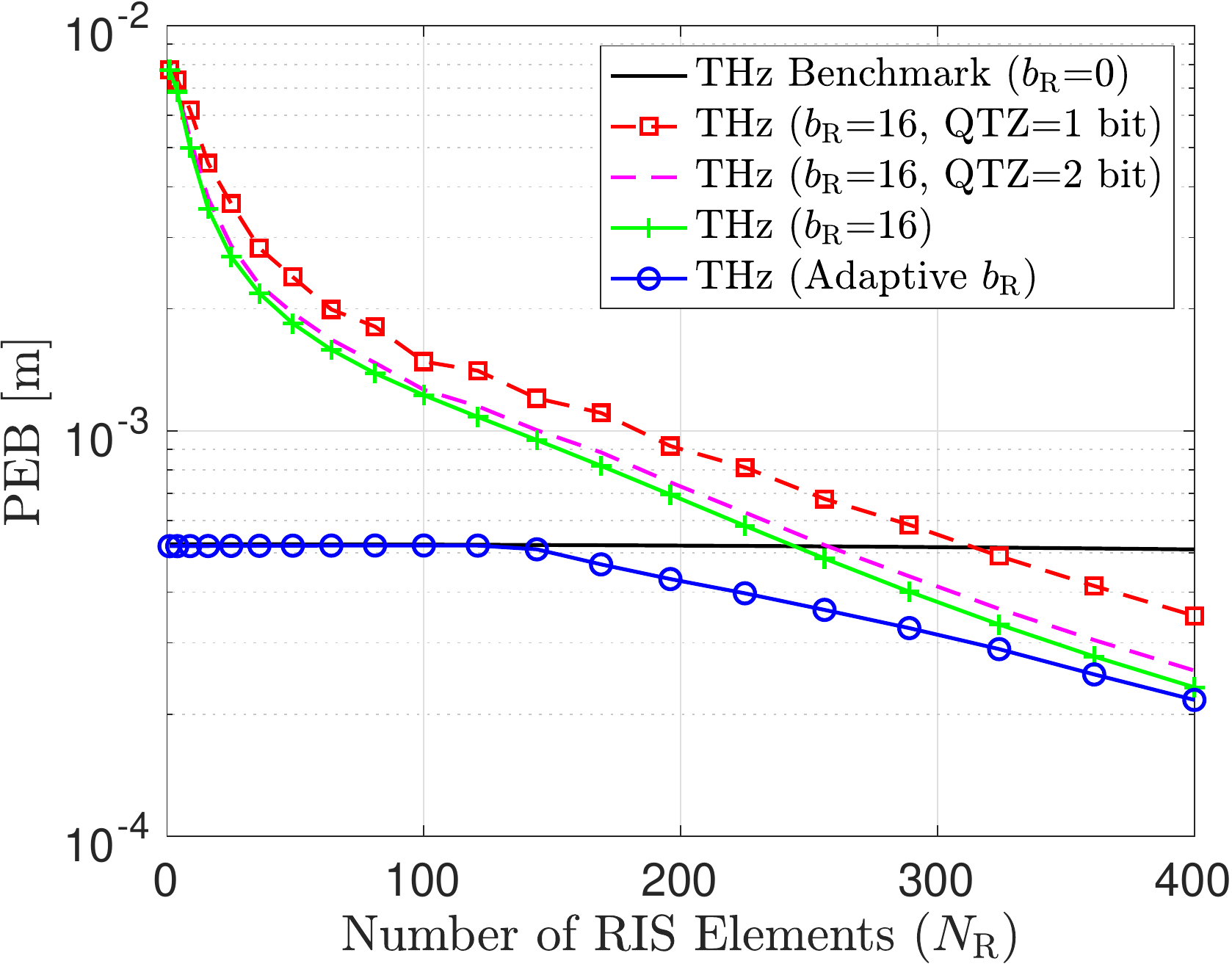}
\caption{CRB vs. the number of RIS elements ($N_\mathrm{R}$) for different scenarios (`QTZ' is short for `quantization'). `$b_\mathrm{R}$=0' means the SAs at BS/UE are beamforming to UE/BS with prior information; `$b_\mathrm{R}$=16' indicates all the $N_B/N_U$ SAs at BS/UE are beamforming to the RIS; `Adaptive' utilizes an optimal beam assignment with grid search).}
\label{fig:sim_crlb_RIS_elements}
\end{figure}

We simulate the effect of RIS dimensions on the PEB of a THz system. For a better visualization of the convergence, we use scaled positions of BS/RIS/UE as $[0, 0, 0]^T/[0.5, 0.5, 0.1]^T/[0.5, 0.4, 0.05]^T$, and keep other parameters as in~Table~\ref{table:Simulation_parameters}. Assuming prior information is known, the RIS coefficients can be optimized using the method in~\cite{he2020large} to maximize the SNR of the received signal. The beamforming angles of each SA at the BS (UE) are set to the directions of either the RIS or UE (BS). For a BS with $N_\mathrm{B}$ SAs, the total beams assigned to RIS ($b_\mathrm{R}$) and UE ($b_\mathrm{U}$) satisfies $N_\mathrm{B}=b_\mathrm{R}+b_\mathrm{U}$.
Note that such directional beams and SNR-based RIS coefficient optimization is not optimal for localization purposes; however, we use it as a benchmark for further optimization algorithms. 
Other scenarios also include quantized coefficients (1-bit and 2-bit quantization on the RIS coefficients), and the corresponding simulation results are shown in Fig.~\ref{fig:sim_crlb_RIS_elements}.

The figure shows that a large RIS with optimized coefficients improves the performance, and a 2-bit quantization on the RIS coefficient is sufficient to assist localization (PEB is close to the setup with continuous phases).
Without beamforming to the RIS, an increased number of RIS elements has {almost no} effect on the PEB of the system. However, with the AOSA structure, the PEB is also affected by the SA beamforming angles (e.g., assigned beams to RIS $b_\mathrm{R}$) and more efforts are needed to jointly optimize active and passive beamforming.

\subsection{The Effect of NLOS Paths}
\begin{figure}[t]
\begin{minipage}[b]{0.45\linewidth}
  \centering
  \centerline{\includegraphics[width = 1\linewidth]{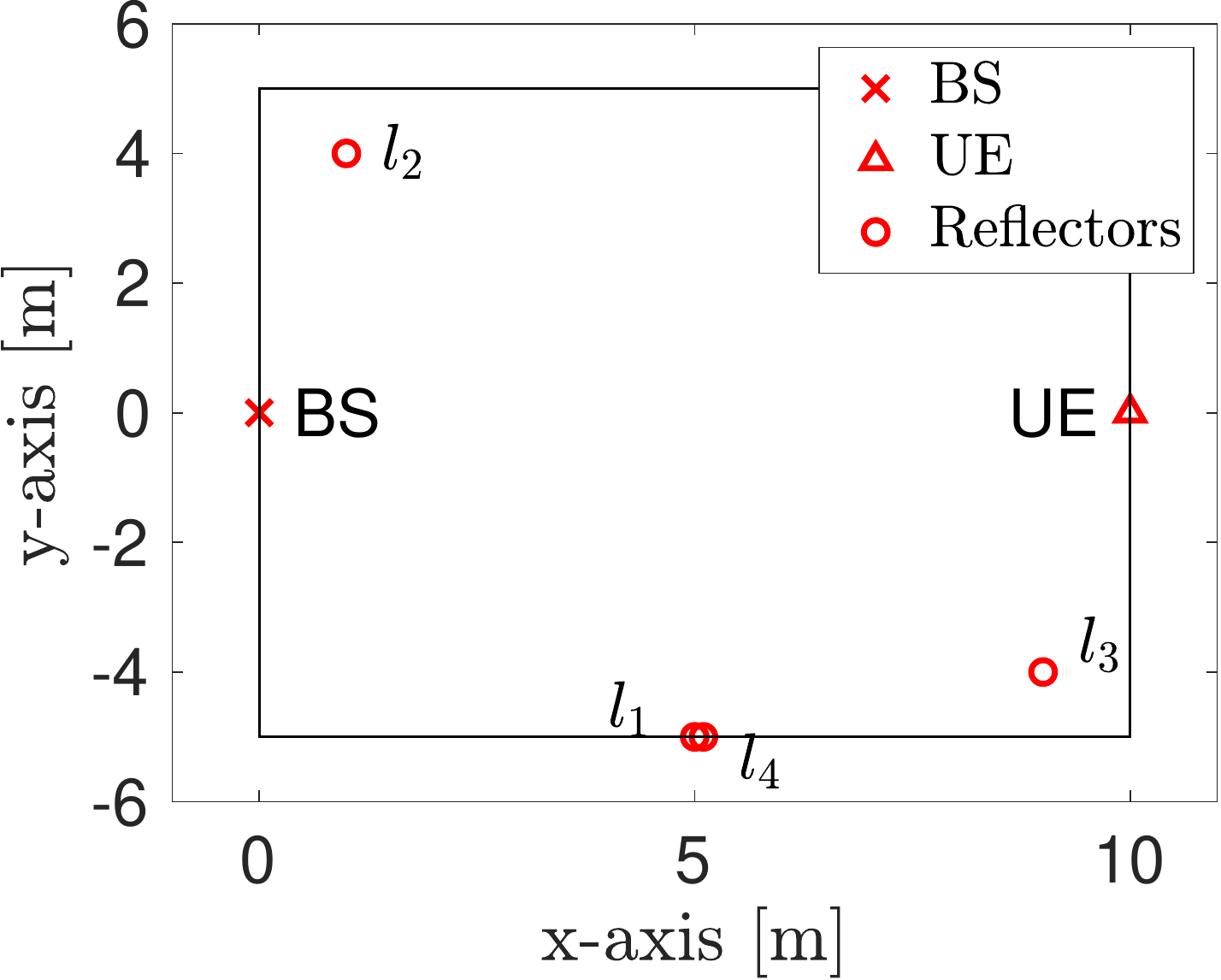}}
\centerline{(a) Layout}\medskip
\end{minipage}
\hfill
\begin{minipage}[b]{0.49\linewidth}
  \centering
  \centerline{\includegraphics[width = 1\linewidth]{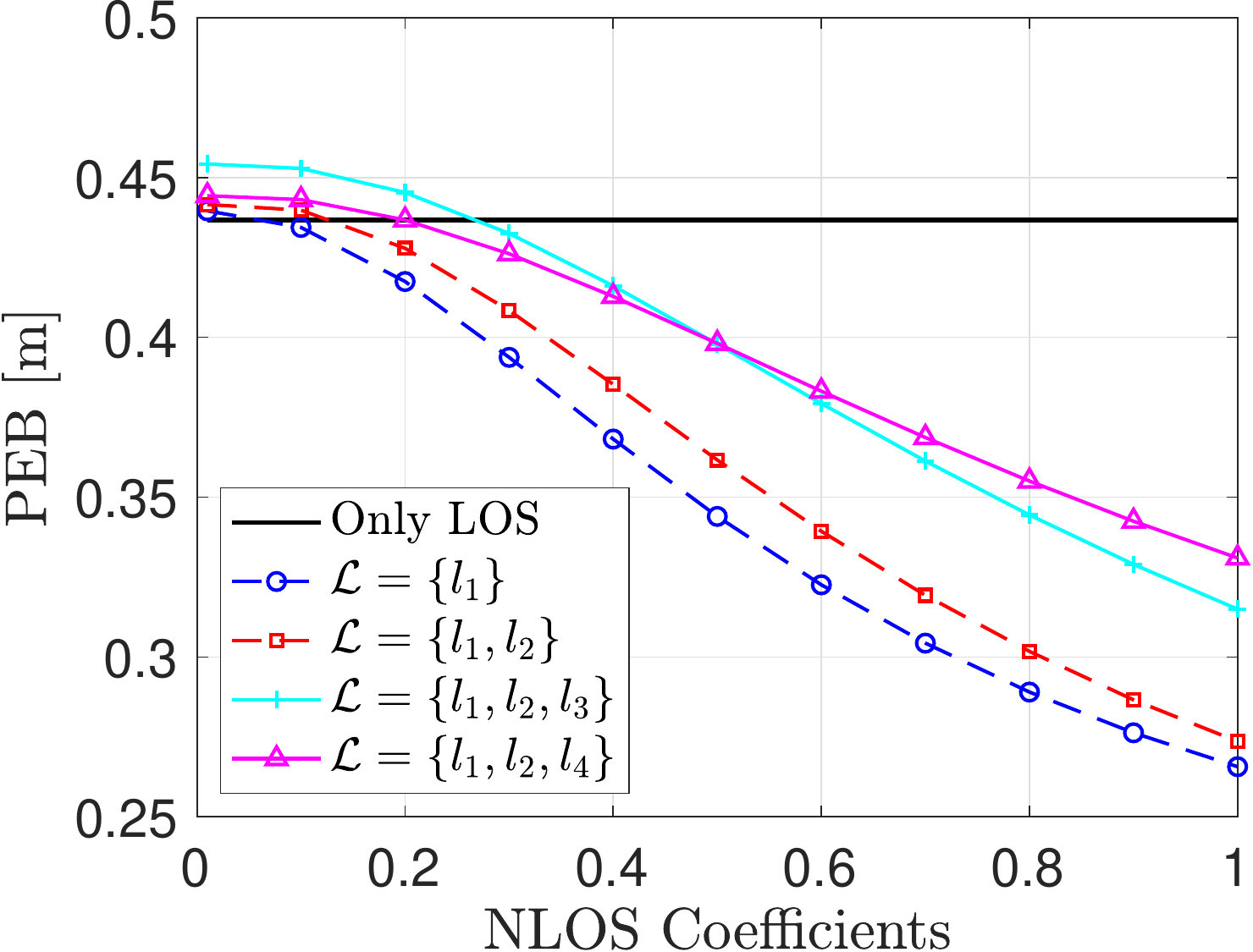}}
  \centerline{(b) PEB} \medskip
\end{minipage}
\hfill
\begin{minipage}[b]{0.48\linewidth}
  \centering
  \centerline{\includegraphics[width = 1\linewidth]{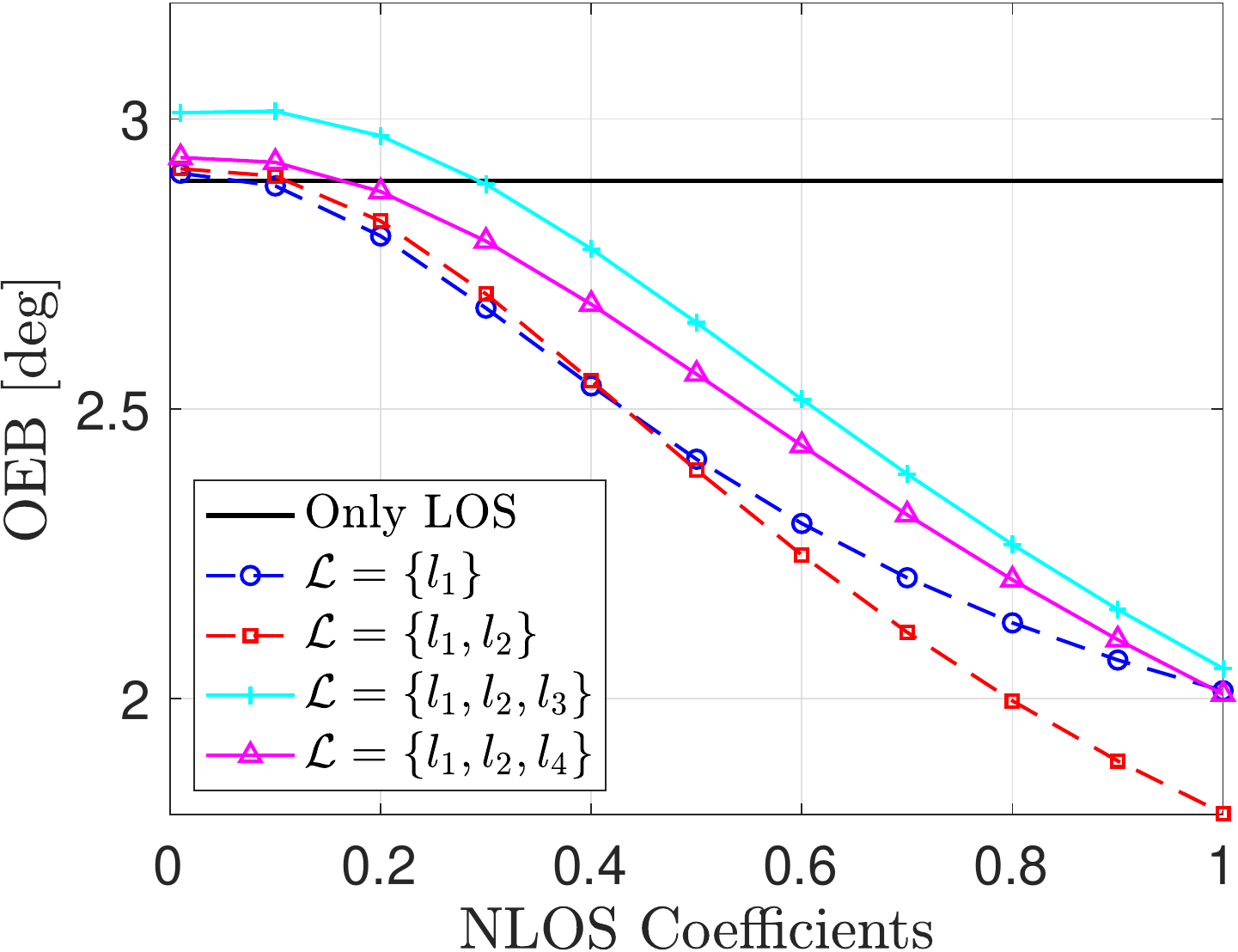}}
\centerline{(c) OEB}\medskip
\end{minipage}
\hfill
\begin{minipage}[b]{0.49\linewidth}
  \centering
  \centerline{\includegraphics[width = 1\linewidth]{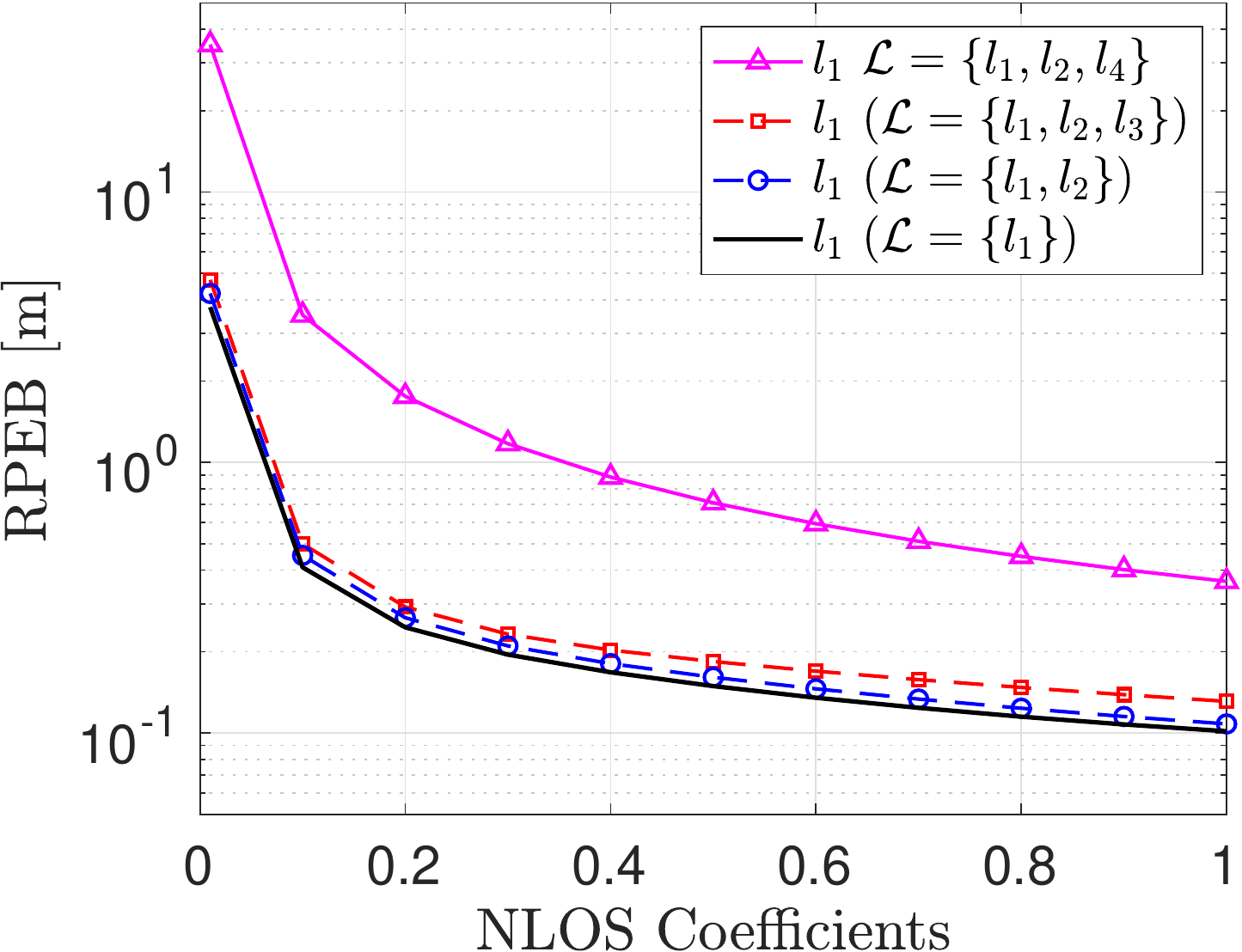}}
\centerline{(d) RPEB}\medskip
\end{minipage}
\caption{CRB vs. NLOS coefficients. {The layout of the BS, UE and possible reflectors is shown in (a), where $l_1$ and $l_4$ are not resolvable.} The NLOS paths are harmful to the CRB when the reflection coefficients are small. When increasing the coefficients, PEB, OEB, and RPEB get lower. We also notice that even if the positions of the reflectors are not resolvable ($\mathcal{L}=\{l_1, l_2, l_4\}$), the OEB is still lower than in a setup with a weaker NLOS path ($\mathcal{L}=\{l_1, l_2, l_3\}$). However, the RPEB of $l_1$ in the unresolvable scenario is much higher than in other layouts.}
\label{fig:sim_crlb_with_nlos_paths}
\end{figure}

We also evaluate the effect of NLOS signals on the CRB with different number of reflectors ($L_N=0,1,2,3$) as shown in Fig.~\ref{fig:sim_crlb_with_nlos_paths}. Four layouts of the landmarks (incident points) are considered, namely, $\{l_1\}$, $\{l_1, l_2\}$, $\{l_1, l_2, l_3\}$ and $\{l_1, l_2, l_4\}$, where $l_1$-$l_4$ are located at $[5, -5, 0]^T$, $[1, 4, 0]^T$, $[9, -4, 0]^T$ and $[5.1, -5, 0]^T$, respectively. By changing the reflection coefficients of all the reflectors (assumed to be equal) from 0 to 1, the PEB/OEB of the UE and the reflector position error bound (RPEB) are shown in Fig.~\ref{fig:sim_crlb_with_nlos_paths} (b)-(d).

This figure shows that the NLOS paths are helpful if they are resolvable and the reflection coefficient is large (which depends on the shape and the material of an object).
THz channels are expected to have fewer NLOS paths due to the high path loss and narrow beamwidth. With fewer NLOS paths, the number of localization parameters (e.g., the position of reflectors) and non-resolvable paths are reduced. {Fewer NLOS paths reduce the computational complexity, but lose geometrical diversity. Hence, more transmission times are needed to improve the localization performance and to map the whole environment.}

\subsection{The Visualziation of PEB for Different UE Positions}
\label{sec:PEB_visualization}
We visualize the 2D PEB ($z_\mathrm{U} = 0$) for different setups by changing the position of UE within a $\unit[5\times5]{m^2}$ area. The positions of the BS/RIS/reflector are $[0, 0, 0]^T$, $[2.5, 2.5, 0]^T$ and $[2.5, -2.5, 0]^T$, respectively. The number of the BS/UE elements is $N_\mathrm{B}=4\times4$/$N_\mathrm{U}=2\times2$, and
the reflection coefficient is set as $0.9$. {The PEB is obtained with a single snapshot ($\mathcal{G}=1$), which only works for systems with multiple RFCs.}

\begin{figure}[t]
\begin{minipage}[b]{0.48\linewidth}
  \centering
  \centerline{\includegraphics[width = 0.98\linewidth]{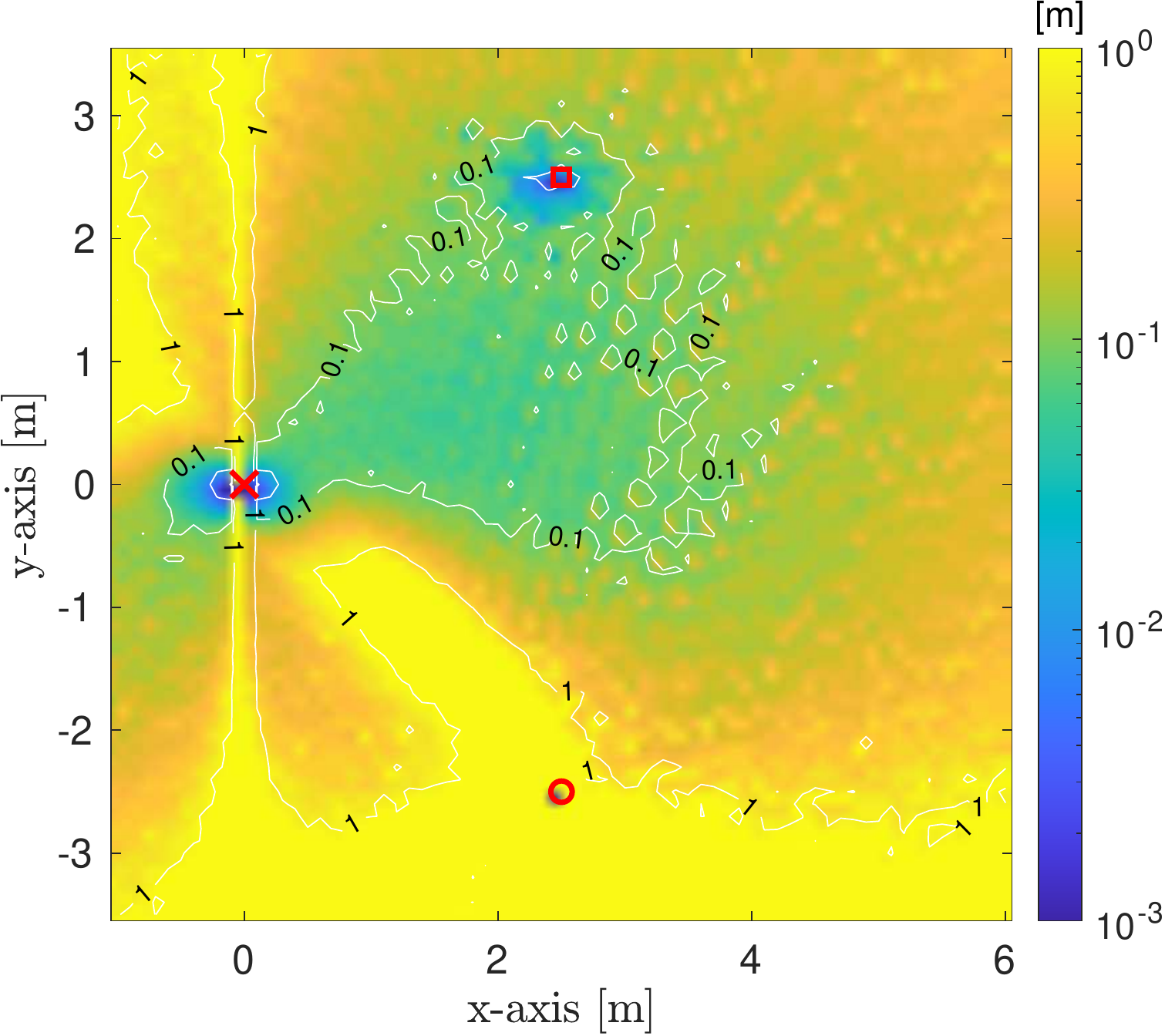}}
  \centerline{(a) mmWave (no prior)} \medskip
\end{minipage}
\hfill
\begin{minipage}[b]{0.48\linewidth}
  \centering
  \centerline{\includegraphics[width = 0.98\linewidth]{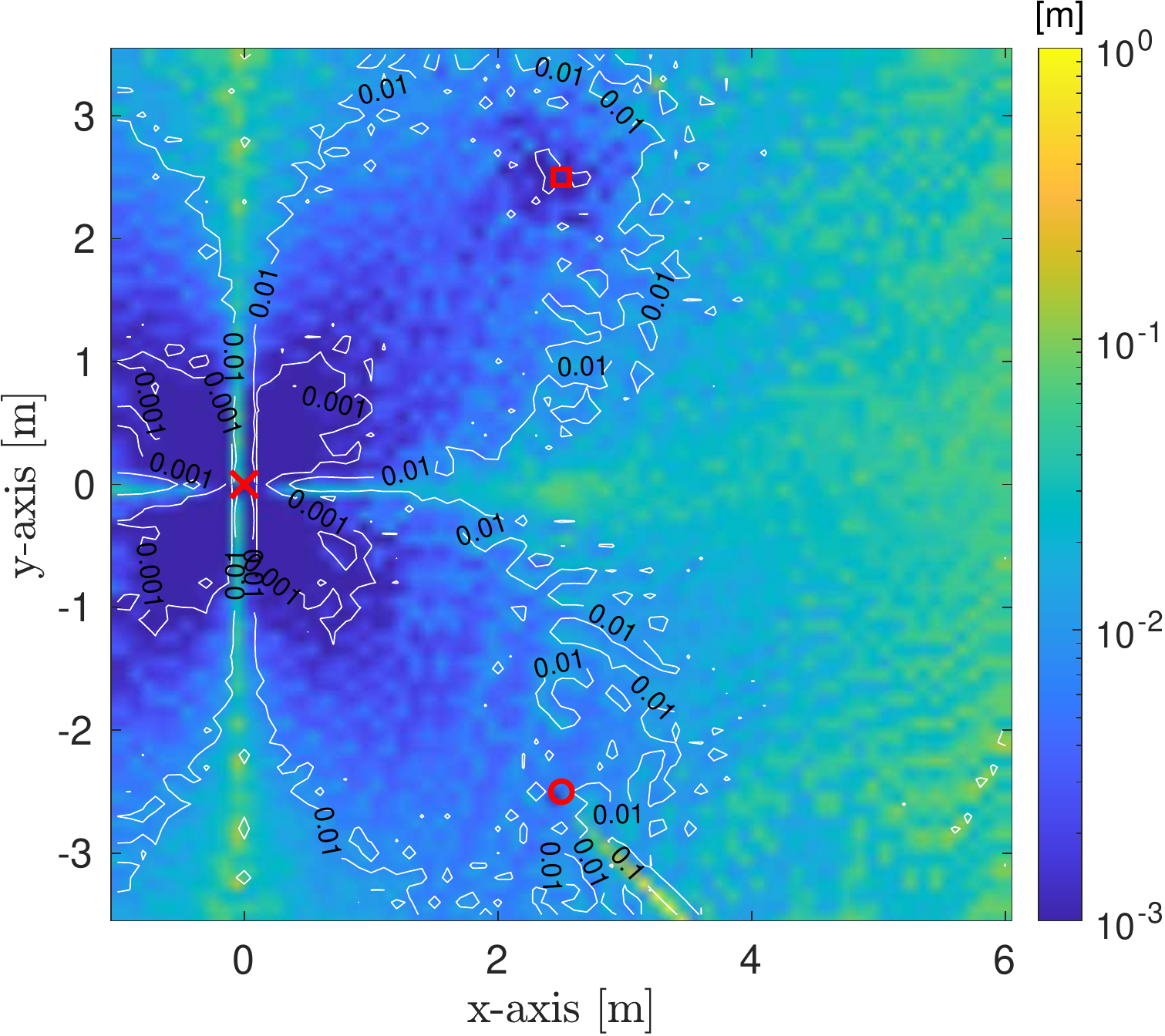}}
\centerline{(b) THz-1 (no prior)}\medskip
\end{minipage}
\hfill
\begin{minipage}[b]{0.48\linewidth}
  \centering
  \centerline{\includegraphics[width = 0.98\linewidth]{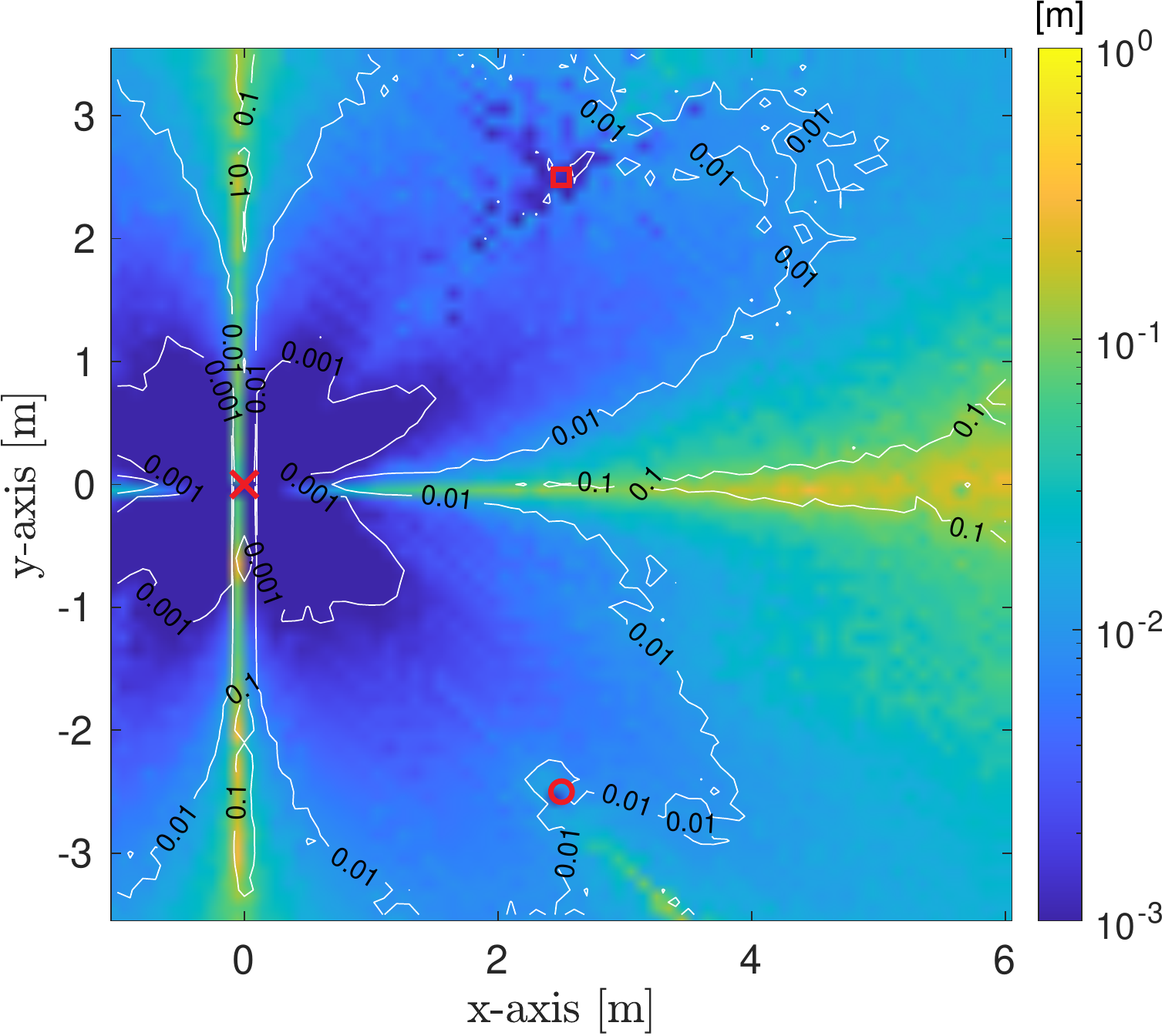}}
  \centerline{(c) THz-1 (with prior)} \medskip
\end{minipage}
\hfill
\begin{minipage}[b]{0.48\linewidth}
  \centering
  \centerline{\includegraphics[width = 0.98\linewidth]{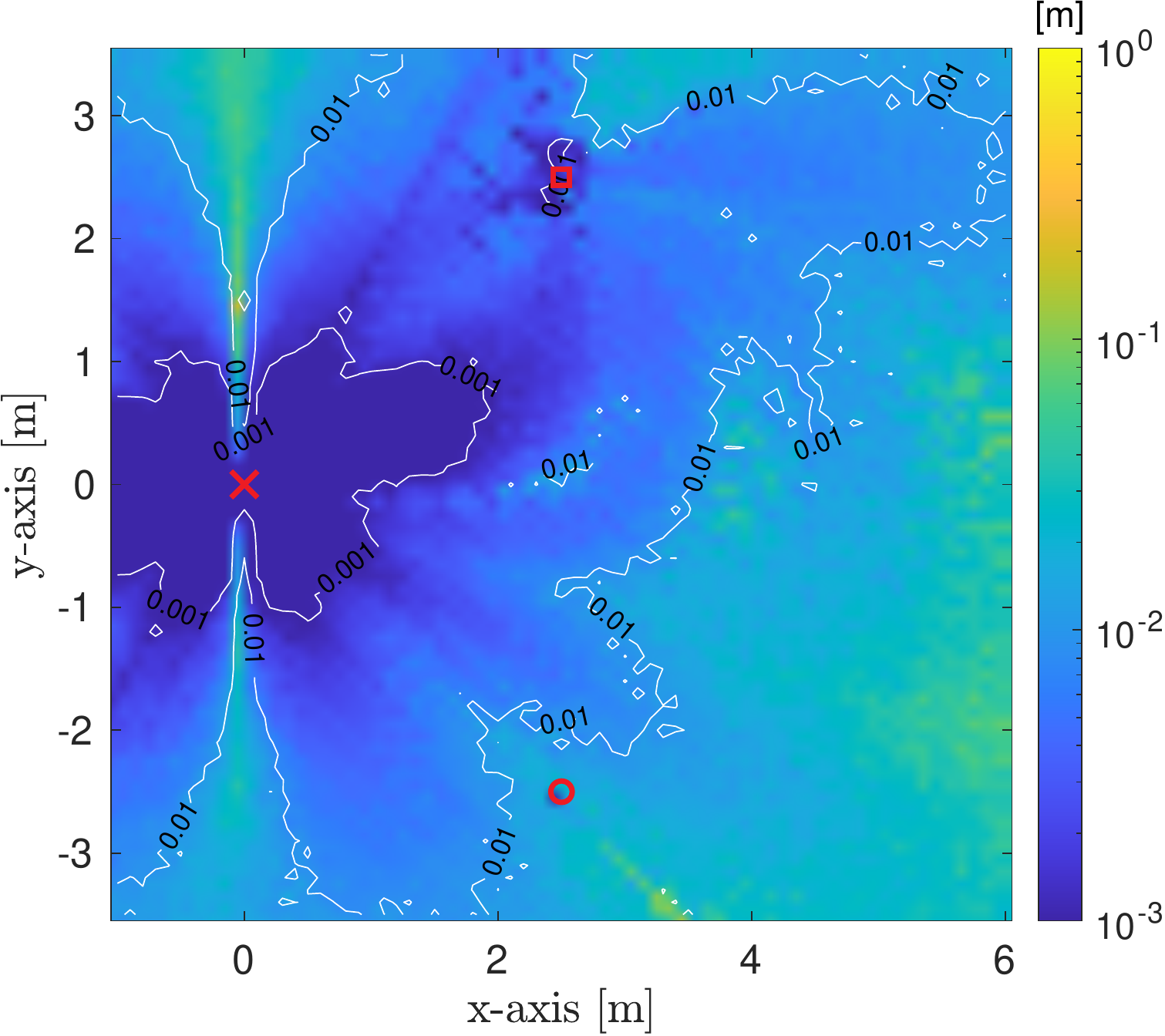}}
  \centerline{(d) THz-2 (with prior)} \medskip
\end{minipage}
\hfill
\caption{2D PEB visualization for different setups. (a) A conventional $4\times 4$ MIMO system and a $20\times 20$ RIS ($\unit[5\times5]{cm}$ footprint); (b) An AOSA-based THz system with $20\times20$ antennas ($\mathring N_\mathrm{B}=\mathring N_\mathrm{U} = 5\times 5$) and an $100\times 100$ RIS (same footprint as scenario (a)); (c) The same parameters as in (b) are used. By assuming the prior information of the UE position is known, the coefficients of the RIS elements are chosen to maximize the SNR, and 1/4 of the SA beams at the BS/UE are set to point to the RIS; (d) The same setup as (c) THz-1 by changing the UE orientation from $\ov_\mathrm{U}=[0, 0, 0]^T$ to $\ov_\mathrm{U}=[5\pi/6, 0, 0]^T$.}
\label{fig:PEB_OEB_visualization}
\end{figure}

In Fig.~\ref{fig:PEB_OEB_visualization} (a) and Fig.~\ref{fig:PEB_OEB_visualization} (b), the transmitted symbols, beamforming angles, and RIS coefficients are chosen randomly.
The THz system generally shows a lower PEB, and even lower PEBs appear when the UE is close to the BS or RIS. Due to the implementation of analog beamforming in the AOSA structure, there exist a `blind area' in (b) (e.g., $y_\mathrm{U}=0$).
In Fig.~\ref{fig:PEB_OEB_visualization} (c) and Fig.~\ref{fig:PEB_OEB_visualization} (d), fixed beamforming angles are utilized with prior information. This beam allocation strategy is obviously not optimal, and efficient beamforming optimization algorithms are needed.

\subsection{Summary}
In this section, we perform extensive simulations to illustrate the potential of THz systems in localization and sensing. With the incorporation of the RIS, a better localization performance is expected. However, joint optimization of AOSA active beamforming and RIS components is a challenging problem that requires more investigation in the future.
Next, we discuss several potential research directions for THz localization, which can assist in algorithm and system design and further improve localization performance.

\section{Lessons Learned and Future Directions}
\label{sec:future_directions}
{
Until now, we have discussed important topics of THz localization and performed extensive simulations with some interesting results observed. In this section, we would like to highlight the lessons learned from the simulations in Section~\ref{sec:simulation_and_evaluation}, and discuss the future directions from the aspects of channel modeling, localization, and optimization in Sections~\ref{sec:thz_system_model_and_properties}-\ref{sec:system_design_and_optimization}.
\subsection{Lessons Learned}
\label{sec:lessons_learned}
\begin{enumerate}
    \item Deal with SWM and PWM wisely in channel modeling and performance analysis. SWM requires high computational complexity since the phase change cannot be described using a simple steering vector, and there could be amplitude variations across the array. As a result, PWM is usually preferred, with possible performance loss in the near-field. From the CRB analysis point of view, the SWM and PWM are also different. In the far-field, the local AOA and AOD can be estimated directly, whereas the UE state is integrated into the channel model as shown in~\eqref{eq:measurement_vector_mmwave} and \eqref{eq:measurement_vector_Thz}.
    \item A similar concept of AOSA can be used in the RIS channel realization. A large number of RIS elements is needed to combat the high path loss of the RIS channel. This is even more challenging in high-frequency signals (e.g., $200\times 200$ RIS elements can be fitted into a $\unit[10\times10]{cm^2}$ area in a $\unit[0.3]{THz}$ system). With segmented sub-RIS and equivalent array response, the complexity can be reduced.
    \item Be aware of the model mismatch. We have seen the CRB mismatch between the SWM and the PWM models. Since PWM is an approximation, the model is inaccurate in the near field and should be avoided if possible. In addition to the channel model mismatch, the mismatch caused by the mobility of the UE and \ac{hwi} should also be considered. The \ac{mcrb} could be used as a tool to analyze such mismatches.
    \item Tradeoff between coverage and beamforming gain needs to be taken care of in system design. When designing a system, the directionality of the antenna and the size of SA should be considered to achieve a high beamforming gain. However, the gain in SNR will affect the coverage of the system, and severe misalignment will occur. These two aspects need to be well-designed depending on the application scenarios.
    \item Performance (CRB) analysis is an important step for algorithm evaluation and system optimization. However, the lower bound may not be reached in some scenarios, for example, low SNR scenarios and the existence of multiple non-resolvable paths. In addition to the CRB, other types of bound such as CCRB, and MCRB are also important in different scenarios.
\end{enumerate} 
}

\subsection{Future Directions}
\label{sec:future_direction_subsection}
{By moving from mmWave to THz, a better localization performance is expected. However, new issues and challenges need to be rethought to benefit from the system in this band. 
First, new KPIs may need to be defined for specific applications (e.g., quality of service rather than localization error). {Also, position integrity\footnote{Positioning integrity: is a measure of the trust in the accuracy of the position-related data provided by the positioning system and the ability to provide timely and valid warnings to the location service client when the positioning system does not fulfill the condition for the intended operation~\cite{tr22872}.} and availability\footnote{Availability: is defined as the fraction of the time that the estimated localization error is below the alert limit~\cite{hexax_d31}.} may become more important in localization, especially for the scenarios that need highly reliable position information.} In addition, we need to have methods that are scalable, given a large number of antennas/RIS elements and a large volume of data.
Regarding the RIS, this new enabler brings topics such as the synchronization to other network elements, information-sharing between operators, and the roles at different frequencies (should RIS operate in the same way at mmWave-band and THz-band or not). More problems will need to be tackled when we are moving to a higher frequency.

The research on THz localization is still at the early stage, with many directions to be explored. We list a few directions from the aspects of the channel model (1-2), localization performance analysis and algorithm design (3-6), and system optimization (7-9):

\begin{enumerate}
    \item Stochastic model analysis: We have formulated a deterministic channel model in this work. In realistic scenarios, however, the AOAs and amplitudes of scattered signals are stochastic. {The effect of randomized NLOS signals needs to be modeled, and the effect of scatters on the localization performance can be evaluated. The stochastic model is also helpful for object/reflector detection and classification in sensing tasks.}
    \item Accurate channel modeling: Channel model is the foundation of geometry-based localization. Currently, we use an extrapolation of mmWave models by introducing features in high-frequency systems such as atmosphere attenuation, SWM, wideband effect and AOSA structure. However, the effects of HWIs and other THz-specific aspects may not be captured in the channel model (including the LOS, RIS, and NLOS channel models). These model mismatches degrade the localization performance, and thus, a more accurate channel model is important.
    \item BS/RIS calibration. In most of the localization tasks, we are interested in the position and orientation of UEs by assuming the known anchor information (e.g., BS/RIS position and orientation). For the scenarios with more than one anchor, there could be calibration errors in the rest of the anchors compared to the reference anchor. In this case, jointly UE localization and BS/RIS calibration would be of great interest.
    \item Doppler estimation: In addition to position and orientation, the Doppler effect of the UE is not discussed in this work, which is crucial for mobile scenarios. This additional type of channel parameter can contribute to the tracking and SLAM tasks. In addition, by introducing the Doppler effect, localization can be done within a longer integration time.
    \item Cooperative localization. Coverage is one of the challenging issues in high-frequency communication and localization. As a result, D2D communication and cooperative localization can help even if there is an outage between the UE and the BS. As for SLAM, the collaboration between UEs can complete the mapping tasks within a shorter period of time.
    
    \item 
    Advanced performance analysis tools: We have discussed various types of CRBs (e.g., CRB, CCRB) for position and orientation estimation. Nevertheless, other bounds should be studied and enter more widespread use to account for phase ambiguities (e.g., in carrier phase-based localization), low-SNR operation, model mismatch, and with prior information. These advanced performance analysis tools will enable the algorithm development and system design for THz-band localization.
    
    \item Scene-aware localization: From the SLAM algorithm, surrounding map information could be available. In addition to the map, the probability density functions of past access locations can also be used to perform scene-aware localization. Beamforming vectors at the BS/UE and RIS element coefficients can be optimized to avoid obstacles and take advantage of the strong reflectors for localization purposes. In this topic, how to maintain a map with minimum resources and update the map with time is a problem to be solved.
    
    \item Dynamic deployment optimization. It is straightforward to optimize the deployment of static BSs for coverage or accuracy purposes. In temporal high traffic situations as in stadiums or conference halls, the BS could also be dynamically deployed, e.g., attached to UAVs. The location and route of the UAVs need to be optimized to meet the communication and localization requirements of the UEs. However, the connectivity in dynamic THz UAV networks is also challenging, which should be addressed carefully.
    
    \item AI-based methods: Model-based methods are easy to analyze, but when unknown model mismatches exist, AI-based methods are more preferred to learn or to mitigate the effect of such mismatches. In the latter case, access to common databases is needed to compare and evaluate different approaches. Furthermore, the collection, sharing, and storage of a large amount of data, transfer a learned model into another domain to reduce the training duration, and the protection of user privacy are urgent issues that need to be solved.
\end{enumerate}
}

{In summary, we need to put more effort in three aspects in order to improve THz localization accuracy and efficiency: (a) develop a more accurate system model (directions 1, 2, 3), (b) utilize other types of information (directions 4, 5, 7, 8), and (c) develop more advanced tools for analysis and optimization (directions 6, 9).}

\section{Conclusion}
\label{sec:conclusion}
This work explores the potential of the 6G THz system from a localization point of view, emphasizing comparisons with 5G mmWave localization systems. Comparisons include system and signal properties, channel modeling and assumptions, localization problem formulation, and system design and optimization. {Preliminary simulations are carried out to show the potential of THz localization compared with mmWave systems, in terms of the PEB and OEB.}
This tutorial outlines recommendations on efficient and practical localization algorithm design for RIS-assisted AOSA-based MIMO systems, providing insights into other research directions. With joint localization and communication systems operating at the terahertz band, data-hungry and high localization accuracy demanding applications such as intelligent networks, autonomous transportation, and tactile internet are anticipated in future communication systems.


\section*{Abbreviations}
\label{sec:abbreviations}
\small
\printacronyms[
display = used, 
heading = none, 
template = supertabular
]
\normalsize

\ifCLASSOPTIONcaptionsoff
\fi
\bibliographystyle{IEEEtran}
\bibliography{IEEEabb, reference}

\textbf{Hui Chen} received his B.S. degree in electrical engineering from Beijing Forestry University, Beijing, China, in 2013, the M.S. degree in computer application technology from the University of Chinese Academy of Sciences (UCAS), Beijing, China, in 2016, and the Ph.D. degree in electrical \& computer engineering from King Abdullah University of Science and Technology
\end{document}